\documentclass[useAMS,usenatbib,twocolumn]{mn2e}
\usepackage[dvipdfmx]{graphicx}
\usepackage{epstopdf}
\usepackage{txfonts}

\usepackage[T1]{fontenc}
\usepackage{ae,aecompl}
\usepackage{color}
\usepackage[FIGTOPCAP,nooneline,sc]{subfigure}
\usepackage{here}
\usepackage{multirow}

\usepackage{graphicx}	
\usepackage{amssymb}	

\def\aap{A$\&$A}

\newcommand{\beq}{\begin{equation}}
\newcommand{\beqa}{\begin{eqnarray}}
\newcommand{\eeq}{\end{equation}}
\newcommand{\eeqa}{\end{eqnarray}}

\newcommand{\lsim}{\la}

\newcommand{\lmk}{\left(}
\newcommand{\rmk}{\right)}

\newcommand{\III}{{I\hspace{-.1em}I\hspace{-.1em}I}}
\newcommand{\II}{{I\hspace{-.1em}I}}
\newcommand{\IV}{{I\hspace{-.1em}V}}
\newcommand{\VI}{{V\hspace{-.1em}I}}

\title{
Probabilistic eccentricity  bifurcation  for  stars around shrinking massive black hole binaries}

\author[M. Iwasa and N. Seto ]{Mao Iwasa \thanks{E-mail:
iwasa@tap.scphys.kyoto-u.ac.jp}
 and Naoki Seto
\\
Department of Physics, Kyoto University, 
Kyoto 606-8502, Japan
}

\date{Accepted XXX. Received YYY; in original form ZZZ}

\pubyear{2016}

\begin{document}

\maketitle
\begin{abstract}

Based on the secular theory, we discuss the orbital evolution of stars in a nuclear star cluster to which a secondary massive black hole is infalling with vanishing eccentricity.  We find that the eccentricities of the stars  could show sharp transitions, depending strongly on their initial conditions.   By examining the phase-space structure of an associated Hamiltonian,  we show that these characteristic behaviors are partly due to a probabilistic bifurcation at a separatrix crossing, resulting from the retrograde 
apsidal precession by the cluster potential.   We also show that  separatrix crossings are closely related to  realization of a large eccentricity and could be important for  astrophysical phenomena such as tidal disruption events or gravitational wave emissions.


\end{abstract}


\section{Introduction}
The Kozai-Lidov (KL) mechanism is a well known effect for hierarchical 
triple systems. It oscillates the inner eccentricity and inclination, as a result of the angular momentum exchange between the inner and outer orbits. The KL mechanism was originally examined for asteroids \citep{kozai1962} and satellites \citep{lidov1962} by using the secular equations that is derived after averaging the mean anomalies of the two orbits. Since then,  the KL mechanism has been applied to various astronomical contexts, such as evolution of triple main sequence stars \citep{ford2000,fabrycky2007,naoz2014,borkovits2016,toonen2016}, orbits of exoplanetary systems \citep{holman1997,ford2000,nagasawa2008,naoz2011,munoz2016}, accelerated evolution of gravitational wave sources for ground based detectors \citep{wen2003,antonini2012,seto2013,antognini2014,antonini2014,silsbee2016}, collisions of stars \citep{perets2009,katz2012,thompson2011,kushnir2013}, evolution of triple massive black hole (MBH) binaries \citep{blaes2002,hoffman2007,iwasawa2011} and so on.
In addition, hierarchical four-body systems have been discussed quite recently \citep{pejcha2013,hamers2016,hamers2017}.

 While the two original works \citep{kozai1962,lidov1962}  were made under relatively simple theoretical framework and  orbital setting, advanced effects have been also studied. For example,  the impacts of the outer eccentricity  (\citealt{naoz2016}, see also \citealt{shapee2013,michaely2014}) and the potential shortcoming of the orbital averaging scheme \citep{bode2014,luo2016} have been extensively discussed in the last five years. These two aspects could be important for highly eccentric inner orbits.

The KL mechanism has been examined also for nuclear star clusters \citep{ivanov2005,wegg2011,chen2011,bode2014,li2014,iwasa2016,stephan2016}.
Nowadays, almost all galaxies are considered to have MBHs in their nuclei \citep{ferrarese2005}. If two galaxies merge, the distance between their two central  MBHs would be continuously decreased by dissipative processes, and the two MBHs are likely to coalesce in the end \citep{merritt2013}.  
Along the way, the orbits of stars in the nuclear star cluster around each MBH would be dynamically affected by the other MBH. Here, the KL mechanism could play a significant role  for enhancing  the tidal disruption rates
 or observable gravitational wave signals. 

For example, \cite{bode2014} examined evolution of such nuclear star clusters by numerical simulations, mainly during the stages when the distance between the two MBHs decreases relatively rapidly.  \cite{li2014} analytically studied the individual orbits of stars, by setting the distance between the two MBHs at various values (without continuous variation). \cite{iwasa2016} analyzed how the slow and continuous contraction of the distance modifies the orbital elements of the stars. They separately included the post-Newtonian effects of the central black hole and the gravitational potential of the nuclear star cluster itself.  Their analysis is based on a geometrical approach with a help of the adiabatic invariant in a time evolving phase-space \citep{landau1969,ssd}. They reported that, when the cluster potential is included, the individual orbits of the stars could show peculiar transitions and the evolved  eccentricities could have an inverted  correspondence to the initial eccentricities \citep[see Fig. 3 in][]{iwasa2016}.

In this paper, we continue our study on the orbital evolution of  nuclear star clusters, now simultaneously including the post-Newtonian effects and the cluster potential.  The resultant phase-space becomes  more complicated.  But, interestingly, we newly identified a probabilistic bifurcation of the inner orbital eccentricities at a separatrix crossing.  Below, still using the geometrical approach, we carefully examine how this bifurcation works.

Here, we briefly mention a  possible implication of this work to theoretical studies on orbital dynamics. 
Analyses for mean motion resonances have been one of the central topics in the field \citep{goldreich1965,sinclair1972,yoder1973,henrard1983,peale1987,ssd,lithwick2012,fabrycky2014,goldreich2014,batygin2015}. Indeed, the simple dynamical model around the resonant capture is an impressive achievement in the theory of orbital dynamics \citep[see {\it e.g.}][]{henrard1982,borderies1984,ssd}.  Even though the phenomenon discussed in this paper are purely based on the secular theory without depending on mean anomalies, the underlying physics have similarities to the dynamics of resonant capture. In fact, our geometrical approach owes much to its successful applications to the mean motion resonances. We expect that our detailed study would inversely help us to better understand the mean motion resonances and the related theoretical techniques, from a wider point of view. 

This paper is organized as follows. In \S 2, we describe our astronomical model and present the orbitally averaged Hamiltonian. We also  discuss our system from astronomical viewpoints for nuclear star clusters, rather than the orbital dynamics.  In \S 3, we present numerical examples to demonstrate the characteristic features at  separatrix crossings.  In \S 4, we analyze the  phase-space structure, paying special attentions to the evolution of fixed points and separatrixes. Then, in \S 5, we discuss the probabilistic bifurcation at a separatrix crossing and also analyze the large eccentricities observed  during orbital evolutions. \S 6 is a short summary  of this paper.

\section{Description of our model}

\subsection{assumptions and settings}

\begin{figure*}
\begin{center}
\includegraphics[width=12cm]{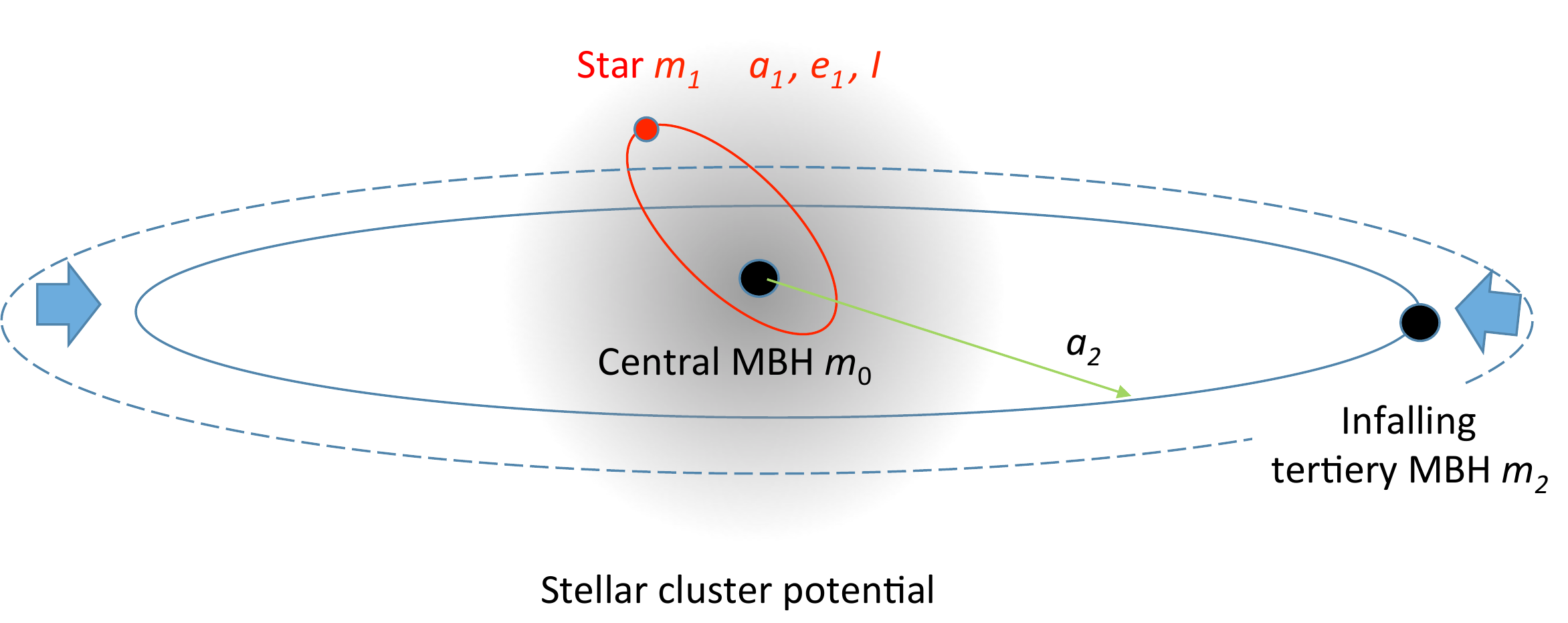}
\caption{Configuration of our system. A star $m_1$ moves around
the primary MBH $m_0(\gg m_1)$, perturbatively affected by (i) the  1PN effect of the primary MBH $m_0$ and 
(ii) the gravitational potential of the nuclear star cluster and (iii) the tidal field of the tertiary MBH $m_2$.
 We denote its orbital elements by $a_1$ (the
semimajor axis) and $e_1$ (the eccentricity). The secondary MBH
$m_2$ slowly inspirals to the primary MBH with vanishing
eccentricity $e_2=0$. The parameter $I$ represents the
inclination between the inner and outer orbits.
 The subscripts 1 and 2 denote the
inner and outer orbital parameters.
}
\label{fig:sys}
\end{center}
\end{figure*}

As shown in Fig. \ref{fig:sys}, we deal with a system composed
of the following three elements; (i) the primary MBH $m_0$, (ii)
the associated nuclear star cluster, and (iii) the infalling secondary MBH
$m_{2}$  with
vanishing eccentricity. Our main interest is the evolution of individual stars
in the cluster, during the inspiral of the secondary MBH  (see also \citealt{merritt2013} for a potential role of the star clusters surrounding the perturber MBH $m_2$).

For simplicity, we assume that the star cluster is stationary and  spherical, and has
 an isotropic velocity distribution. In addition, we ignore
the direct gravitational interaction between stars, and only
include their mutual interaction through a smooth and stationary stellar
potential (see \cite{iwasa2016} and Appendix A for the stationarity).
Then we can  examine the evolution of individual stars separately, as
if we merely deal with a hierarchical triple system formed by two 
MBHs and a single star (of course, including the
cluster potential).  The dynamics of the star $m_1$ is mainly controlled by
the Newtonian potential of the primary MBH $m_0$, but is perturbatively affected 
by its post-Newtonian effect and the gravitational potentials of the stellar cluster   
as well as the tertiary MBH $m_2$.  These three perturbative effects will appear
as different terms in our secular Hamiltonian, and their competition generates
interesting effects.

Here, we briefly summarize our notations. We apply the suffix j
for the inner (j$=1$) and the outer (j$=2$) orbital elements. We denote
the semimajor axes by $a_{\rm j}$, the eccentricities by
$e_{\rm j}$, and the arguments of pericenters by $g_{\rm j}$.
The angle $I$ represents the inclination between the inner and
outer orbits.
We also define the dimensionless inner angular momentum  
\beqa
G_1\equiv\sqrt{1-e_{1}^{2}},
\eeqa
and its component orthogonal to the outer orbital plane 
\beqa
J_1\equiv G_1\cos I.
\eeqa
Due to the symmetry, we can fix the outer orbital plane, during its orbital decay. 
Meanwhile, in our
analysis based on the secular theory, the inner semimajor axis
$a_1$ as well as the projected angular momentum $J_1$ stay constant (as explained in \S 2.2).
As mentioned earlier, we put $e_2=0$.
 
Next, we discuss the density profile of the spherical 
cluster.
For the star $m_1$ with a fixed semimajor axis $a_1$,
we only need the density profile around the distance $r=a_1$ from 
the primary MBH.
We adopt  a power-law model that is parameterized as 
\beqa
\rho(r)=\rho_1 (r/a_1)^{-\beta},
\eeqa
where $\rho_1$ is the cluster density at $r=a_1$ and the index
$\beta$ is in the range $0\le \beta< 3$ \citep{merritt2013}. To be concrete, we
hereafter put $\beta=3/2$, as the fiducial value.

The stellar potential causes the apsidal precession of a star in
the retrograde direction, and its characteristic timescale
$T_{\rm SP}$ \citep{merritt2013} is given by
\beqa
T_{\rm SP}&\equiv& \frac{(3-\beta)m_0}{4\pi
\sqrt{1-e_{1}^{2}}\rho_1 a_{1}^{3}}P_1\label{eq:sp}
\eeqa
with the inner orbital period $P_1\simeq 2\pi \sqrt{a_1^3/Gm_0}$.

The star $m_1$ is also affected by the first post-Newtonian (PN) effect by $m_0$. 
It causes the apsidal precession in the prograde direction \citep{holman1997,ford2000,merritt2013}
with a characteristic timescale 
\beqa
T_{\rm1PN}&\equiv&
\frac{a_{1}^{5/2}(1-e_{1}^{2})}{3m_{0}^{3/2}}.
\eeqa

The outer MBH has a perturbative effect on the inner orbit (for
$a_1\ll a_2$). Without the stellar potential and the 1PN
correction, we can reproduce the traditional KL mechanism. More
specifically, the inner eccentricity and inclination oscillate
with the characteristic timescale $T_{\rm KL}$
\citep{holman1997,kinoshita1999,ford2000,fabrycky2007,antognini2015},
\beqa
T_{\rm
KL}&\equiv&\frac{2}{3\pi}\frac{(m_{0}+m_{2})}{m_2}\frac{P_{2}^{2}}{P_{1}}.
\eeqa
Here, $P_2$ is the outer orbital period.

For the initial conditions of our system, we take a large outer distance $a_2$ to
completely suppress the KL mechanism with 
$T_{\rm KL}\gg{\rm min}[T_{\rm SP},\,T_{\rm 1PN}]$. But, along with the contraction
of
the outer orbit, the KL mechanism gradually becomes stronger. We will find
various interesting phenomena in midstream.

\subsection{Averaged Hamiltonian}

In this paper, we focus on the long-term evolution of the inner
orbit. To this end, we apply the standard secular theory to our
triple systems. By the von-Zeipel canonical transformation, we
can take the orbital averages with respect to the inner and
outer mean anomalies \citep{harrington1968,ford2000,blaes2002}.
After some algebra including an appropriate scaling, we obtain
the dimensionless Hamiltonian
\beqa 
\mathcal{H}_{\rm T}(g_{1}, G_{1})=\mathcal{H}_{\rm
qp}+\mathcal{H}_{\rm SP}+\mathcal{H}_{\rm 1PN}
\label{ham}
\eeqa
for the inner orbital elements $g_1$ and $G_1$ (composing the
conjugate variables).
Other dynamical variables such as the inner mean anomaly and the
longitude of the inner ascending node do not appear in our Hamiltonian.
Both $a_1$ and $J_1$ are conjugate to these two  cyclic variables
 and conserved in our study.

The three terms in the Hamiltonian (\ref{ham}) are given by
\beqa
\mathcal{H}_{\rm qp}&\equiv&
-3G_{1}^{2}-15\frac{J_{1}^{2}}{G_{1}^{2}}-15(1-G_{1}^{2})\left(1-\frac{J_{1}^{2}}{G_{1}^{2}}\right)\cos
2g_{1},\label{hq}\\
\mathcal{H}_{\rm SP}&\equiv&-12\eta
(1-G_{1}^{2})\left[1+\beta(-1+\beta)\left(\frac{1-G_{1}^{2}}{16}+\mathcal{O}(e_{1}^{4})\right)\right],\label{hs}\\
\mathcal{H}_{\rm1PN}&\equiv&\frac{24}{G_{1}}p\eta.\label{h1}
\eeqa
Here, $J_1$ is an integral of motion and $\beta$ is the power-law index
of the stellar cluster (see Eq. (3)). 
Since the parameter $J_1$ appears only through the form $J_1^2$, we can limit
$J_1\ge 0$ without loss of generality, and the variable $G_1$
is bounded by $ J_1\le G_1\le 1$.

In addition
 to the dynamical variables $(g_1,G_1)$,  our Hamiltonian  contains two important parameters $\eta$ and $p$
defined by \footnote{Our definition for $\eta$ is different from \cite{iwasa2016} by a factor of $-12$. }
\beqa
\eta&\equiv&-\frac{4\pi\rho_{1}}{3m_{2}}a_{2}^{3}<0, \\
p&\equiv&\frac{3m_{0}^{2}}{2\pi\rho_{1}a_{1}^{4}}>0.\label{eta}
\eeqa
We will shortly explain their physical meanings.

In our Hamiltonian (\ref{ham}),
the first term $\mathcal{H}_{\rm qp}$ represents the quadrupole
coupling between the outer MBH and the inner orbit
\citep{fabrycky2007}. We neglect the higher order
couplings, because the triple system is hierarchical and the
outer orbit is assumed to be circular.

In Eq. (\ref{ham}), the second term $\mathcal{H}_{\rm SP}$ originates from the Newtonian
potential of the spherical stellar cluster \citep{merritt2013} and is obtained by expanding a
hypergeometric function with the variable $e_1=\sqrt{1-G_{1}^2}$, as in Eq. (\ref{hs}). With respect to this expansion, we include the higher order terms
$O(e_{1}^2)$ for our numerical calculation in \S 3, but we
only keep the leading-order term $-12\eta (1-G_1^2)$ for our
analytical arguments in \S 4 and 5 (thus $\mathcal{H}_{\rm T}$ is independent of $\beta$).
 Actually, as demonstrated in \S 3,   this truncation works quite well for physically relevant range $0\le \beta \le 3$.

Our Hamiltonian depends on the  outer semimajor
axis $a_2$ through the parameter $\eta$.
Note that  we have $\eta \propto T_{\rm KL}/T_{\rm SP}$ (ignoring the 
$e_1$-dependence).
Indeed, the parameter $\eta$ represents the strength of the cluster potential
relative to the quadrupole coupling between the star and the outer MBH.
This parameter increases with time from $\eta=-\infty$ (at
$a_2=\infty$) to $\eta=0$ (formally at $a_2=0$). In our study, the parameter
$\eta$  works as an effective time variable showing the contraction
stage of the outer orbit.
Therefore, we hereafter express the total  Hamiltonian (\ref{ham}) by
 $\mathcal{H}_{\rm T}(g_{1}, G_{1};\eta)$.

The last term $\mathcal{H}_{\rm 1PN}$ is the first order PN term
\citep{blaes2002}. In addition to $\eta$, the
parameter $p$ plays important roles in our study. We have
$p\propto T_{\rm SP}/T_{\rm 1PN}$ (again ignoring the $e_1$-dependence). 
Therefore, this parameter shows the strength of the 1PN effect
relative to the star cluster potential. As we have $a_1=const$
in our secular analysis, it does not evolve with
time ($p=const$).

The Hamiltonian approach has been a powerful method to study the dynamics of mean motion resonances  \citep{henrard1982,borderies1984,ssd}. In that case, usually, the primary orbital parameters ({\it e.g.} eccentricity and inclination) are perturbatively handled with a help of disturbing function to evaluate the interaction between orbits \citep{ssd}. But, in our framework based on the expansion parameter $a_1/a_2$, we can  deal with a large eccentricity $e_1$ and a large relative inclination $I$.  Indeed, as we see below, the maximum eccentricity corresponding to $G_1=J_1$ plays a critical role for our phase-space evolution. This non perturbative handling of the basic orbital elements is a notable advantage of the present study. 

In our preceding paper \cite{iwasa2016}, we separately added the terms  $\mathcal{H}_{\rm 1PN}$ and  $\mathcal{H}_{\rm sp}$ to the quadrupole term  $\mathcal{H}_{\rm qd}$. Namely, we examined the two Hamiltonians $\mathcal{H}_{\rm qp}+\mathcal{H}_{\rm SP}$  and $\mathcal{H}_{\rm
qp}+\mathcal{H}_{\rm 1PN}$.  But, as we see below, the competitions between the three terms enrich the phase-space structure, resulting in notable evolution of orbital elements.

\subsection{Physical scales and Relaxation effects}

In this paper, we basically proceed our study with the
scaled Hamiltonian (\ref{ham}). Therefore, our main arguments
are somewhat abstract.
But, before going into details, in this subsection, we evaluate
the actual magnitude of the dimensionless parameters $p$ and
$\eta$, using fiducial astrophysical systems. We also make a brief
discussion about the relaxation effects on the inner orbit.

\begin{figure}
\begin{center}
\includegraphics[width=8cm]{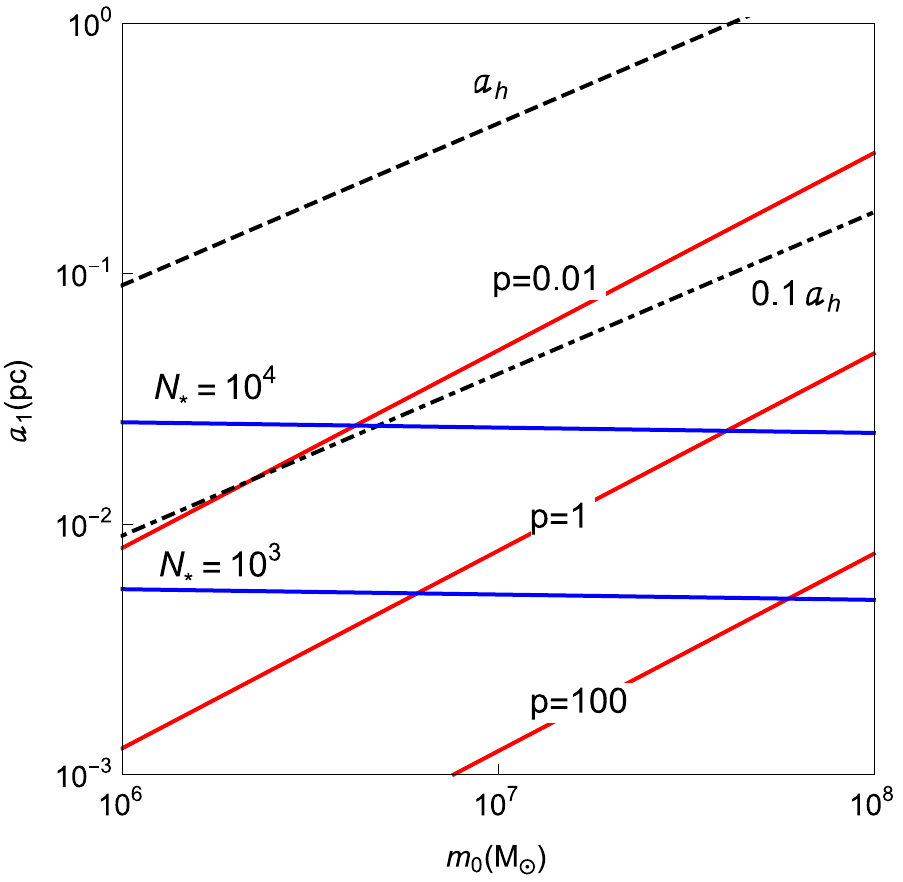}
\caption{The contour plots for the parameter $p$ (red lines) and the number of enclosed
stars $N_{\ast}$ (blue lines). We also show the semimajor axis $a_1$ in units of
the hard binary separation  $a_h$ 
(for the mass ratio $q=1$).}
\label{fig:pcont}
\end{center}
\end{figure}

First, based on standard references (see {\it e.g.} Merritt 2013),  we re-express the  density profile of the nuclear star cluster as follows
\beqa
\rho(r)=\frac{3m_0}{4\pi r_{\rm h}^{3}}\left(\frac{r}{r_{\rm
h}}\right)^{-3/2},\label{3/2} \label{de1}
\eeqa
where $r_{\rm h}$ is the influenced radius of $m_0$ and is given by
\beq
r_{\rm h}\equiv \frac{Gm_0}{\sigma^2}\label{de2}
\eeq
 with the one-dimensional velocity
dispersion $\sigma$. In addition, we apply the
$M_{\bullet}$-$\sigma$ relation \citep{2013ApJ...764..184M}
\beq
M_{\bullet} (=m_0) =2.09\times10^8\lmk \frac\sigma{200{\rm
km/s}}\rmk^{5.64}M_{\odot}. \label{de3}
\eeq
Using Eqs. (\ref{de1})-(\ref{de3}), we can fix the density profile $\rho(r)$ for a given MBH mass
$m_0$.

For our fiducial cluster model, the parameter $p$ is written as
\beqa
p=\frac{G}{c^2}\frac{3m_{0}^2}{2\pi
\rho(a_1)a_{1}^4}=5.04\times10^{-4}\left(\frac{a_{1}}{1{\rm
pc}}\right)^{-5/2}\left(\frac{m_0}{10^8 M_{\odot}}\right)^{1.97}.
\eeqa
In Fig.2,  we present a contour  plot for $p$.

In order to provide a rough idea about the number of stars
corresponding to a given parameter $p$, we define the integral 
\beqa
N_{\ast}(<a_1)=\frac{1}{m_\ast}\int_{0}^{a_1} 4\pi r^2 \rho(r)
dr,
\eeqa
which approximately represents the total number of stars with
semimajor axis less than $a_1$.
Here, $m_\ast$ is the typical mass of stars in the
cluster.
In Fig.2, we show the enclosed numbers  $N_{\ast}(<a_1)$,  by 
setting $m_\ast=1M_\odot$.

Meanwhile, the parameter $\eta$ depends also on the outer semimajor axis $a_2$.
 In order to specify its typical range relevant  for our
study, we briefly introduce  the standard arguments on the
orbital decay of a MBH binary \citep{merritt2013}.

After the merger of two galaxies, the distance between their
central MBHs decreases due to  dynamical friction and
sling-shot ejections of  stars.
But, when the binary separation decreases down to the so-called
hard binary separation
\beqa
a_{\rm h}&\equiv&\frac{m_{2}}{m_0+m_2}\frac{r_{\rm
h}}{4}\nonumber\\
&=&3.51\frac{q}{q+1}\left(\frac{m_0}{10^8M_{\odot}}\right)^{0.645}{\rm
pc},
\eeqa
($q \equiv m_2/m_0$), the infall time is considered to increase
significantly, though its actual value is highly uncertain (also
depending strongly on the environment around the binary).

In our theoretical framework based on the adiabatic invariant 
(explained in \S 5),
the slow contraction of the outer orbits is essential.
Therefore, below, we consider the range $a_{2}\lsim a_{\rm h}$
for the outer orbit.

Once the outer distance $a_2$ is given, the dynamical stability
of the triple system imposes the hierarchical
orbital configuration $a_1\lsim0.1a_2$ for the inner orbit, assuming comparable MBH masses $q=m_2/m_0\sim 1$ \citep{mardling2001}.
Therefore, in Fig.2, our target star should have  semimajor axis $a_1\lsim0.1 a_{\rm h}$. 

Next, we examine the parameter $\eta$. 
It is written in terms of $a_2$ and $a_1/a_2$ as
\beqa
\eta&=&-\frac{m_0}{m_2}\left(\frac{a_2}{r_{\rm
h}}\right)^{3/2}\left(\frac{a_1}{a_2}\right)^{-3/2}\nonumber\\
&=&-44.3 \lmk \frac{F(q)}{F(1)}\rmk
\left(\frac{a_2}{a_{\rm
h}}\right)^{3/2}\left(\frac{a_1/a_2}{0.01}\right)^{-3/2},
\eeqa
with $F(q)\equiv \sqrt{q/(1+q)^3}$.  We should recall that
the inner semimajor axis $a_1$ stays constant in our secular
analysis. With this expression, we can read how the parameter
$\eta(<0)$ increases (from a large negative value), along with
the contraction of the outer radius $a_2$.

Finally, we comment on the relaxation processes that will not be
handled in our main arguments.
 Here,   following our
previous paper \citep{iwasa2016}, we concentrate on
the resonant relaxation.
This is because, in the nuclear star clusters, the timescale of resonant relaxation
  is generally much smaller than that of the
two-body relaxation.

The resonant relaxation is a
diffusion process in the angular momentum space \citep{1996NewA....1..149R,alexander2005,2011MNRAS.412..187K}. It is
classified into two categories; the scalar and vector types. While
the scalar resonant relaxation changes both the magnitude and
orientation of the angular momentum, the vector resonant
relaxation changes only the orientation of the angular momentum.

In our previous paper \citep{iwasa2016}, we examined
the impact of the outer MBH on these two types. We
showed that the vector type is not effective around the
evolutionary phase in interest ({\it e.g.}  at the separatrix
crossing) due to the overall precession around the symmetry axis normal to the outer orbital plane.
 In contrast, the scalar resonant relaxation could
become effective, if its characteristic timescale $T_{\rm rr,s}$
is smaller than the infall time $T_{\rm
inf}=|a_{2}/\dot{a}_{2}|$.
 When the inner apsidal precession is
dominated by the cluster potential, the timescale $T_{\rm rr,s}$ is explicitly given by
\beqa
T_{\rm rr,s}&=&\frac{m_{0}}{m_{1}}P_{1}\nonumber\\
&=&1.6\times10^{9}\left(\frac{1M_{\odot}}{m_1}\right)\left(\frac{m_{0}}{10^{8}M_{\odot}}\right)^{1/2}\left(\frac{a_1}{0.3
{\rm pc}}\right)^{3/2}{\rm yr}. \nonumber\\
\label{rrs}
\eeqa
We can ignore the resonant relaxation, if the infall time satisfies the following condition
\beq
T_{\rm inf}<T_{\rm rr,s}.
\eeq

\begin{figure}
\begin{center}
\includegraphics[width=8cm]{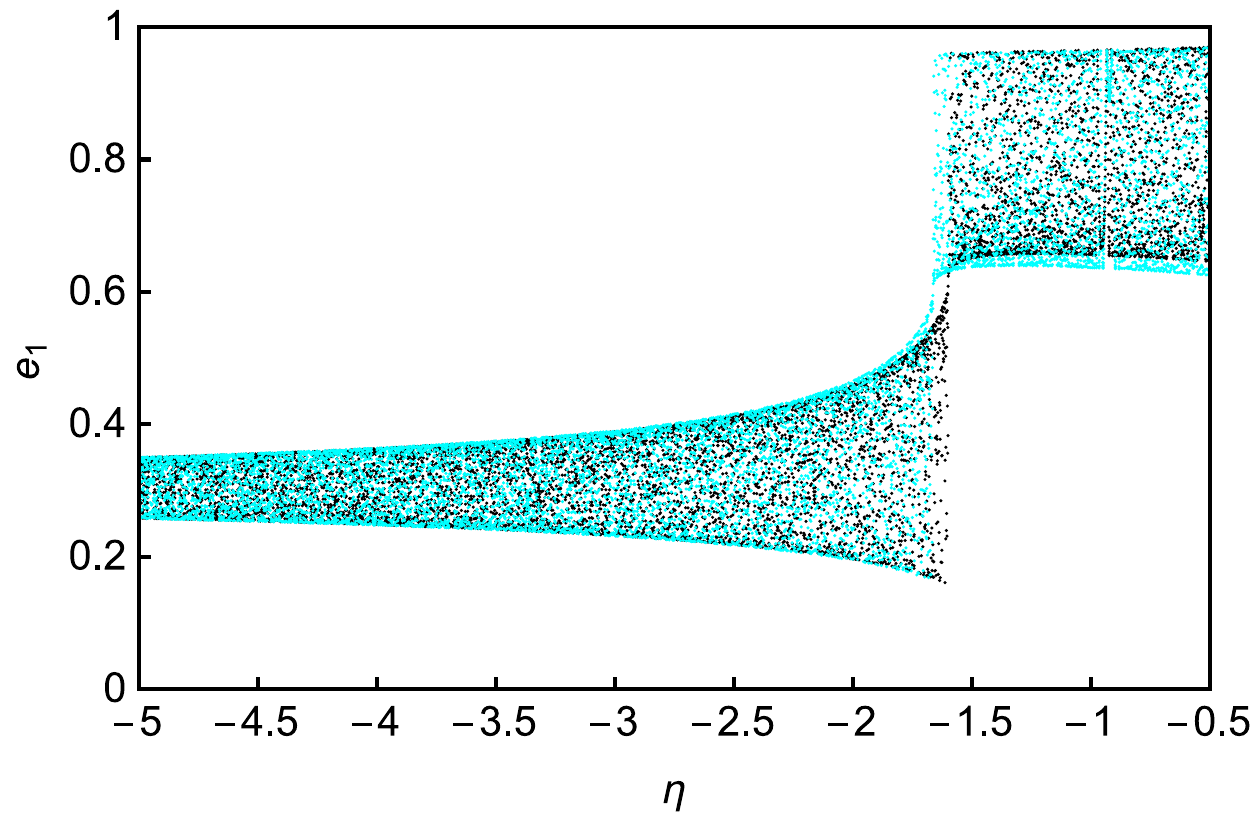}
\includegraphics[width=8cm]{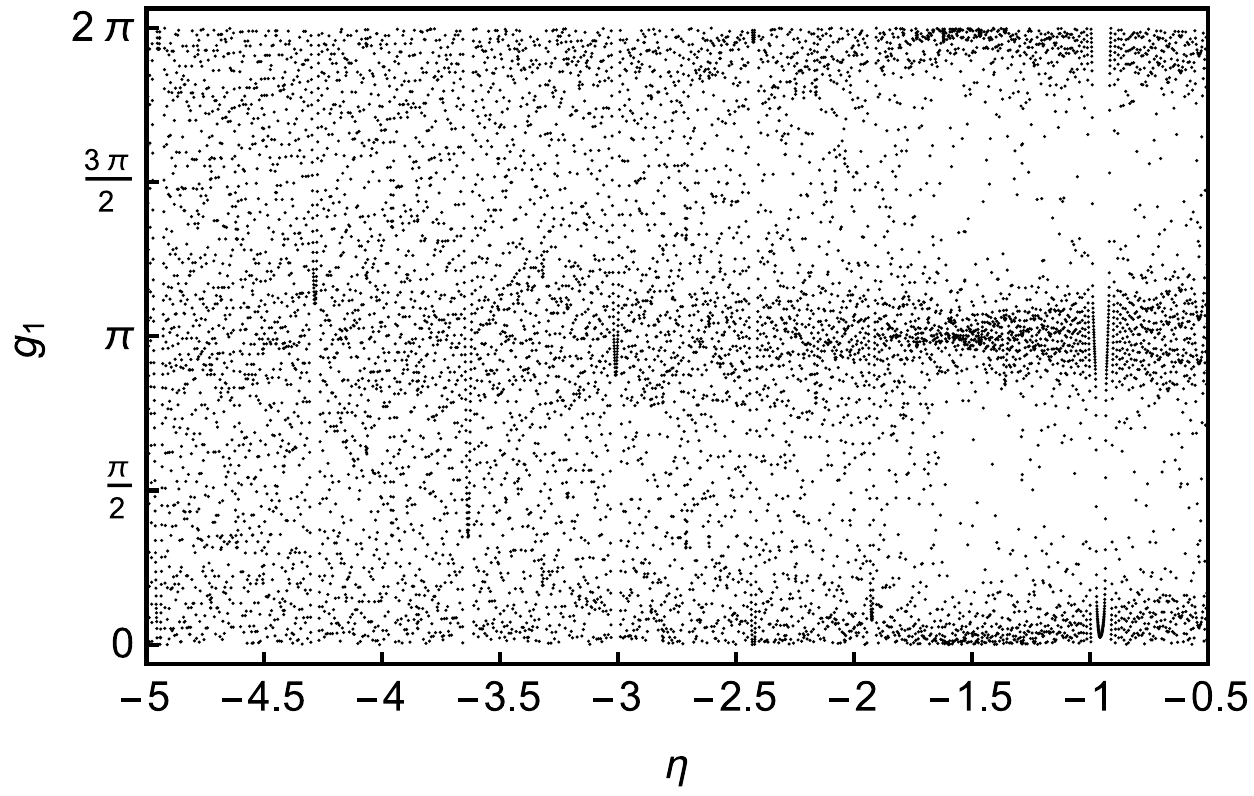}
\caption{(Black points) Numerical results for the run R1, including the higher order terms for $\mathcal{H}_{\rm SP}$ in Eq. (\ref{hs}) with $\beta=1.5$. The initial    condition is
 $(g_{1},\, G_{1} )=(0.03,\, 0.950)$ at $\eta=-20$ with 
the time-independent parameters
$(J_1,\,p)=(0.2,\,0.2)$. We show the inner eccentricity $e_1$  and the
 argument of
pericenter $g_1$  as a function of the effective time $\eta$.
We can observe the single transition at $\eta\simeq-1.65$.
The short-term vertical patterns ({\it e.g.} around $\eta=-3.60$) are artificially caused 
by our data sampling scheme.
(Cyan points) Numerical results for the run R3, only with the lowest order terms for $\mathcal{H}_{\rm SP}$ (corresponding to $\beta=1$). The initial condition is   $(g_1,G_1)=(0.04,0.950)$ at $\eta=-20$.
}
\label{fig:evo3}
\end{center}
\end{figure}

\begin{figure}
\begin{center}
\includegraphics[width=8cm]{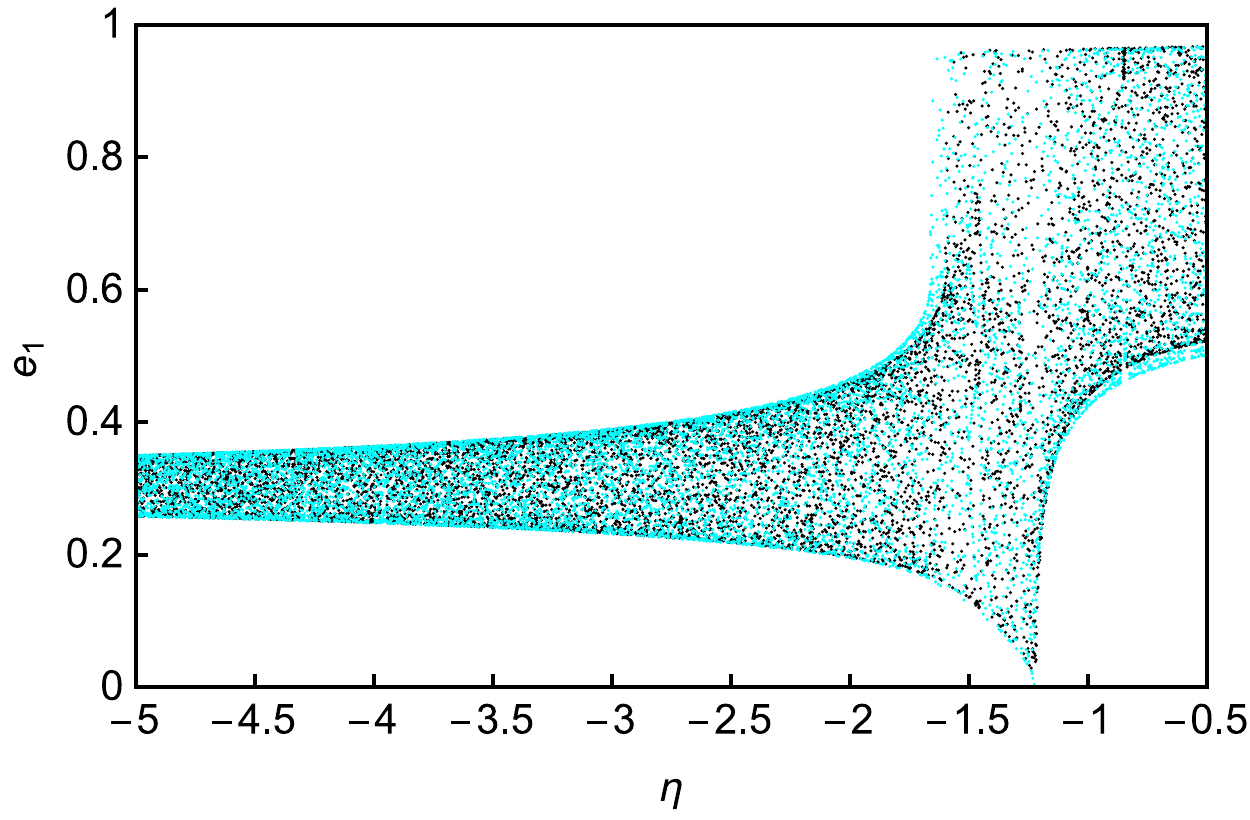}
\includegraphics[width=8cm]{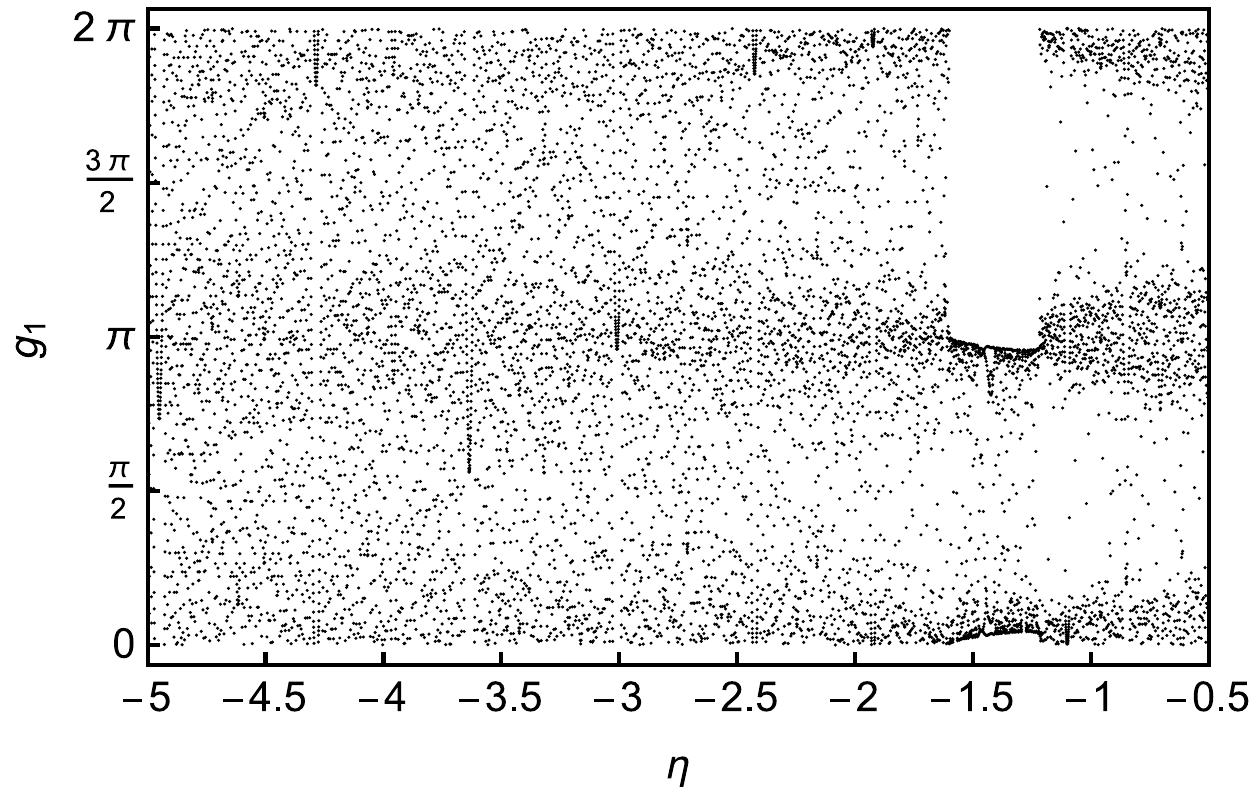}
\caption{(Black points) Numerical results  for the run R2, including the higher order terms for $\mathcal{H}_{\rm SP}$.   The initial condition  $(g_1,G_1)=(0,0.950)$ is slightly different from R1 (shown in  Fig. \ref{fig:evo3}). 
We have two transitions at $\eta\simeq-1.65$ and $-1.20$.
Between them, the
angle $g_1$ librates around $\pi/2$. (Cyan Points) Numerical results for the run R4, only with the lowest order terms for $\mathcal{H}_{\rm SP}$. The initial condition is   $(g_1,G_1)=(0.01,0.950)$.}
\label{fig:evo4}
\end{center}
\end{figure}

\section{Numerical Examples}
 In this paper,   our primary objective is to analytically examine the Hamiltonian
(\ref{ham})
and related dynamics.
But, numerical demonstrations would be also helpful to provide intuitive pictures of
our targets. 
In this section, we first show results for the two numerical runs R1 and R2 that have nearly identical initial conditions but later evolve in entirely different
ways.

The two runs have the same time-independent  parameters $(p,J_1)=(0.2,0.2)$.  At the effective time
$\eta=-20$, we set their initial conditions
$(g_{1},\, G_{1} )=(0.03,\, 0.950)$ for R1 and $(0,\, 0.950)$ for R2. 
Thus, only the initial phases $g_1$ are slightly  different.

For the two runs R1 and R2, we included the higher order corrections for the terms $\mathcal{H}_{\rm SP}$ in Eq. (\ref{hs}), and  numerically integrated the canonical equations 
\beq
\frac{dg_{1}}{dt}=\frac{\partial \mathcal{H}_{\rm T}}{\partial G_{1}},~~
\frac{dG_{1}}{dt}=-\frac{\partial \mathcal{H}_{\rm T}}{\partial g_{1}}\label{can}
\eeq
with setting the outer decay rate at $d\eta/dt=0.01$. Note that, with our scaled Hamiltonian (\ref{ham}),
the circulation/libration period of the angle $g_1$ is typically   $\mathcal{O}(1)$
and much smaller than 
the variation timescale $\eta/{\dot \eta}$ of the parameter $\eta$.

In Figs. 3 and 4, with the black points, we show the evolution of 
the inner orbital elements $e_1$ and $g_1$ for the two runs R1 and R2,  as a function of the effective time $\eta$.
We only plot the range $\eta>-5$, since the two system evolved  almost identically
 in the earlier stage. 
In Fig. 3, the eccentricity $e_1$ shows a sharp transition around $\eta=-1.65$. 
Its oscillation range discontinuously shifted from (0.18,0.60) to (0.60,0.96).

However, in Fig. 4, around the same epoch $\eta=-1.65$, the eccentricity $e_1$ turned into a wider 
amplitude oscillation  from the original range (0.18,0.60) to  the new one (0.18,0.96).  At the same time, 
 the angle $g_1$ was captured into 
a libration around $g_1=\pi/2$.  This libration state terminated
 around $\eta=-1.20$, and, we concurrently  had 
$e_1\simeq 0$.  Therefore, in distinction from Fig. 3, Fig. 4 has two clear transitions, even though
these two runs  have almost the same initial conditions.

In Figs. 3 and 4, using the cyan points, we show the numerical results for two additional  runs R3 (Fig. 3) and R4 (Fig. 4), now only keeping the lowest order term for $\mathcal{H}_{\rm SP}$ (corresponding to $\beta=1$).  Their initial phases are $g_1=0.04$ (R3) and 0.01 (R4) that are not identical to R1 and R2, reflecting the probabilistic nature of the bifurcation, as explained later. In Figs. 3 and 4, the two runs R3 and R4 reproduce the characteristic features of the original runs R1 and R2 quite well, with small shifts of the characteristic epochs.
We additionally examined the cases with $\beta=0.5$ and $1.75$ (the Bahcall-Wolf profile), and confirmed their time profiles  are also similar to the cyan points in Figs. 3 and 4. More quantitatively, for example,  the first transition epochs $\eta$ (as seen in  Fig.3) for the four slopes $\beta$ are $-1.68$ ($\beta=0.5$), $-1.66$ (1.0), $-1.65$ (1.5) and $-1.55$ (1.75).  Indeed, the shifts are small for the realistic range of $\beta$,  supporting the validity of our truncation. 

 Therefore, below, we only keep the lowest order term for $\mathcal{H}_{\rm SP}$. This considerably simplifies our Hamiltonian, and allows us to develop analytical evaluations. 

As demonstrated in Figs. 3 and 4, the evolution of a system could  depend strongly on its initial condition. 
In the following sections, we show that these interesting results are  due to  
the probabilistic bifurcation at a separatrix crossing. 
The orbital evolution at a separatrix crossing is one of the 
central issues in this paper.

\section{Structure of Phase Space}

Next, we analytically explain the evolution of the phase space structure for the Hamiltonian (\ref{ham}) 
that depends on the effective time variable $\eta$. 
As we see in \S 4.2, the separatrixes determine the basic profile of the phase space, dividing the librating 
and circulating regions.

Meanwhile, a separatrix starts from and runs into unstable fixed points. Therefore, to follow the evolutions of the separatrixes, 
it is crucial to understand the transitions of the fixed points, in response to our time variable $\eta$.
This preparative study is done in \S 4.1.
 
In the following, we limit the angular variable $g_1$ in the
range $[0,\,\pi)$, identifying  $g_1=\pi$ with $g_1=0$, because of the symmetry of the Hamiltonian (\ref{ham}).

\subsection{Transitions of the fixed points}

\begin{figure*}
\begin{center}
\subfigure[$\eta=-200$]{
\includegraphics[width=4cm]{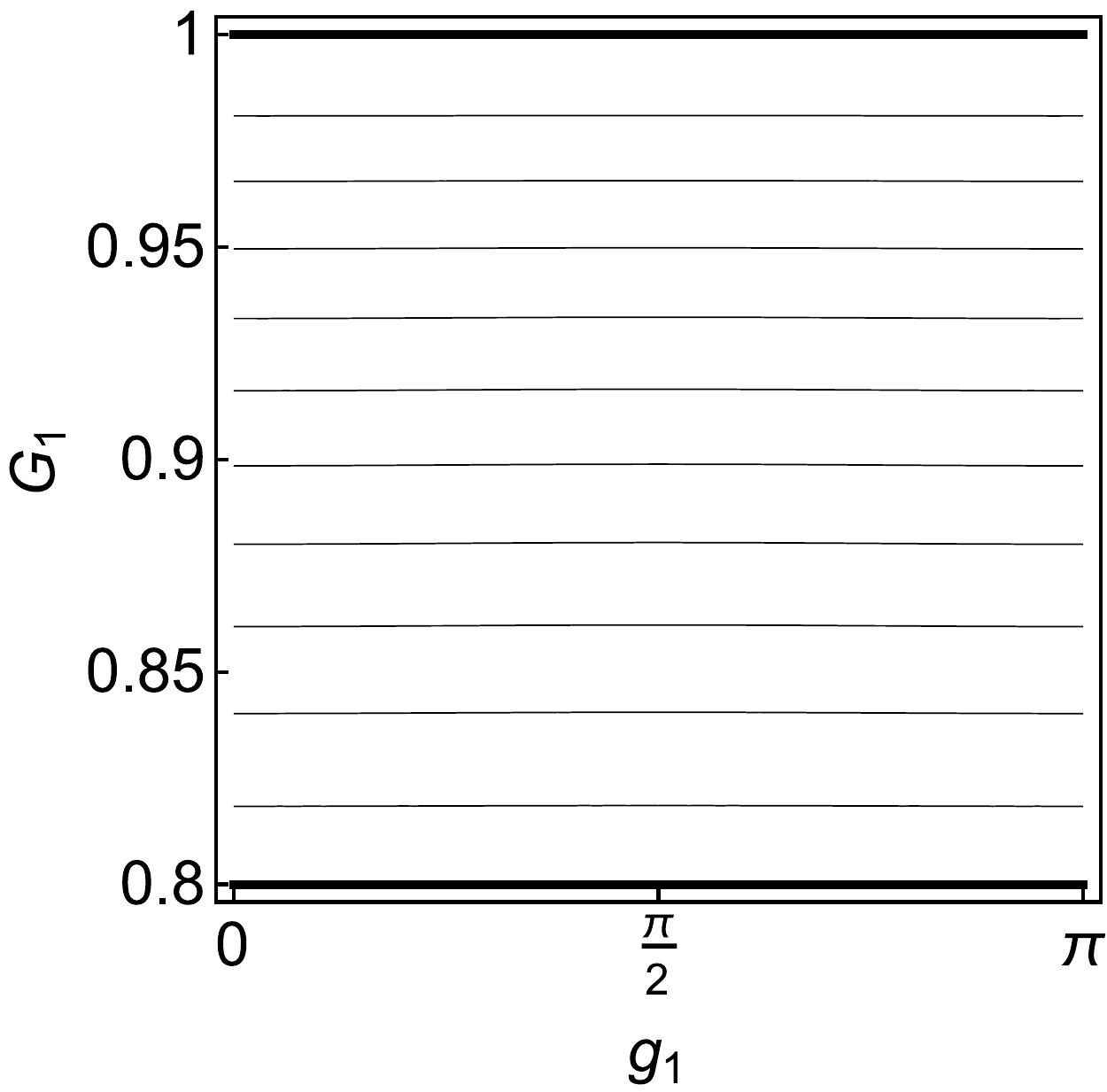}}\,\,
\subfigure[$\eta=-3$]{
\includegraphics[width=4cm]{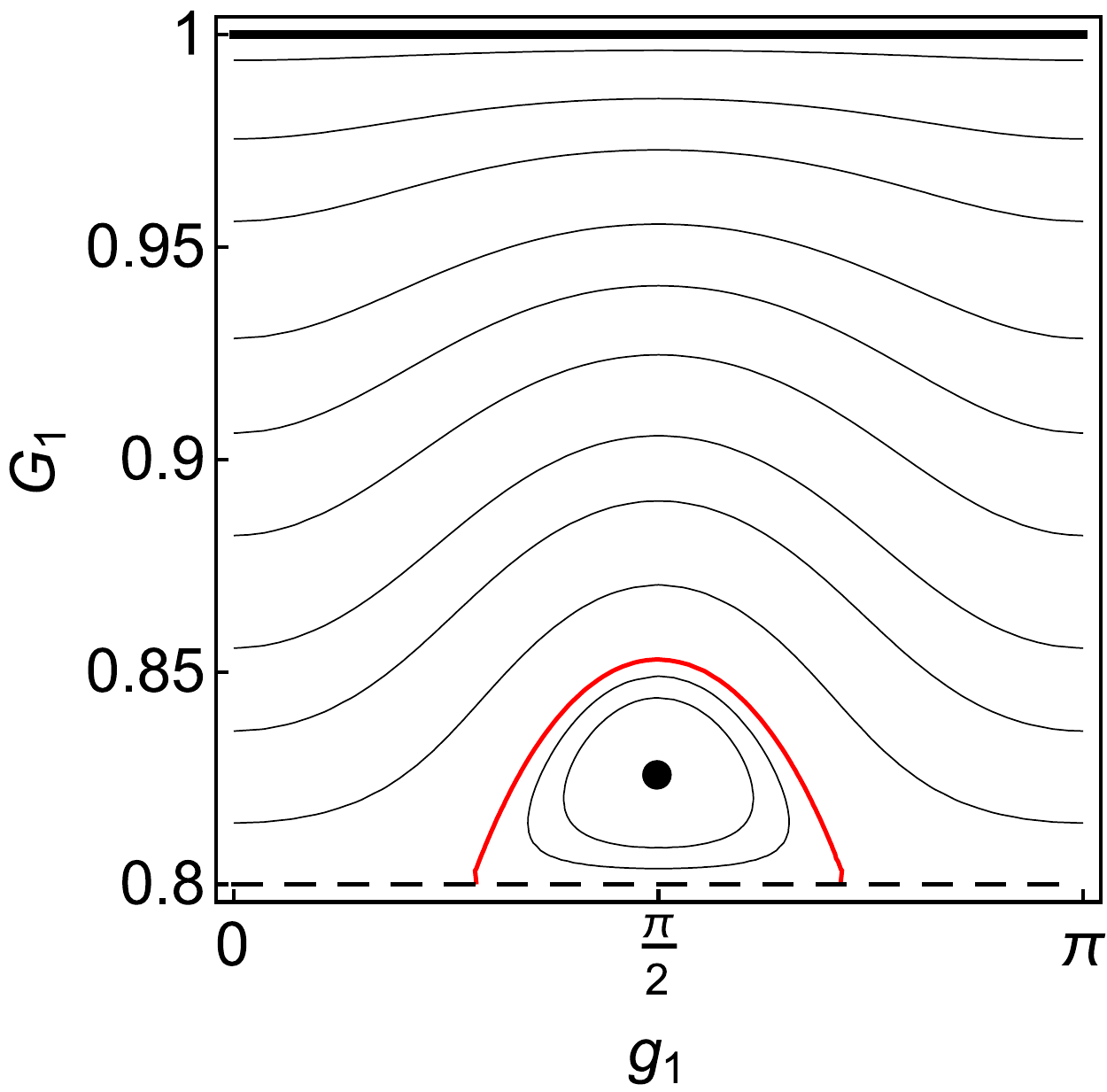}}\,\,
\subfigure[$\eta=-1.641$]{
\includegraphics[width=4cm]{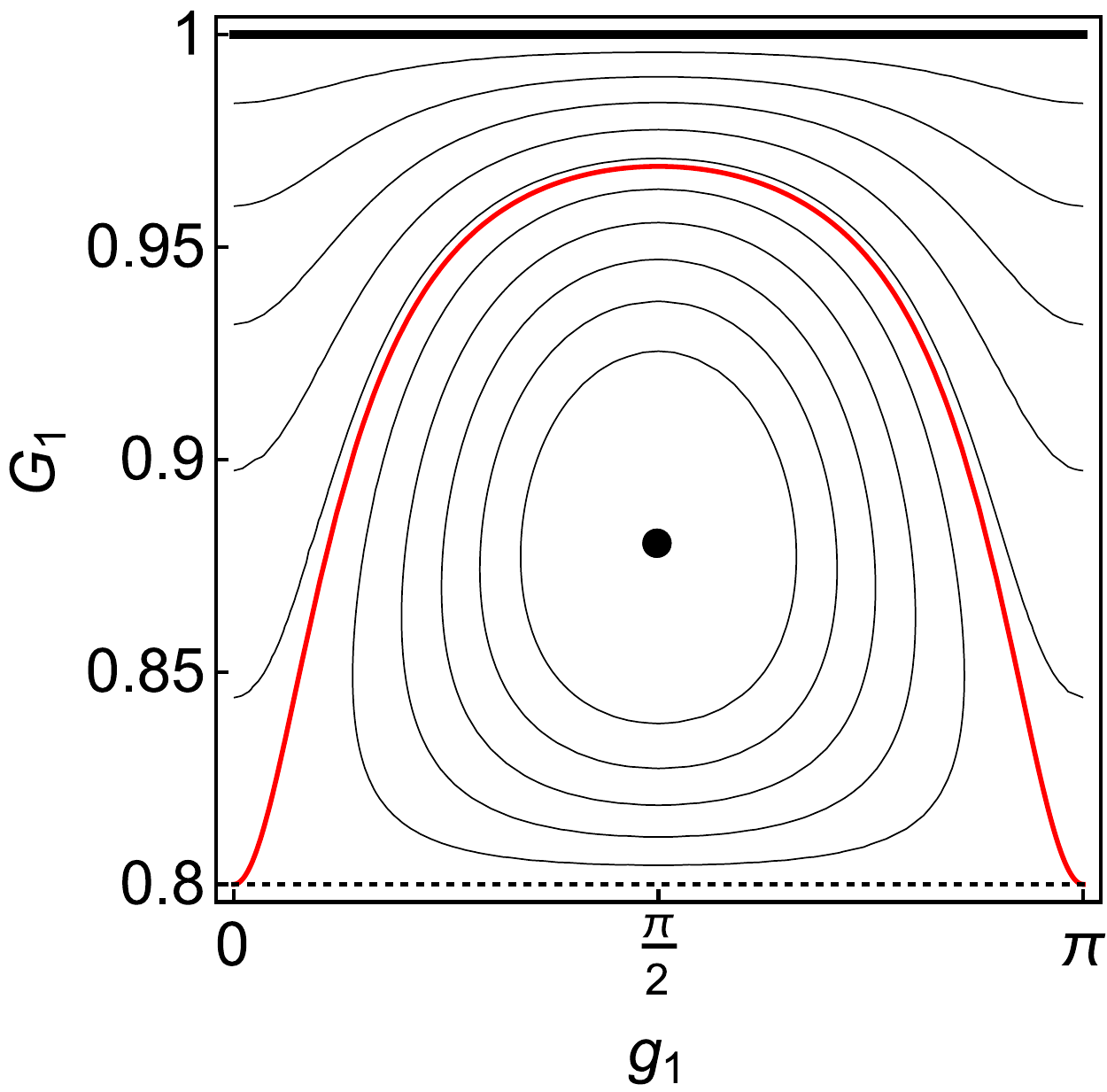}}\\
\subfigure[$\eta=-1.5$]{
\includegraphics[width=4cm]{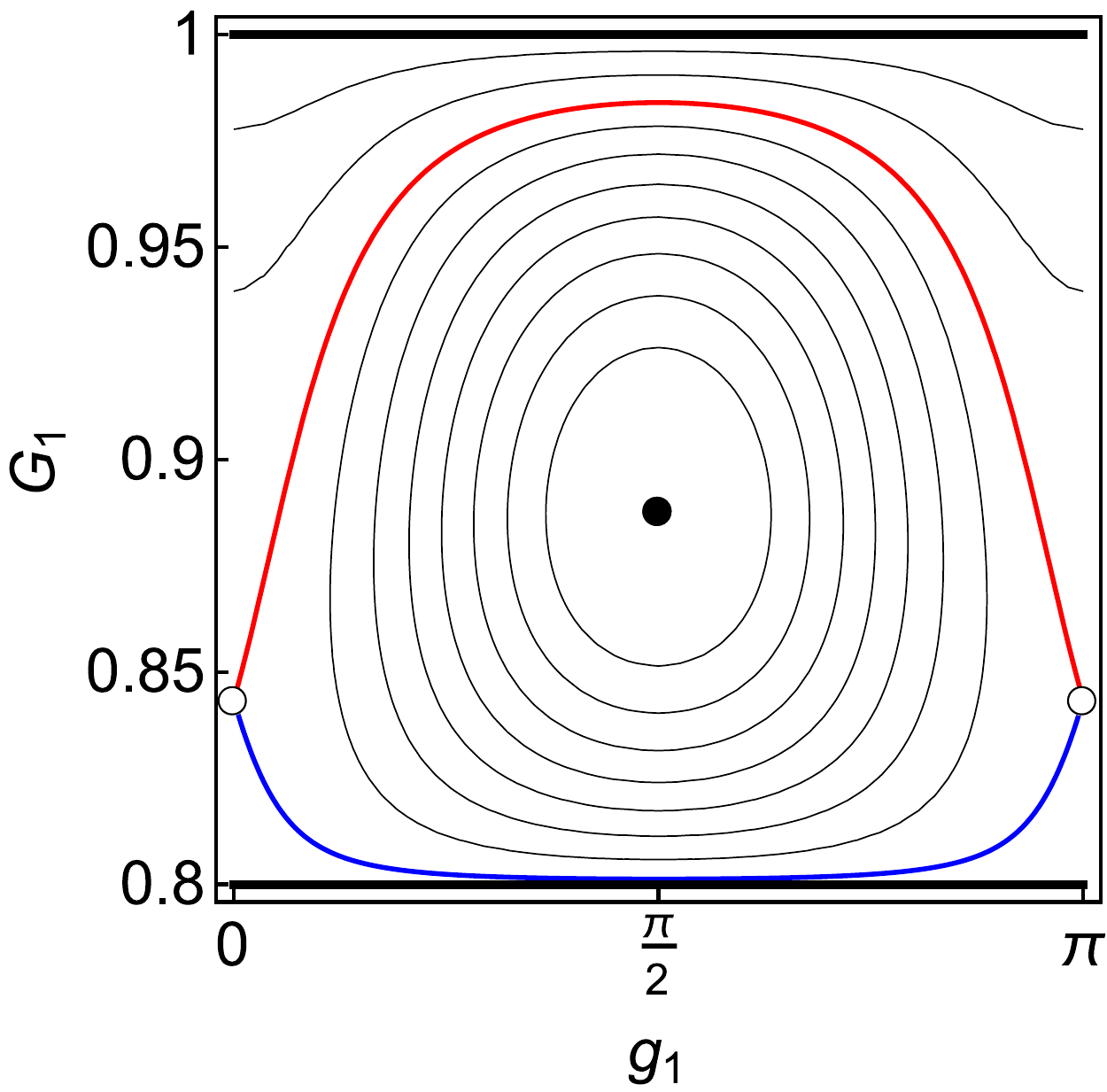}}\,\,
\subfigure[$\eta=-1.375$]{
\includegraphics[width=4cm]{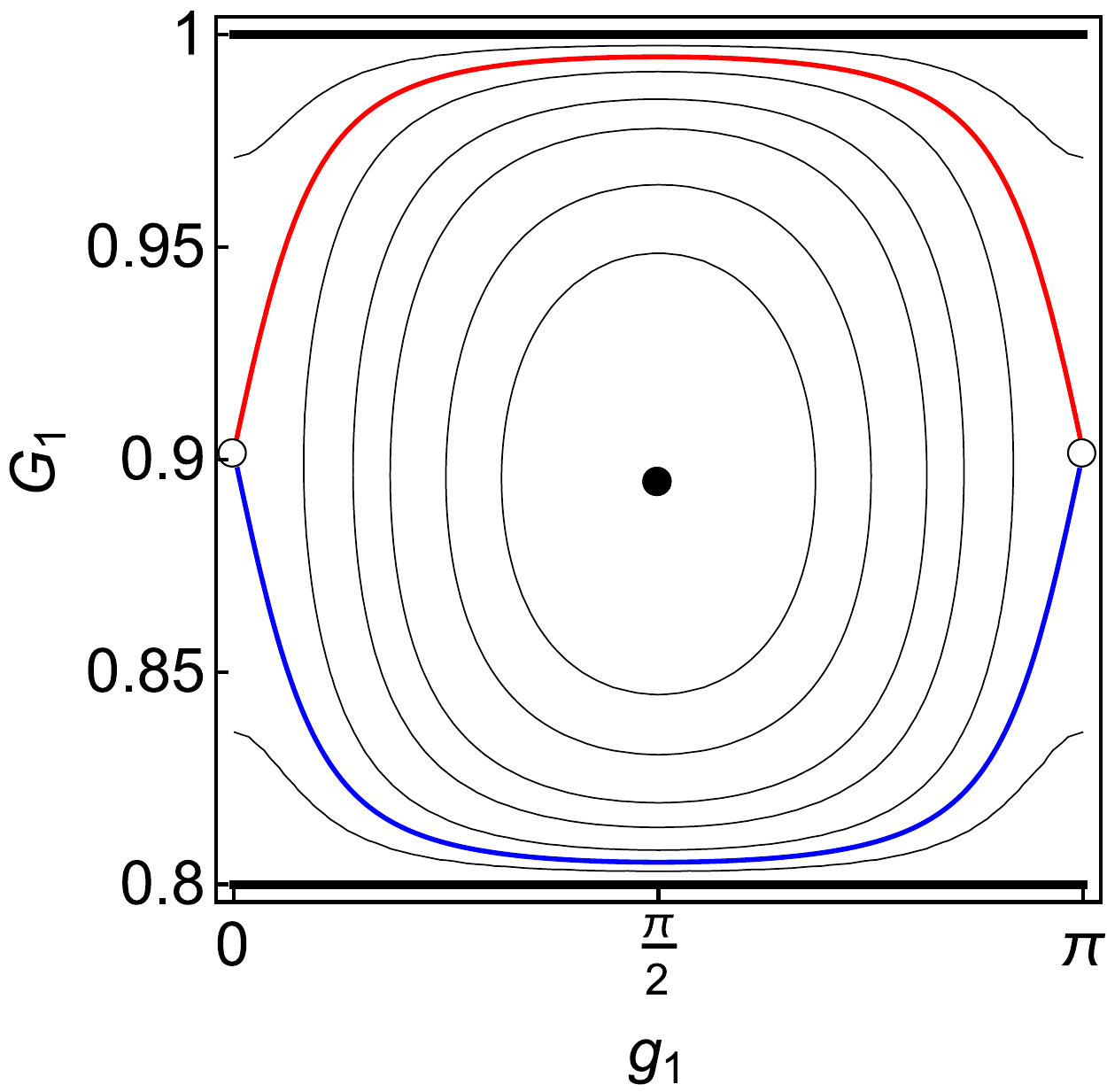}}\,\,
\subfigure[$\eta=-1.25$]{
\includegraphics[width=4cm]{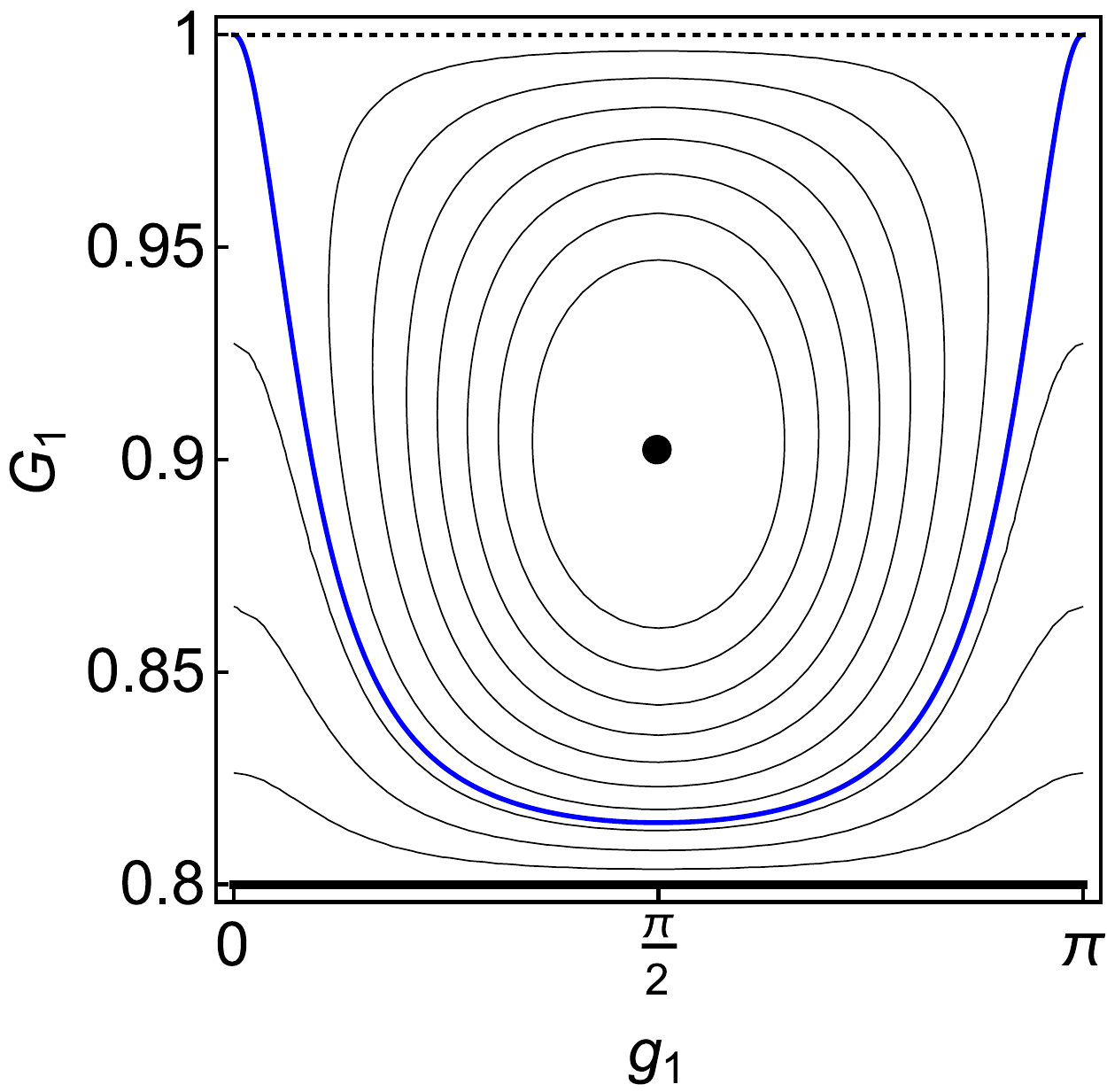}}\\
\subfigure[$\eta=-0.5$]{
\includegraphics[width=4cm]{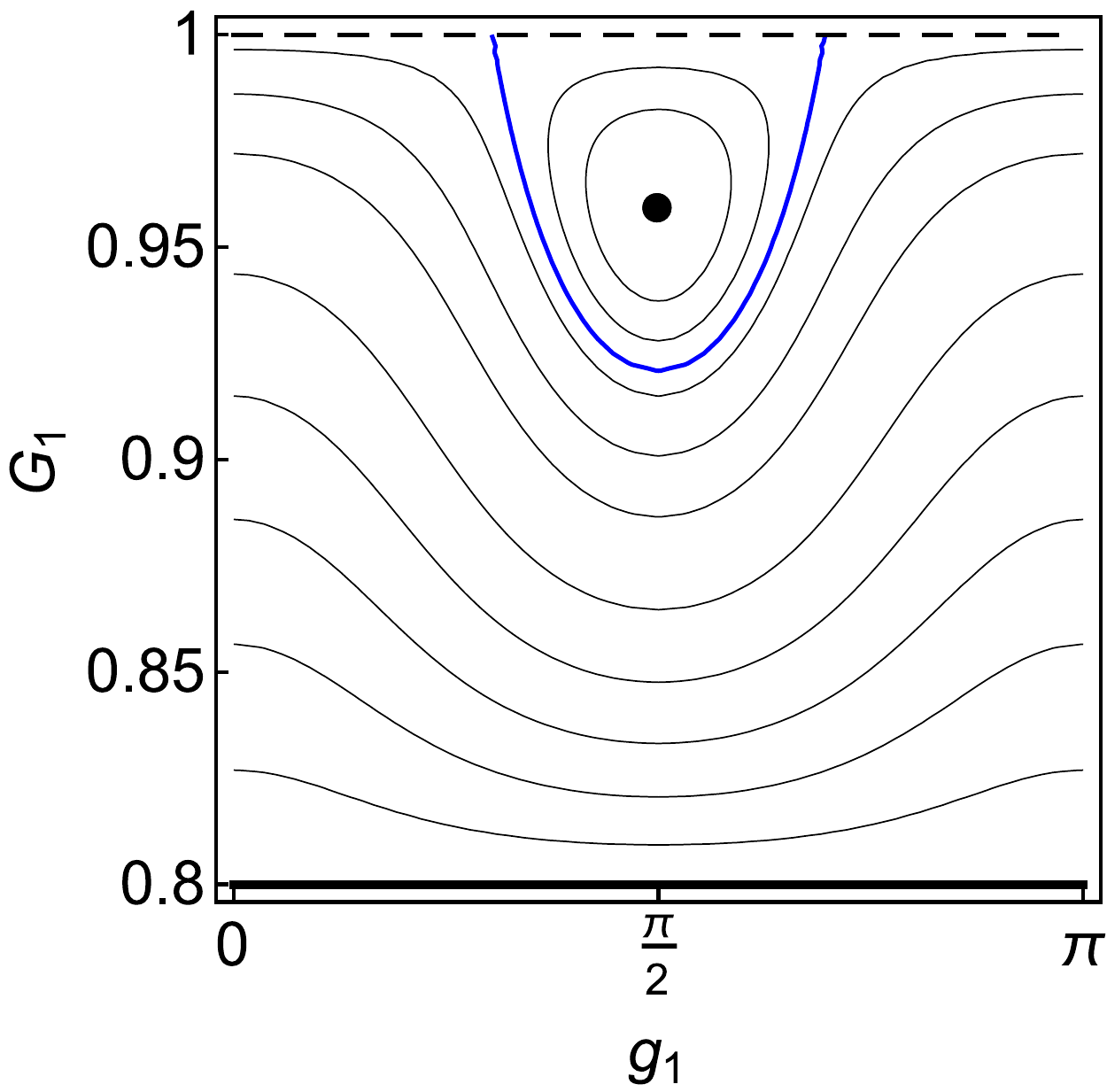}}\,\,
\subfigure[$\eta=-0.05$]{
\includegraphics[width=4cm]{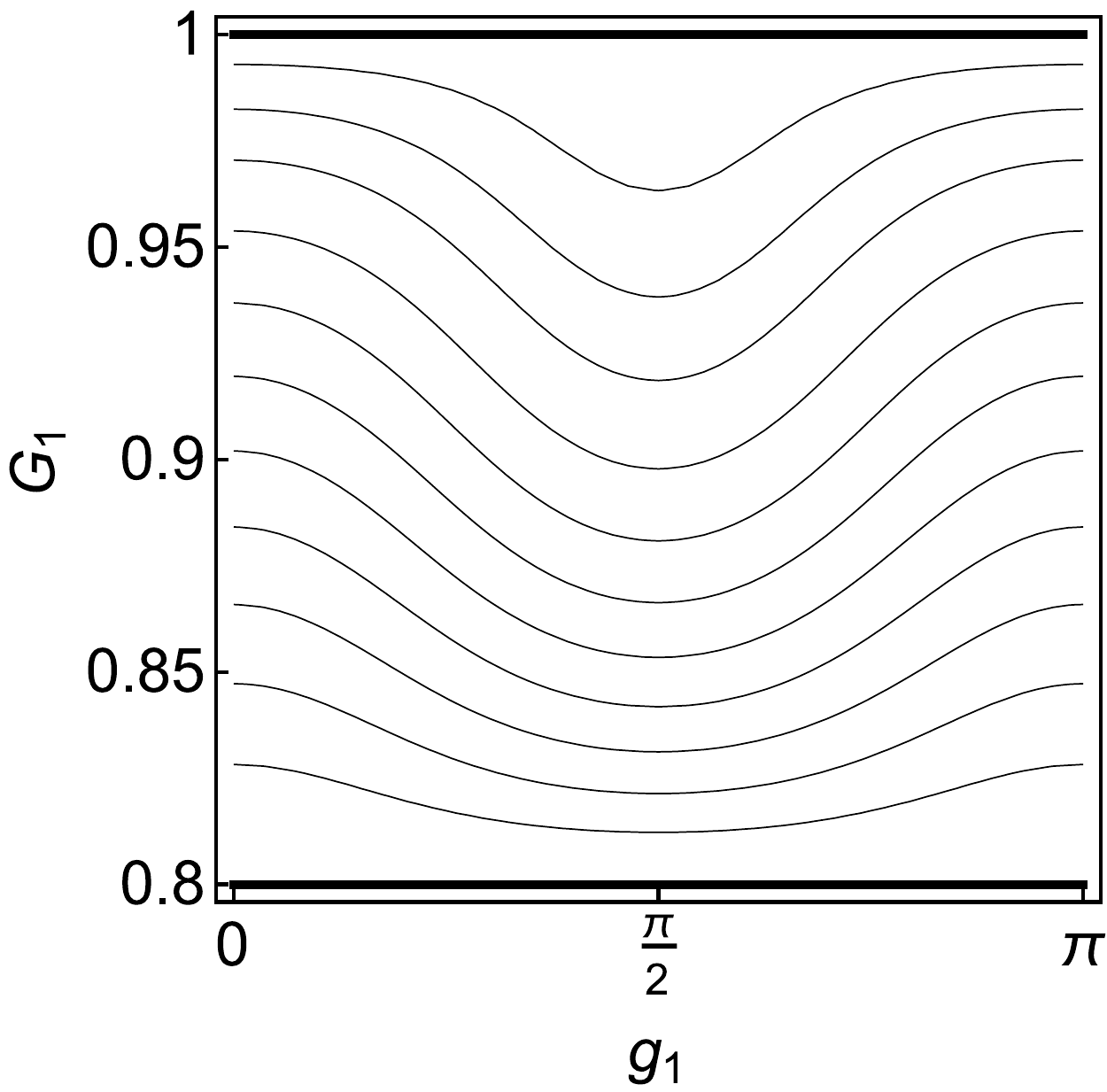}}
\caption{Evolution of the phase space for $(J_1,p)=(0.8,0.2)$. We present the snapshots from $\eta=-200$ to -0.05. The thin black lines are contours of Hamiltonian (\ref{ham}). 
The red and blue lines represent separatrixes. The 
black dots and open circles are stable and unstable fixed points respectively. At the upper and lower boundaries $G_1=1$ and $G_1=J_1$, the angular variable $g_1$ has coordinate singularities and these two boundaries should be regarded as two points (though artificially represented as lines). They are actually fixed points, and  we show their stabilities by using different types of lines; stable (thick solid lines), 
unstable (dotted lines) and marginally stable (dashed lines).  The legends are summarized in Table 1.
}
\label{fig:phase3}
\end{center}
\end{figure*}

\begin{table}
\caption{Definitions of symbols in our phase space}
\label{tab:1}
\begin{tabular}{lcc}
\hline
Legend & definition \\
\hline
\hline 
 filled circle & stable fixed point\\
open circle & unstable fixed point\\
\hline 
\hline 
    &    for the fixed points at $G_1=J_1$ and 1\\
thick solid line & stable fixed point \\
dashed line &  unstable fixed point\\
dotted line &  marginality stable fixed point \\
\hline
\hline 
thin black  line & contour of Hamiltonian \\
red line & upper separatrix \\
blue line & lower separatrix \\
\hline
\end{tabular}
\end{table}

In this section, we identify the five types of  transitions T1-T5 when the basic properties  of the fixed points ({\it e.g.} their total number, stabilities)  change in our phase space. As we explain below, these transitions are  accompanied by  merger or split of multiple fixed points, and classified as the pitchfork bifurcation  in the literature \citep[e.g.][]{Strogatz}.

To begin with, we should point out that, in the $(g_1,\, G_1)$ coordinate, the inner argument of pericenter $g_1$ becomes singular at  $G_1=J_1$ and $1$. This is because the angle $g_1$ loses its geometrical meanings there. More specifically, the inner orbit is circular for $G_1=1$ and is coplanar with 
the outer orbit for $G_1=J_1$ (namely $\cos I =1$), both making the angle $g_1$ ill-defined.
But, we can overcome these coordinate singularities at $G_1=J_1$ and 1,  by applying the following canonical transformations respectively \citep{ivanov2005};
\beqa
(x', y')&=&\sqrt{2(G_{1}-J_{1})}(\sin g_{1}, \cos g_{1}),\label{ct1}\\
(x, y)&=&\sqrt{2(1-G_{1})}(\cos g_{1}, \sin g_{1}).
\eeqa
In these regular coordinates $(x',\,y')$ or $(x,\,y)$, we can readily find that the points $(x',y')=(0,0)$ and $(x,y)=(0,0)$ (thus $G_1=J_1$ and 1) are always fixed points.

Below, we continue to use the original coordinate $(g_1,G_1)$ in which
 the two fixed points $G_1=J_1$ and 1 are stretched into two horizontal lines  (as demonstrated below in Fig. \ref{fig:phase3}). We distinguish their stabilities by using the following three types of lines; thick solid lines (stable), dotted lines (unstable) and dashed lines (marginally stable).

For the standard KL mechanism (i.e. $\eta=0$ for Eq. (\ref{ham})), the fixed point $G_1=J_1$ is always stable, and the stability of another fixed point $G_1=1$ is solely determined by $J_1$ 
($J_1>\sqrt{3/5}$ : stable, $J_1<\sqrt{3/5}$ : unstable, \citealt{kozai1962,lidov1962}). 
By stark contrast, in our study, both of the fixed points $G_1=J_1$ and $1$ change their stabilities, depending on $p,\eta$ and $J_1$, as we see below.

In Fig.\ref{fig:phase3},  for $(J_1,\,p)=(0.8,\,0.2)$,  we show the evolution of contours of the Hamiltonian (\ref{ham}) in the phase space $(g_1, \,G_1)$. We show the stable and unstable fixed points with the black dots and the  open circles respectively (in addition to the lines for the two fixed points $G_1=J_1$ and 1 mentioned above).
As discussed in the next subsection, the separatrixes can be divided into the upper and lower parts. 
They are shown with the  red and blue lines whose definitions are given in the next subsection. We summarized the legends of
our phase-space figures in Table 1.

\begin{table*}
\caption{The five transitions when the fixed points  change  their basic properties.  Column 2; the point where an additional  fixed point is created or annihilated. The $g_1$ coordinate is the asymptotic value. Columns 3 and 4;   the stabilities of fixed points $G_1=J_1$ and 1 around the transition (S: stable, U: unstable). Column 5; the effective time parameter $\eta_i$ at each transition T$_i$. Column 6;   the parameter region ($J_1,\,p$) to realize the valid sign $\eta_i<0$.}
\label{tab:2}
\begin{tabular}{lcccccccc}
\hline
 &fixed point creation/annihilation & Stability of $G_1=J_1$ & Stability of $G_1=1$ & epoch $\eta$ & corresponding region\\
\hline
 T1 & creation at $(g_1,G_1)=(\pi/2,J_1)$ & S $\rightarrow$ U & S & $\eta_1=-\frac{-3J_{1}^{3}+5J_1}{2(J_{1}^3-p)}$ &  $p<J_{1}^3$\\
T2 &creation at $(0,J_1)$ &  U $\rightarrow$ S & S & $\eta_2=-\frac{J_{1}^3}{J_{1}^3-p}$ &  $p<J_{1}^3$\\
T3 &annihilation at $(0,1)$ & S &  S $\rightarrow$ U & $\eta_3=-\frac{1}{1-p}$ &  $p<1$\\
T4 &annihilation at $(\pi/2,1)$ & S & U $\rightarrow$ S  & $\eta_4=-\frac{-3+5J_{1}^2}{2(1-p)}$ & $p<1$ and $J_1>\sqrt{3/5}$\\
T5 &creation at $(\pi/2,1)$ & S &  S $\rightarrow$ U  & $\eta_5=-\frac{-3+5J_{1}^2}{2(1-p)}$  & $p>1$ and $J_{1}<\sqrt{3/5}$\\
\hline
\end{tabular}
\end{table*}

In Fig. \ref{fig:phase3}, as $\eta$ increases due to the outer orbital decay, 
we can see the following four transitions T1,T2,T3 and T4, when the fixed points change their basic properties. We explain them one by one, using Fig. 5.

T1: This transition occurs 
 between Figs.\ref{fig:phase3}a and \ref{fig:phase3}b. The new fixed point (shown by the black dots in Fig. 5) appears at $G_1=J_1$, and starts moving upward along the line $g_1=\pi/2$. The stability of the fixed point $G_1=J_1$ (shown by lines) turns from stable to unstable. 

T2: This transition occurs 
 at Fig. \ref{fig:phase3}c. The new fixed point appears at $G_1=J_1$, and starts moving upward along the line $g_1=0$ (shown as the open circles in Fig. 5). The stability of the fixed point $G_1=J_1$ turns from unstable to stable.

T3:  This transition is at 
Fig. \ref{fig:phase3}f. The fixed point on $g_1=0$ (created at T2) disappears at $G_1=1$. The stability of the fixed point $G_1=1$ turns from stable to unstable.

T4: This transition is between Figs. \ref{fig:phase3}g and \ref{fig:phase3}h.  The fixed point on $g_1=\pi/2$ (created at T1) disappears at $G_1=1$. The stability of the fixed point $G_1=1$ turns from unstable to stable. 


\begin{figure*}
\begin{center}
\subfigure[$\eta=-20$]{
\includegraphics[width=4cm]{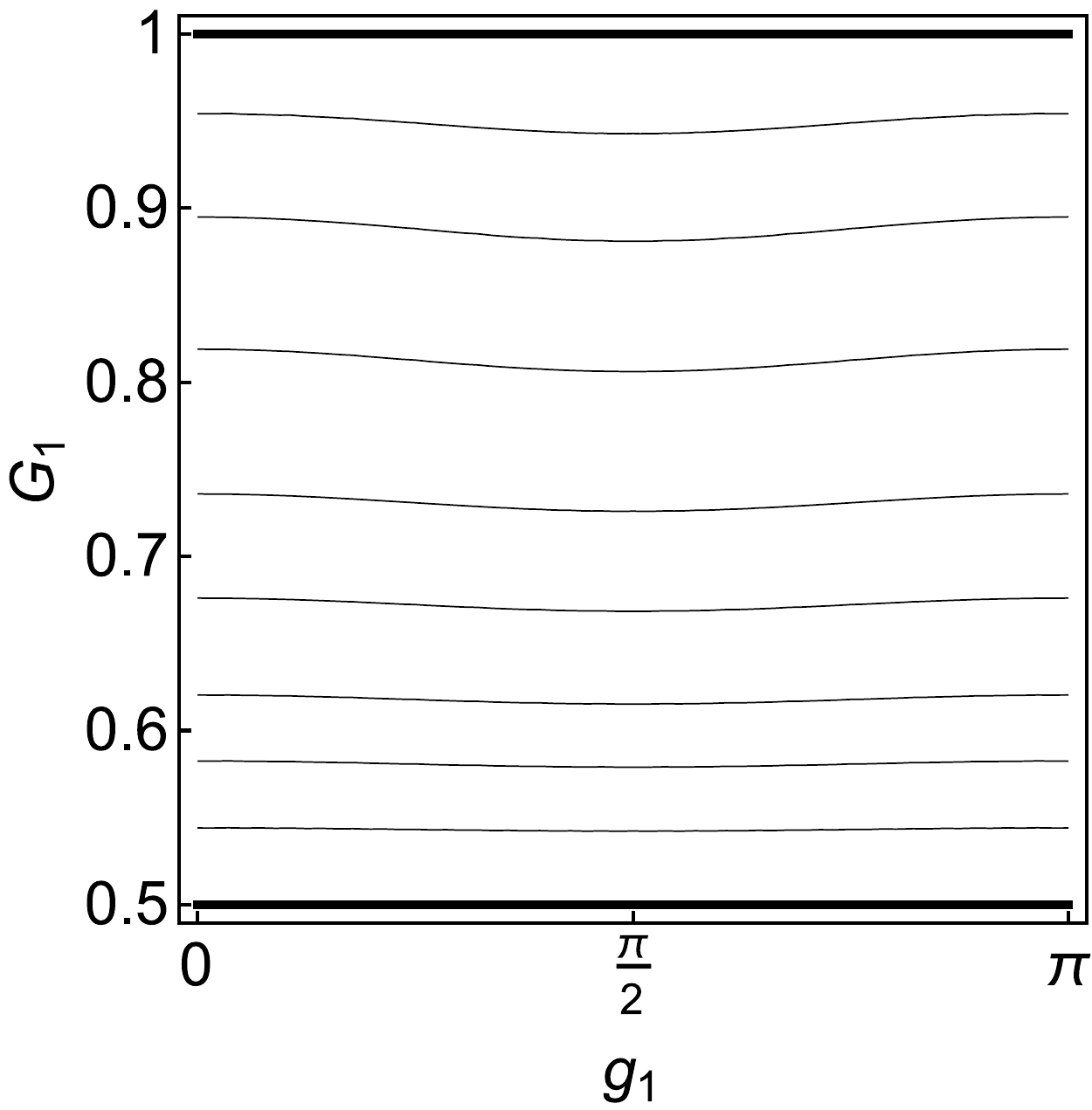}}
\subfigure[$\eta=-2$]{
\includegraphics[width=4cm]{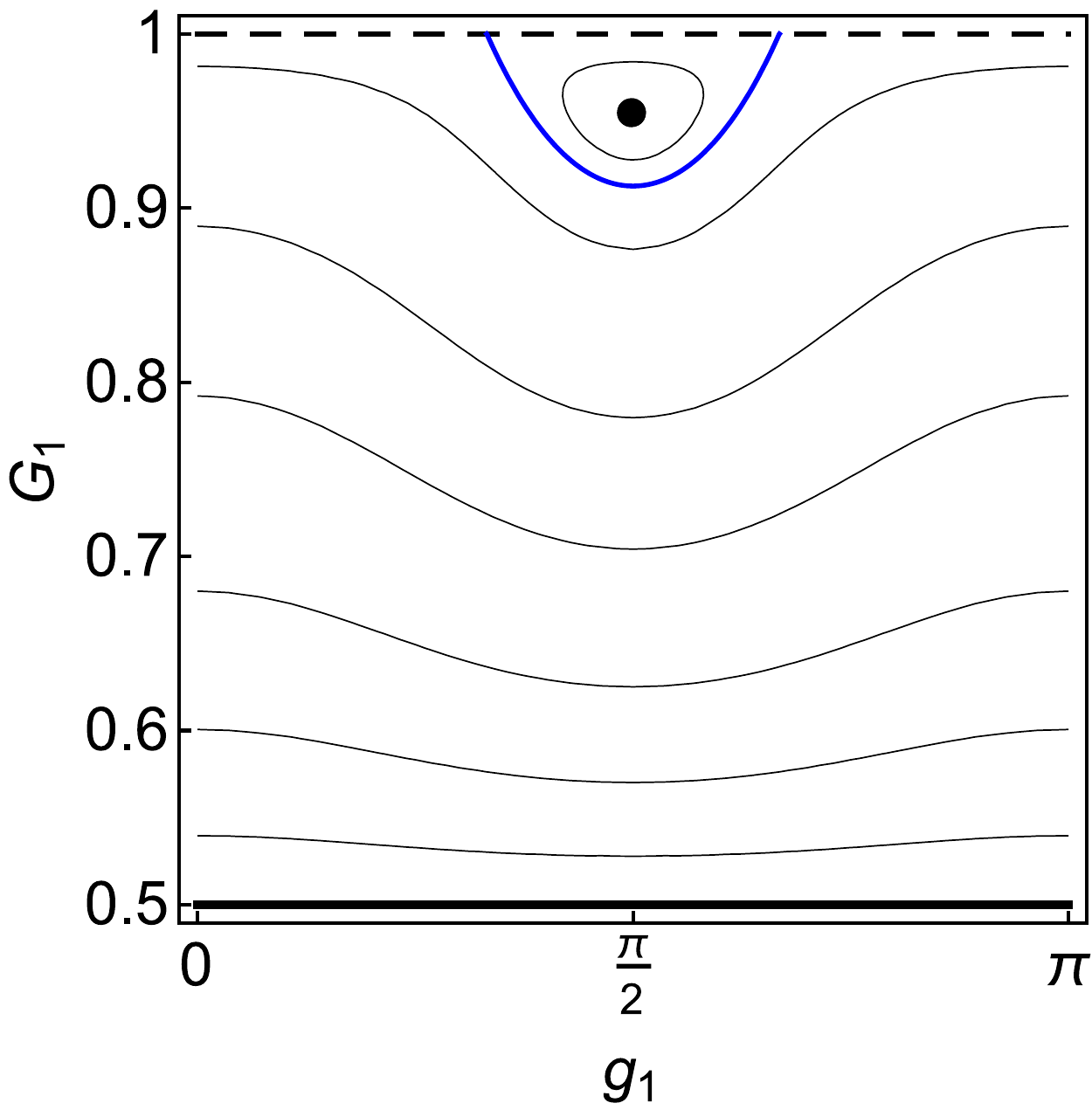}}
\subfigure[$\eta=-1$]{
\includegraphics[width=4cm]{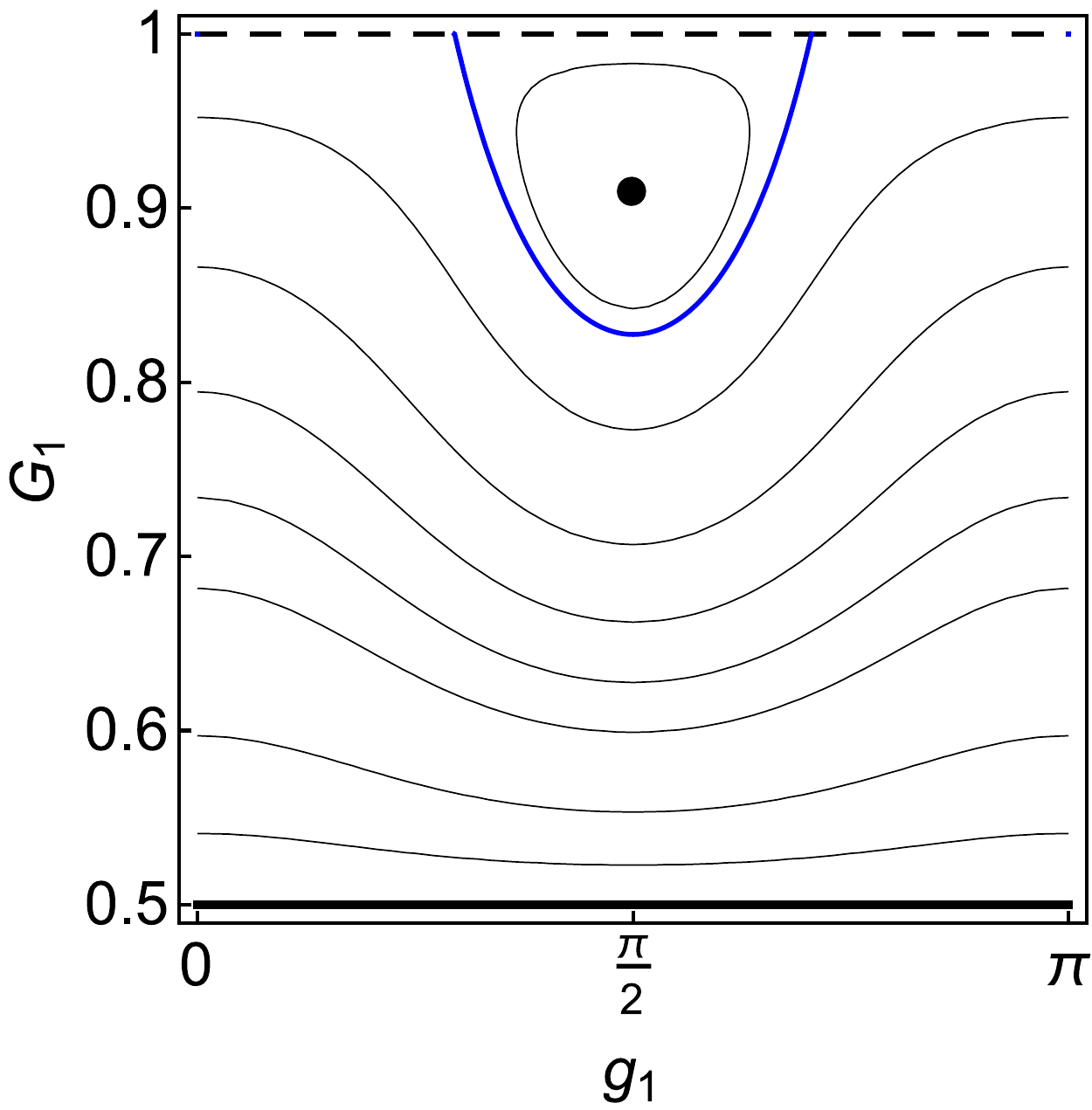}}
\caption{Evolution of the phase space for the region V with  $(J_1,p)=(0.5,1.2)$. We show the snapshots at $\eta=-20,\, -2,\, {\rm and} -1$. }
\label{fig:phase5}
\end{center}
\end{figure*}

We summarize the primary aspects of these four transitions T1,T2, T3 and T4 in Table 2. Here, it important to notice that these  transitions  always accompany the creations (at $G_1=J_1$) or annihilations (at $G_1=1$) of the fixed point that moves upward either along $g_1=0$ or $\pi/2$. Concurrently,  the corresponding fixed  point $G_1=J_1$ or 1 also changes  its stability. These transitions are typical pitchfork bifurcations \citep{Strogatz}.

Now we explicitly evaluate the effective time parameter $\eta _1 $ for the transition T1. Firstly, for the stable fixed point shown with the black dots in Fig. 5 at  $g_1=\pi/2$,  we  derive the relation between the  coordinate value $G_1$ and the time parameter $\eta$.  Then we specify the transition epoch $\eta_1$,  using the condition that, at T1, this fixed point takes the coordinate value  $G_1=J_1$ (see Table 1).

Since we identically have $\partial \mathcal{H}_{\rm T}/\partial
g_{1}=0$ for $g_1=\pi/2$, the desired relation between $G_1$ and $\eta$  is given as $\partial \mathcal{H}_{\rm T,1}/\partial G_1|_{g_1=\pi/2}=0$, or 
\beqa
-3G_1+5\frac{J_{1}^{2}}{G_{1}^3}=-2\eta\lmk G_1-\frac{p}{G_{1}^{2}}\rmk.
\label{eq:3'}
\eeqa
Plugging-in $G_1=J_1$, we obtain the transition epoch for T1 
\beq
\eta_1\equiv-\frac{-3J_{1}^3+5J_{1}}{2(J_{1}^3-p)}.
\eeq
Considering the inequalities $0\le J_1\le 1$,  this solution has the appropriate sign $\eta_1<0$ only for $p<J_1^3$. We present these results in the fifth and sixth columns in Table 2. 

Similarly, we can derive $\eta_2$ for T2. For the unstable fixed point at $g_1=0$ (shown with the open circles in Fig. 5), we  identically have  $\partial \mathcal{H}_{\rm T}/\partial
g_{1}=0$ again, and the relation between $G_1$ and $\eta$ is now given as 
\beqa
G_1=-\eta\lmk G_1-\frac{p}{G_{1}^{2}}\rmk.
\label{eq:4'}
\eeqa
Putting $G_{1}=J_1$ for the transition T2, we have 
\beq
\eta_2\equiv-\frac{J_{1}^3}{J_{1}^3-p}
\eeq
which has the valid sign $\eta_2<0$ for $p<J_1^3$. 
For $(J_1,p)=(0.8,0.2)$, we indeed have $\eta_2=-1.641$ consistent with Fig. 5c.

We can also derive $\eta_4$ and $\eta_3$ by setting $G_1=1$ in Eqs. (\ref{eq:3'}) and  (\ref{eq:4'}) respectively.
The results for T2, T3 and T4  are  summarized in Table \ref{tab:2}.
We should  notice the chronological order of the transitions
$\eta_1<\eta_2<\eta_3<\eta_4<0$ in the parameter region $(J_1,p)$ simultaneously satisfying the inequalities for T1 to T4 listed in Table 1.

\begin{figure}
\begin{center}
\includegraphics[width=7cm]{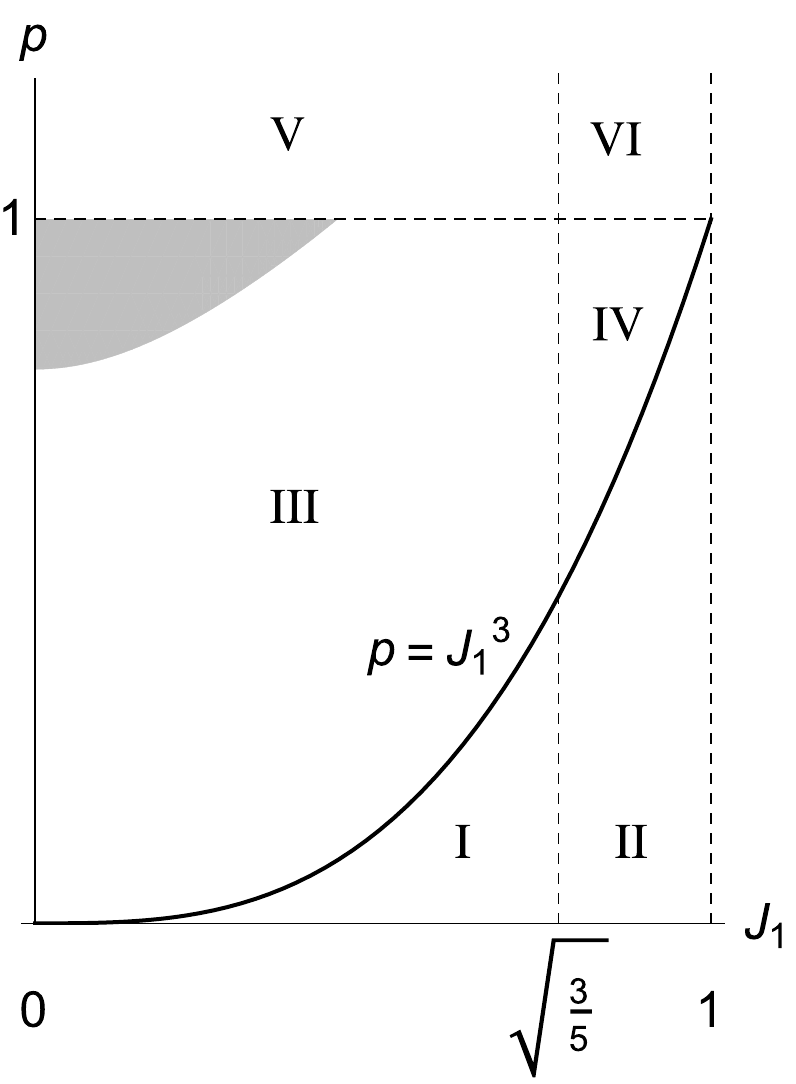}
\caption{The classification of the parameter space
$(J_1,p)$, according to the realized transitions T1 to T5 (see Table 2). In the shaded
area in  \III\, the probabilistic bifurcation does not occur
(explained in  \S 5.4).}
\label{fig:pj}
\end{center}
\end{figure}

So far, we have discussed the four transitions T1, T2,T3 and T4 that are realized for $(J_1,p)=(0.8,0.2)$ when increasing $\eta$ from $-\infty$ to 0.
 However, these are not the complete set of the transitions observed for the valid parameter range $0\le p$ and $0\le J_1\le 1$. 
In fact, we have an additional transition T5, as demonstrated in Fig. \ref{fig:phase5}  for $(J_1,p)=(0.5,1.2)$.
The basic aspects of T5 is summarized in Table 1. This is almost the inverse of the transition T4, and appears only for
 $p>1$ and $J_1<\sqrt{3/5}$. We have the transition epoch $\eta_5$ whose expression is identical to $\eta_4$ (as easily understood from their derivations).
The newly generated stable fixed point (shown with the black dots in Fig. \ref{fig:phase5}) moves downward, as $\eta$ increases from $\eta_5$.

 Given the inequalities in the last column in Table 1, we can divide the parameter space $(J_1,p)$ into the six different regions I to VI, as shown in Fig. \ref{fig:pj}. The transitions observed in each region are summarized in Table 3.

\begin{table}
\caption{The transitions realized  in the regions I - \VI}
\label{tab:3}
\begin{tabular}{lcc}
\hline
Region & Transitions \\
\hline
 I & T1, T2, T3\\
\II & T1, T2, T3, T4 \\
\III & T3 \\
\IV & T3, T4\\
V & T5 \\
\VI & none\\
\hline
\end{tabular}
\end{table}

For simplicity, we have not discussed the situations just on the boundaries of these six regions. But, at this stage, it would be instructive to comment on our previous work \citep{iwasa2016} where we simply put $p=0$ (ignoring relativistic corrections) to examine the effects of the stellar potential.  For $p=0$, the two transitions T2 and T3 are degenerated at the epoch $\eta_2=\eta_3=-1$ where the phase space goes through a drastic transition (see Fig. 3c in \citealt{iwasa2016}). \footnote{For $p=0$ and $\eta=-1$, every point on  the line $g_1=0$ satisfies $\partial \mathcal{H}_{\rm T,1}/\partial G_1=\partial \mathcal{H}_{\rm T,1}/\partial g_1=0$ and can be regarded as a fixed point. But, for $\eta\ne -1$, we do not have an unstable fixed point at $g_1=0$ that is crucially important for the probabilistic bifurcation discussed later.} In contrast, with a finite $p$, we have $\eta_2\ne \eta_3$ for $J_1<1$, and the anomalous behaviors for $p=0$  are \lq\lq{}regularized\rq\rq{} as demonstrated in Fig. 5.

\subsection{Evolution of the phase-space structure}

In the previous subsection, we analyzed the transitions T1 to T5 realized at the specific epochs $\eta_i$ ($i=1,\cdots,5$). 
Now, we discuss the evolution of  the phase-space structure for more general values of $\eta\ne \eta_i$.  We  pay special attention to the separatrixes that play central roles here,  dividing  the librating and circulating regions.  

For a one dimensional Hamiltonian system, 
 an unstable fixed point  is generally   categorized as a saddle, while a stable one as a center \citep{Strogatz}.
This is because the Hesse matrix for the stability analysis is traceless, due to the canonical equations of motion.
But,  for simplicity,  we merely call them unstable and stable fixed points. 

 In our phase space, the separatrixes begin and end at  unstable fixed points, as already demonstrated in Fig. 5. 
Therefore, for our Hamiltonian, the basic structure of the separatrixes can change only  at the five transitions T1 - T5 listed
 in Table 2. Strictly speaking, just at these transitions (see {\it e.g.} Fig. 5f for T4),  the relevant fixed point  becomes marginally stable, namely, an intermediate state between stable and unstable.

\begin{figure*}
\begin{center}
\subfigure[P0 at $\eta=-200$]{
\includegraphics[width=4cm]{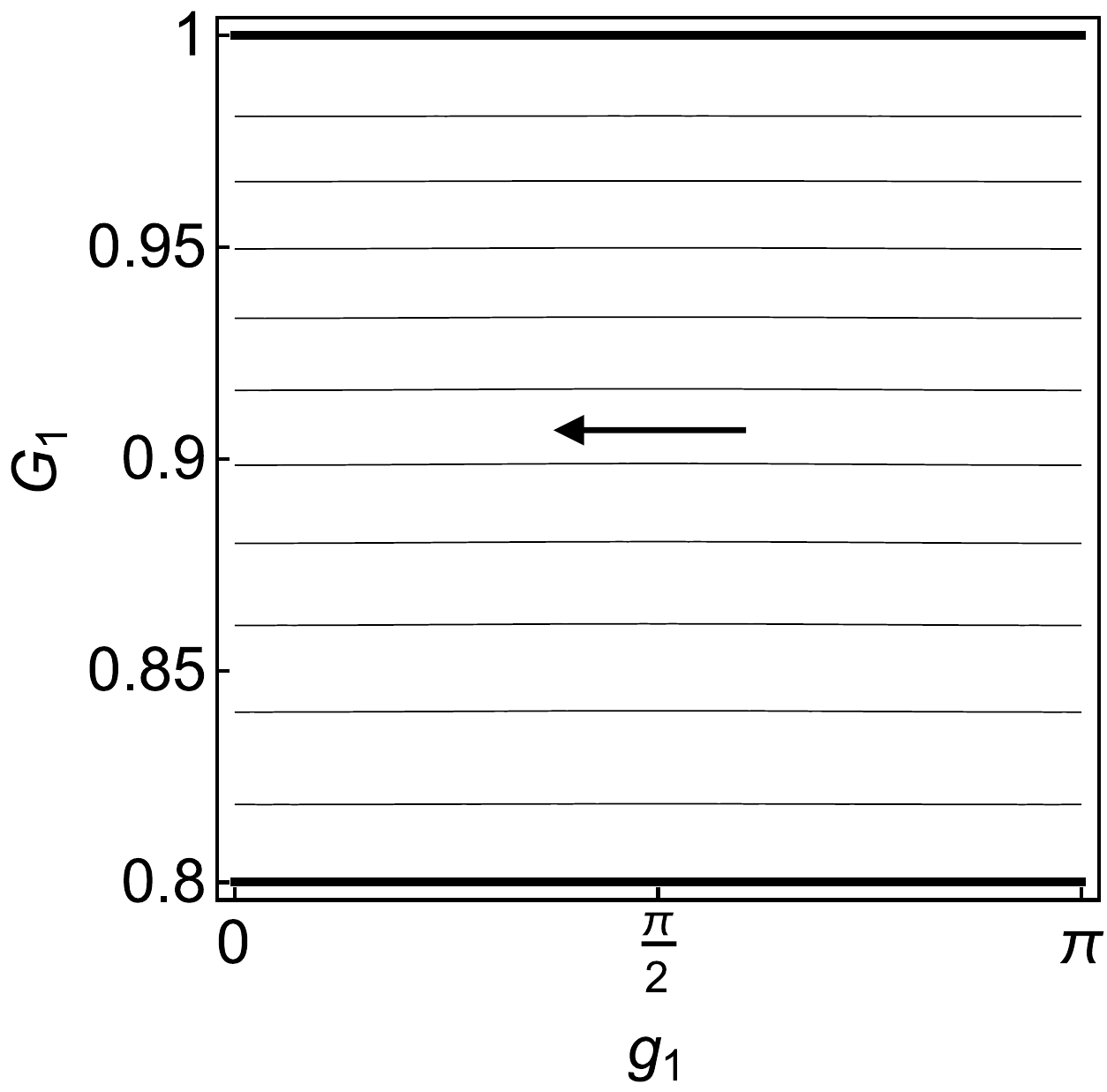}}
\subfigure[P1 at $\eta=-3$]{
\includegraphics[width=4cm]{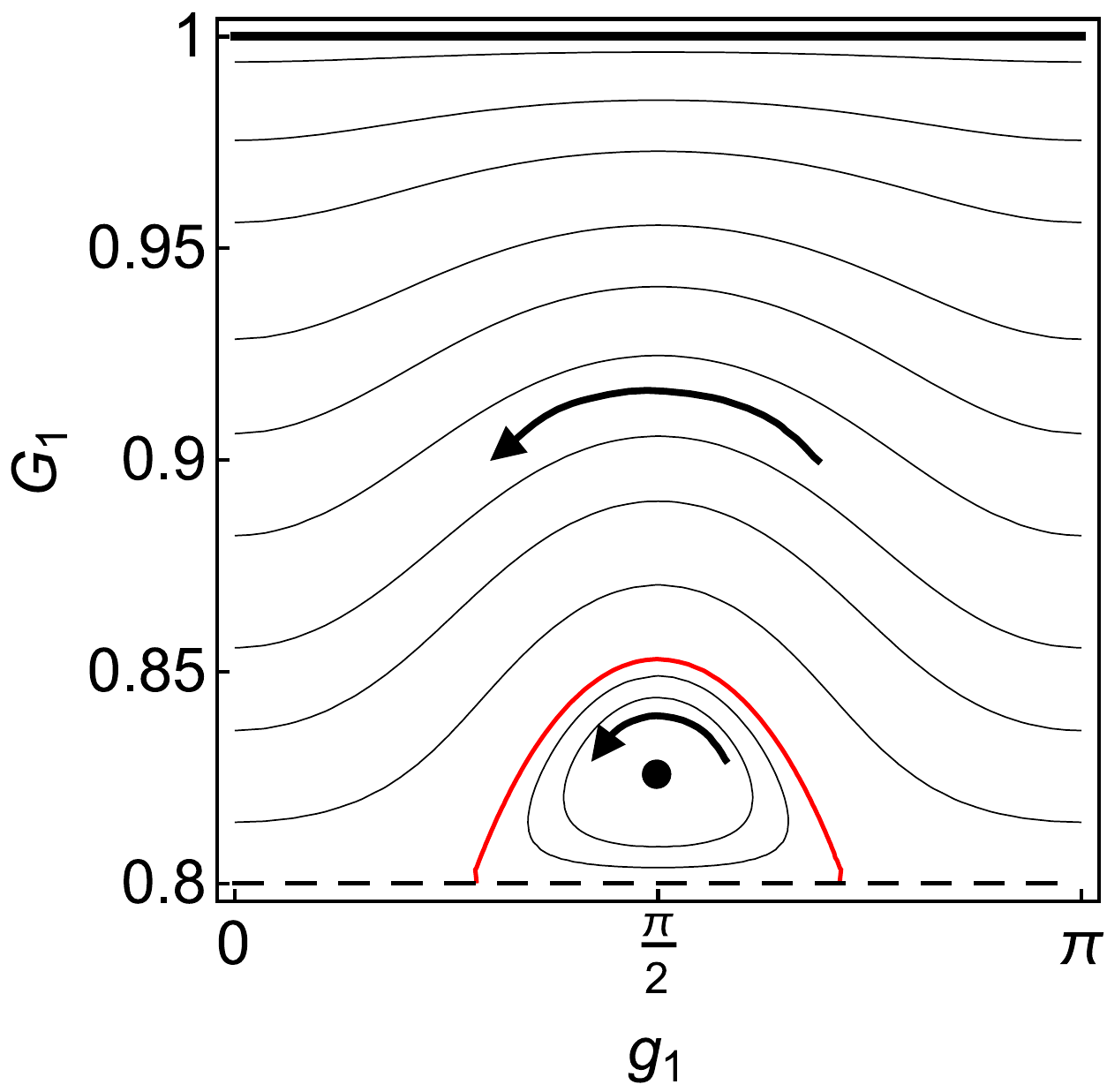}}
\subfigure[P2 at $\eta=-1.5$]{
\includegraphics[width=4.15cm]{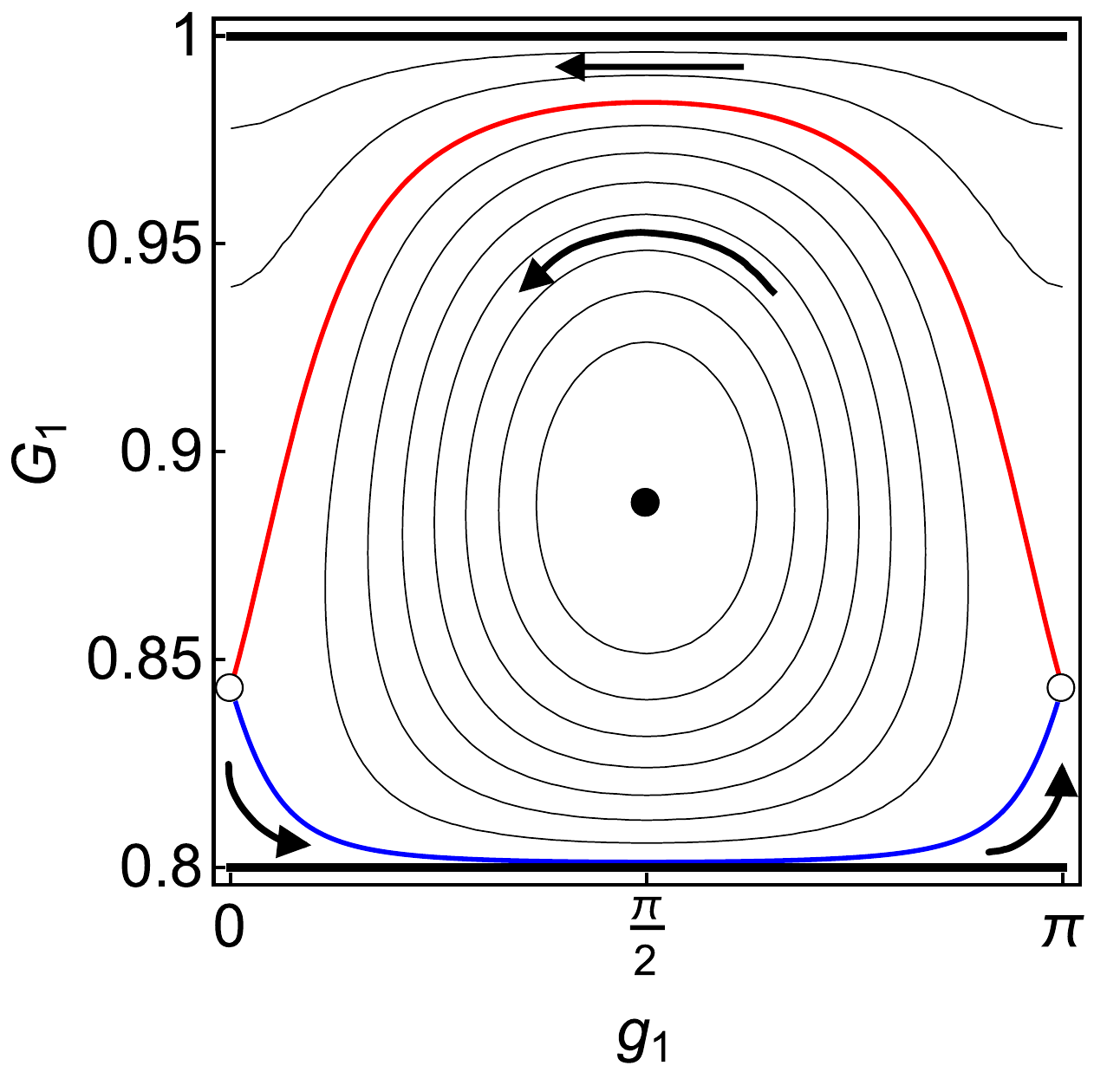}}\\
\subfigure[P3 at $\eta=-0.5$]{
\includegraphics[width=4cm]{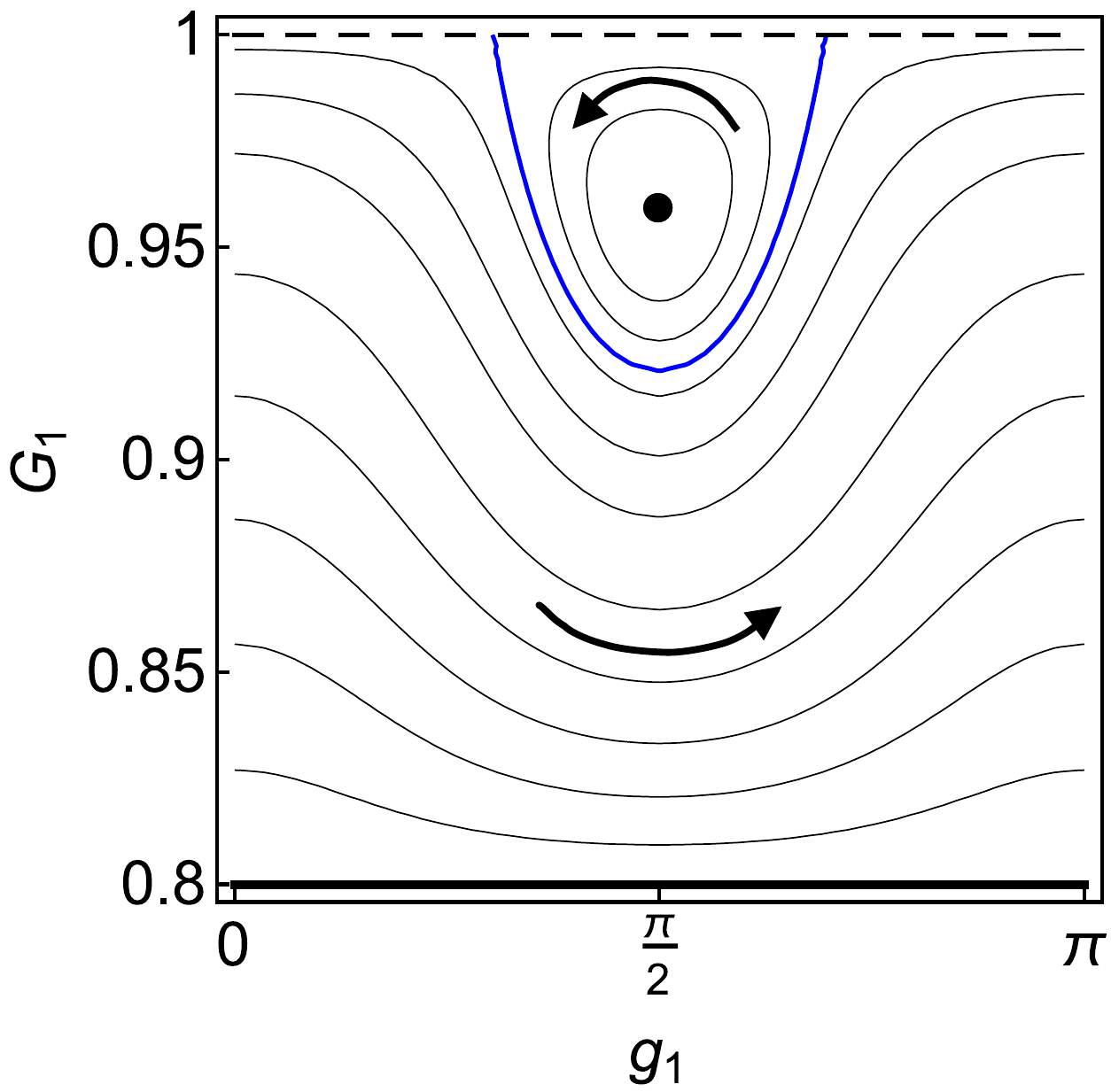}}
\subfigure[P4 at $\eta=-0.05$]{
\includegraphics[width=4cm]{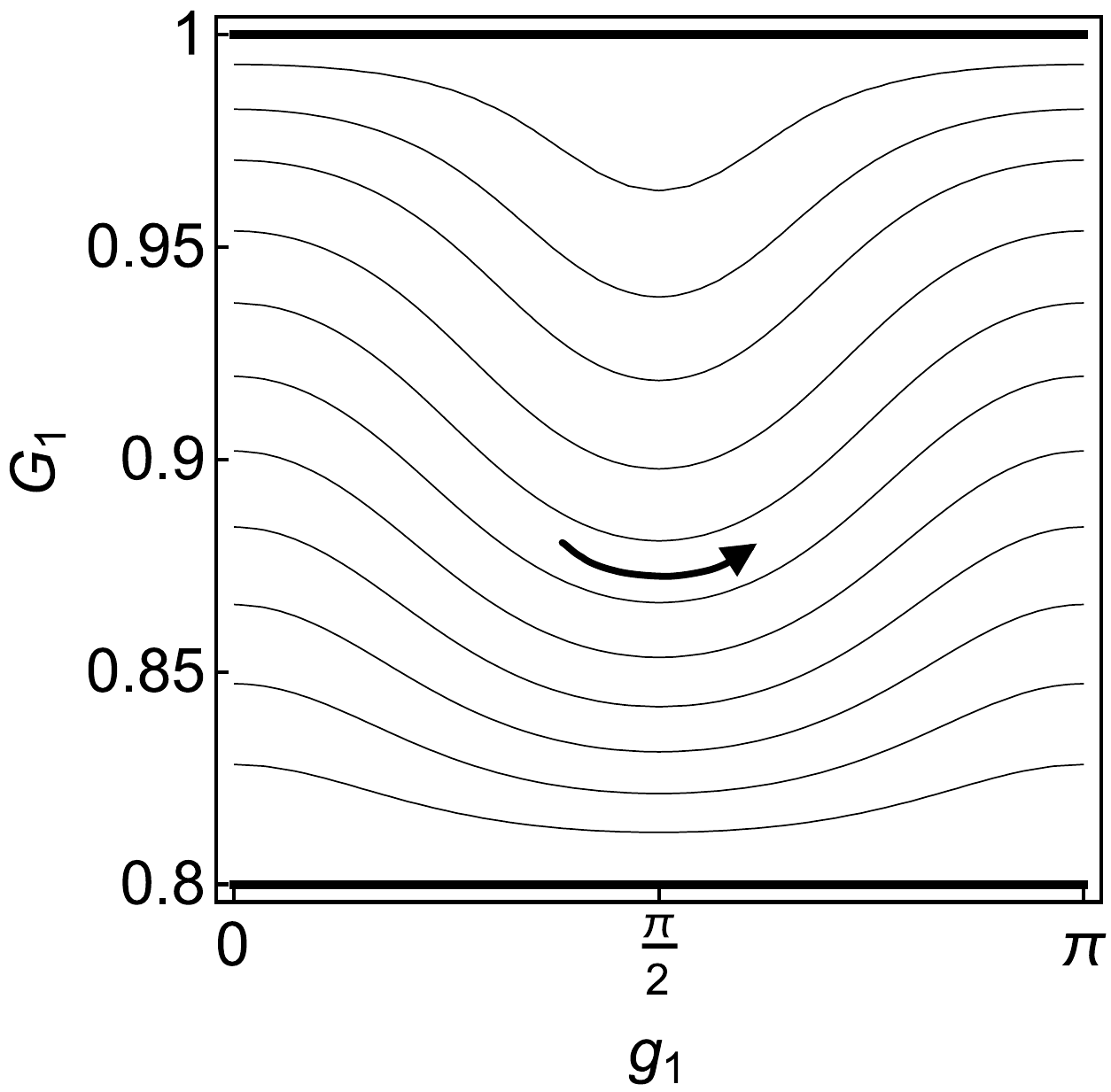}}
\caption{Phase-space evolution  for the region II with $(J_1,p)=(0.8,0.2)$   (same as Fig. \ref{fig:phase3})   that has the four transitions T1 ($\eta_1=-3.94$), T2 ($\eta_2=-1.64$), T3 ($\eta_3=-1.25$) and T4 ($\eta_4=-0.125$). We present the five distinct phase-space patterns P0, P1,P2,P3 and P4 separated by the four transitions.
The separatrixes begin and end at the corresponding unstable fixed points.  The red curves are the upper separatrix $\tilde{G}_{1,+}(g_1,\eta)$, while the blue curves are the lower ones $\tilde{G}_{1,-}(g_1,\eta)$.
 The arrows represent the direction of trajectories.  In panel c, the lower circulating region (below the blue curve) becomes narrow around $g_1=\pi/2$, but the blue curve does not touch the fixed point $G_1=J_1$ shown with the thick sold line.   
}
\label{fig:phase4}
\end{center}
\end{figure*}

To begin with,  we examine a concrete example. In Fig. 7, the point $(J_1,p)=(0.8,0.2)$ belongs to the region II which has the four transitions T1 to T4  as shown in Table 2. Therefore, when increasing $\eta$ from $-\infty$ to 0, its phase space can take the five patterns P0 to P4, divided by the four transitions as follows
$$
{\rm P0 \to T1 \to P1 \to T2  \to P2  \to T3 \to P3 \to T4  \to P4.}  \nonumber
$$
In Fig. 8, we provide the examples of the five patterns P0 to P4 that individually have distinct topological profiles with respect to the separatrixes and the fixed points. Actually, as we see later, these five patterns are the complete set (except for the phase spaces just at the transitions Ti) for the whole regions in Fig. 7, including the region V that has the transition T5 different from T1 to T4 (see Table 2).

In Fig. 8, the separatries are presented with the red and blue curves.
In this paper, we define $G_1=\tilde{G}_{1, +}(g,\eta)$ for the upper separatrix curve above the associated unstable fixed point,\footnote{More precisely, the value of the $G_1$ coordinate is larger than that of the associated unstable fixed point.} and represent it with a red curve.  Here $\tilde{G}_{1, +}(g,\eta)$ should be regarded as a function of $g_1$ and $\eta$ (omitting the dependence on the constant parameters $p$ and $J_1$).  Similarly, we define $G_1=\tilde{G}_{1, -}(g,\eta)$ for the lower separatrix curve below the associated unstable fixed point, showing it with a blue curve.

In Fig. 8b, the pattern P1  has only the upper separatrix curve  $G_1=\tilde{G}_{1, +}(g_1,\eta)$. Since it passes through the unstable fixed point $G_1=J_1$ and satisfies $\mathcal{H}_{\rm T}=$const, the function $\tilde{G}_{1, +}(g_1,\eta)$ is algebraically obtained by solving the following quartic equation for the total Hamiltonian $\mathcal{H}_{\rm T}(g_1,G_1;\eta)$ defined in Eq.(\ref{ham})
\beq
\mathcal{H}_{\rm T}(\cdots, J_1 ;
\eta)=\mathcal{H}_{\rm T}(g_1, \tilde{G}_{1,+}(g_1,\eta) ; \eta),\label{g+}
\eeq
where, in the left-hand side, we explicitly show that our Hamiltonian does not depend on $g_1$ at $G_1=J_1$. 

Meanwhile, the pattern P2 in Fig. 8c has both the upper and lower separatrixes associated with the unstable point at $g_1=0$.
We can obtain the $G_1$-coordinate of the fixed point $G_1^*(\eta)$ from the cubic equation given in Eq.(\ref{eq:4'})
\beqa
G_1^*(\eta)=-\eta\lmk G_1^*(\eta)-\frac{p}{{G_{1}^*}(\eta)^{2}}\rmk
\eeqa
which has a valid solution for $\eta_2<\eta<\eta_3$, as explained in the previous subsection. 
Then, similar to Eq.(\ref{g+}), we can derive the upper $\tilde{G}_{1, +}(g_1,\eta)$ and lower $\tilde{G}_{1, -}(g_1,\eta)$ separatrix curves as the two appropriate solutions for the quartic equation
\beq
\mathcal{H}_{\rm T}(0, G_1^*(\eta) ;
\eta)=\mathcal{H}_{\rm T}(g_1, \tilde{G}_{1,\pm}(g_1,\eta) ; \eta).
\eeq

 For the pattern P3 shown in Fig. 8d, we can derive the expression for the lower separatrix $G_1=\tilde{G}_{1, -}(g_1,\eta)$ by using
\beq
\mathcal{H}_{\rm T}(\cdots, 1 ;
\eta)=\mathcal{H}_{\rm T}(g_1, \tilde{G}_{1,-}(g_1,\eta) ; \eta),
\eeq
as in the case for the pattern P1 (see Eq.(\ref{g+})).

Now we briefly discuss the phase-space structure for the patterns P0 to P4. 
In Fig. 8, a librating region exists for the patterns P1, P2 and P3, and the orientation of the libration is  counter-clockwise. Meanwhile, the region above the upper separatrix (red curve) is always  circulating in the retrograde direction, dominated by the stellar potential. This is also  true for the whole region of P0, as easily expected from the continuity of the system (see Figs. 8a and 8b).
In contrast, the region below the lower separatrix (and also the whole region of P4) has prograde circulation. Here, the quadrupole or 1PN effect dominates the apsidal precession.

In Fig. 8,  only the pattern P2 simultaneously has the three types of motions,  divided by the two separatrixes. This phase space structure is similar to that of a simple pendulum whose Hamiltonian is given by
$
(P^2+\sin 2Q)/2
$
for the conjugate variables $(Q,P)$ (but without the two fixed points corresponding to  $G_1=J_1$ and 1 for our Hamiltonian).

So far, we have studied the evolution of the phase-space structures specifically for the region II in Fig. 7. Below, we discuss other regions. As shown in Table 3, the transitions of the regions I, III, and IV are subsets of those for the region II.

For example, the region II has the single transition T3 and, therefore,  its evolutionary sequence is given as
$$
\rm P2 \to T3 \to P3.
$$
To demonstrate this explicitly, in Fig. 9, we present the snapshots for $(J_1,p)=(0.2,0.2)$ for which we have the transition epoch  $\eta_3=-1.25$. 
Note that, when decreasing $\eta$ down toward $-\infty$, the area of the libration region approaches to 0, and the red and blue curves become more symmetric with respect to the stable fix point shown with the filled circle. The $G_1$ coordinate of the fixed point approaches $p^{1/3}$.

\begin{figure*}
\begin{center}
\subfigure[P2 at $\eta=-200$]{
\includegraphics[width=4cm]{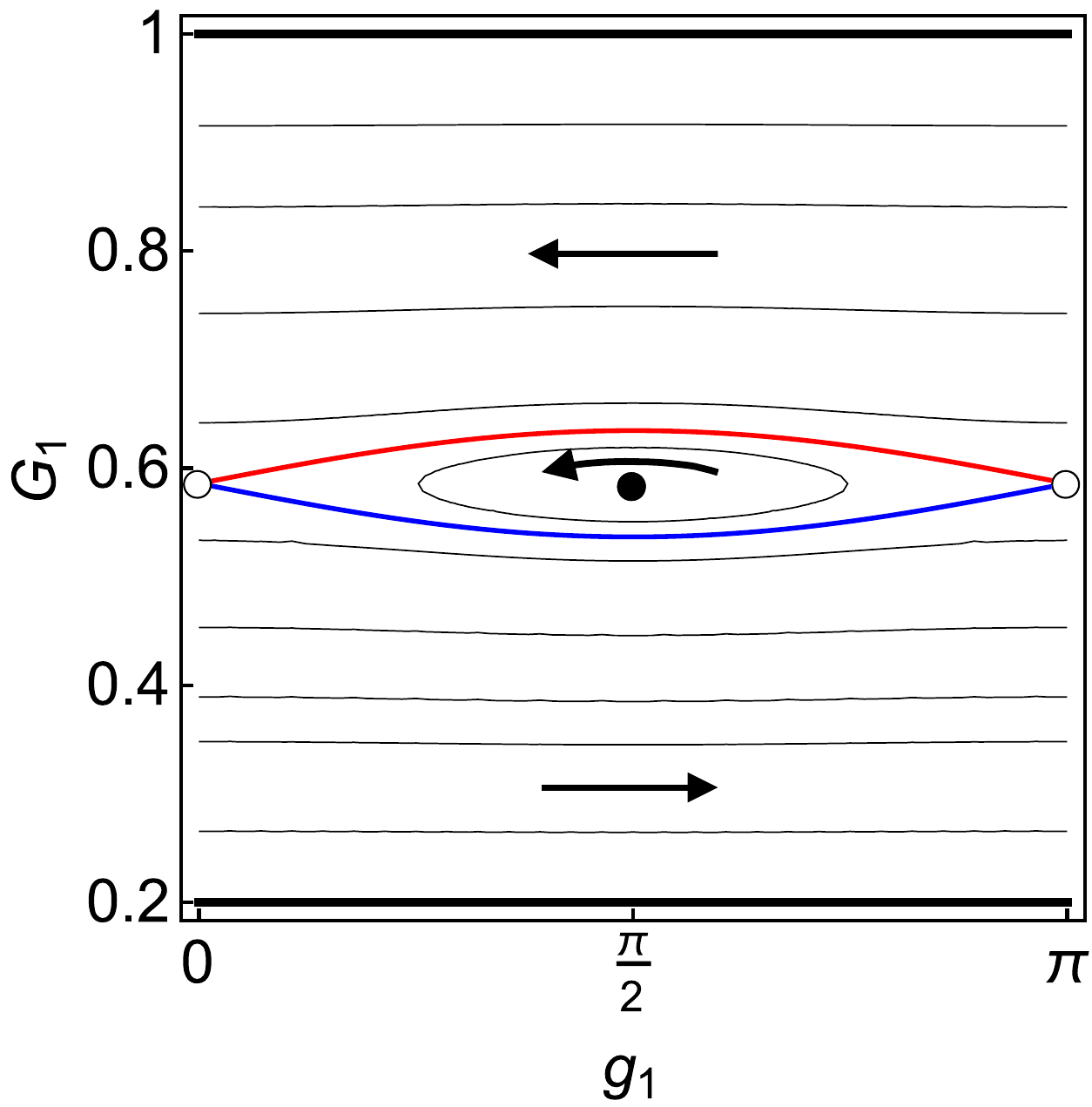}}
\subfigure[P2 at $\eta=-10$]{
\includegraphics[width=4cm]{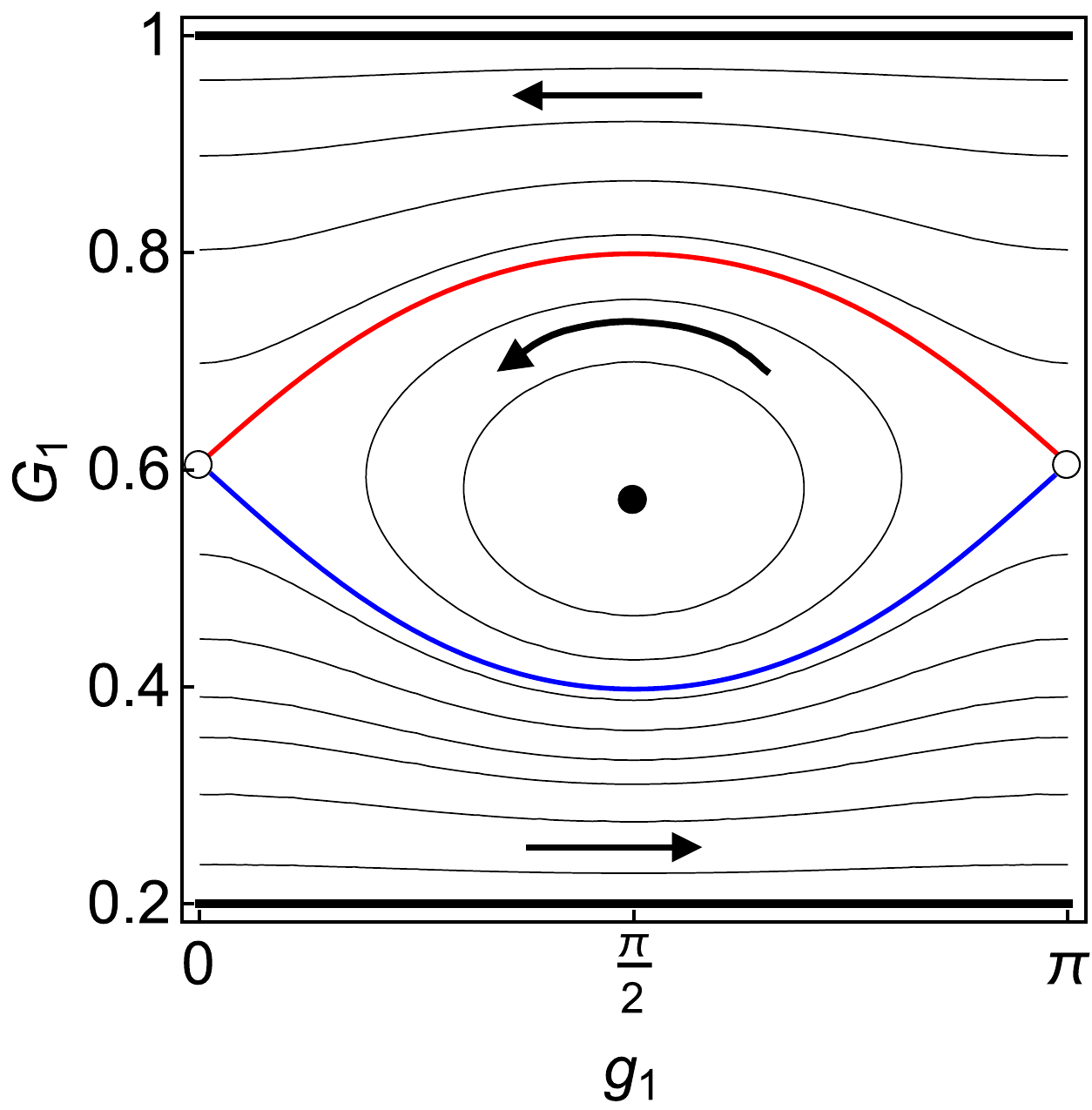}}
\subfigure[P2 at $\eta=-2$]{
\includegraphics[width=4cm]{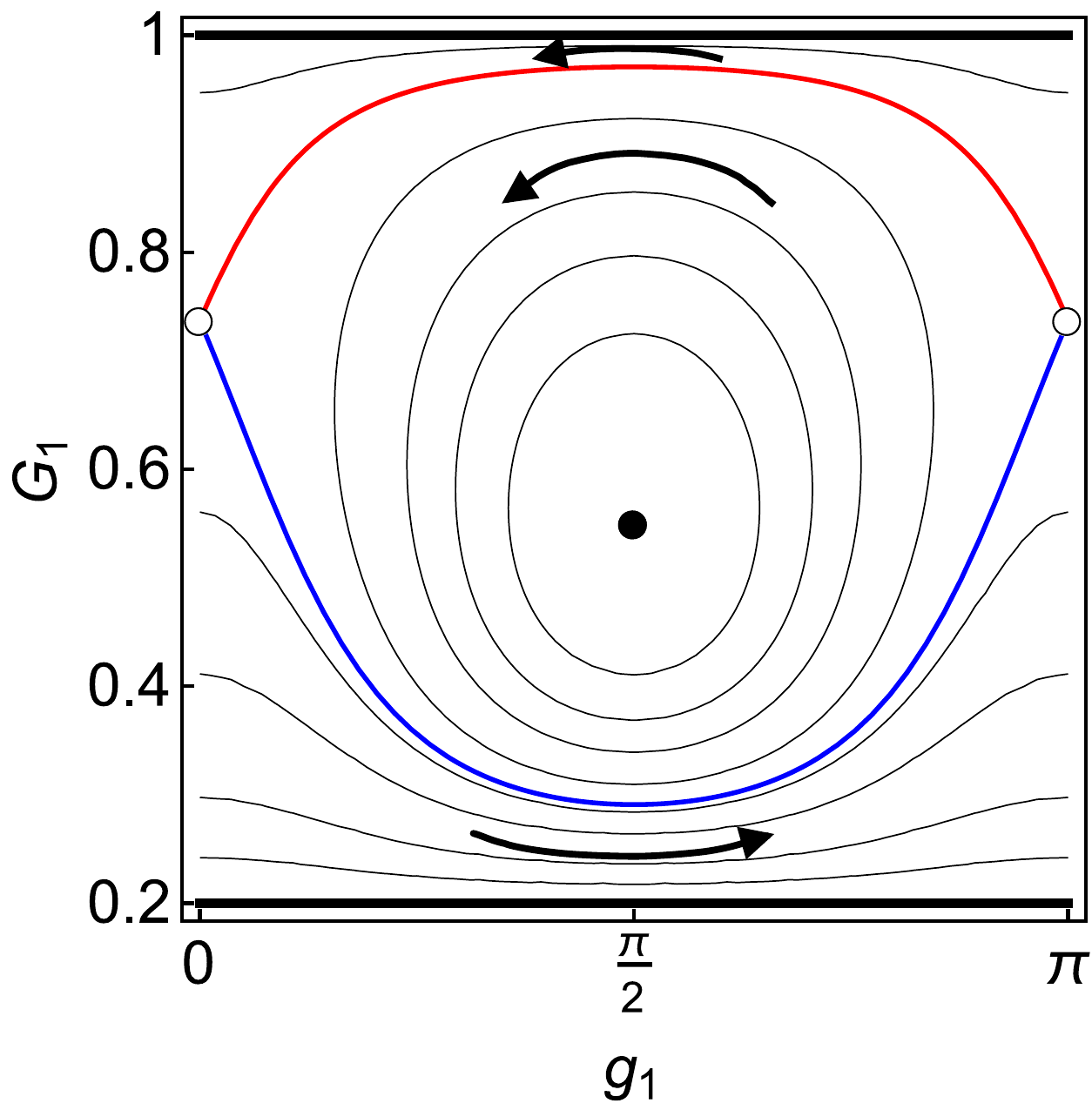}}\\
\subfigure[T3 at $\eta=\eta_3=-1.25$]{
\includegraphics[width=4cm]{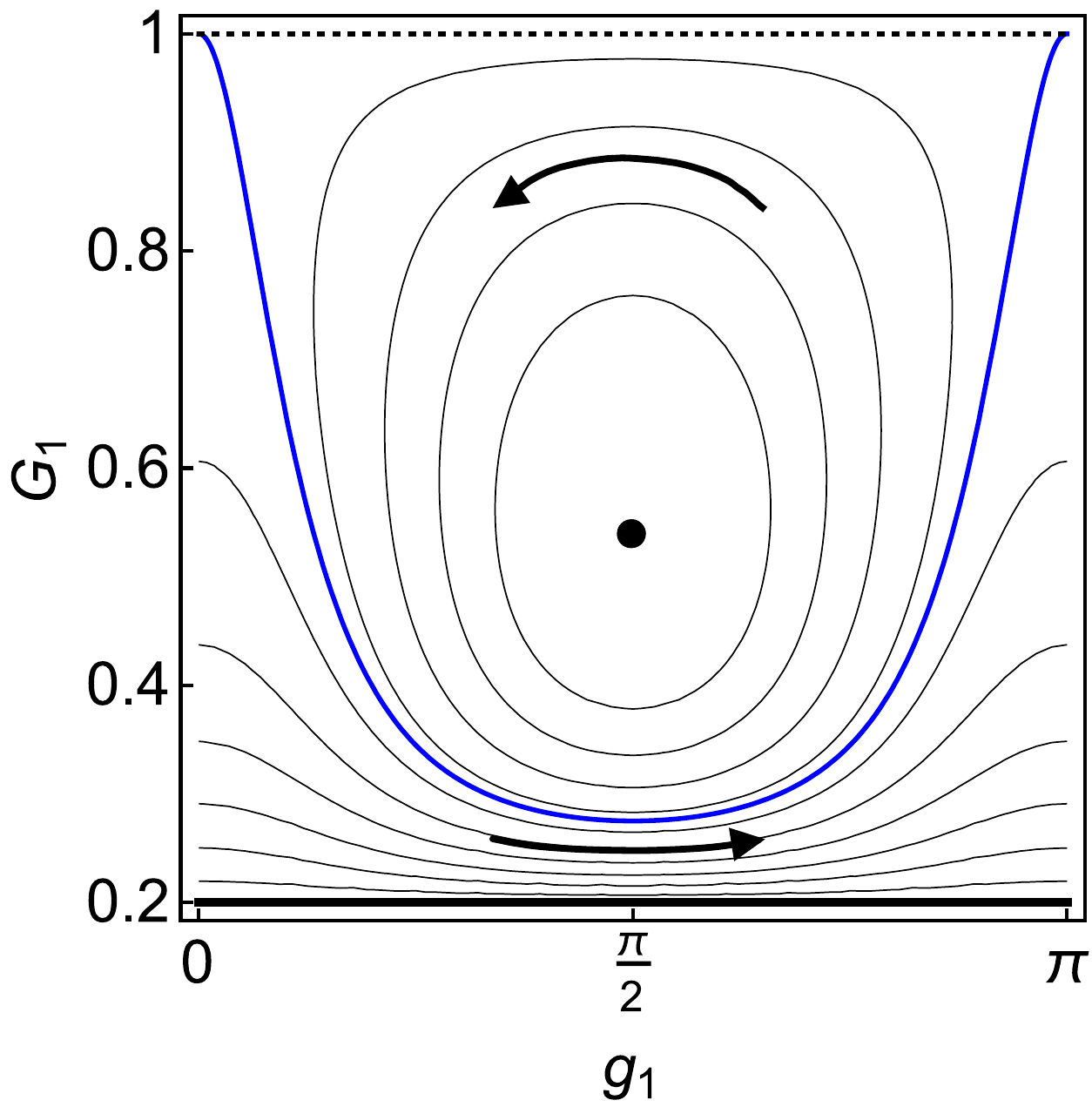}}
\subfigure[P3 at $\eta=-0.5$]{
\includegraphics[width=4cm]{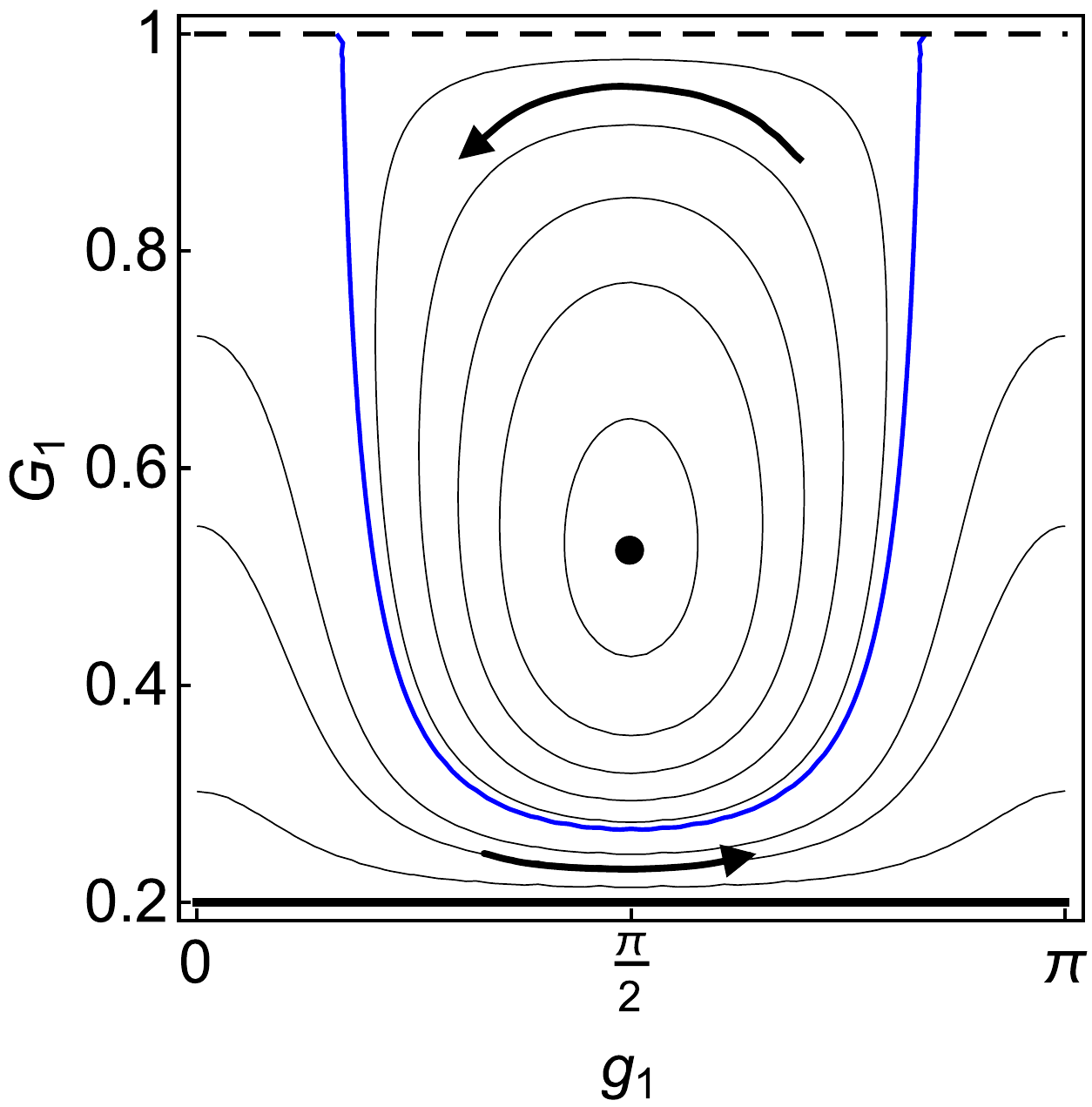}}
\caption{Evolution of the phase space for $(J_1,p)=(0.2,0.2)$ (in the region III) that has only one transition T3 at $\eta_3=-1.25$.  The patterns P2 and P3 are separated by the transition T3.   For P2, the libration region is  between the red and blue separatrixes and its area asymptotically approaches to 0 in the limit $\eta\to -\infty$. }
\label{fig:phase2}
\end{center}
\end{figure*}

In the same manner, for the region I, we have the time sequence
$$
\rm P0 \to T1 \to P1 \to T2 \to P2 \to T3 \to P3,
$$
and
$$
\rm  P2 \to T3 \to P3  \to T4 \to P4
$$
for the region IV.

On the other hand, the region V has the single transition T5 that is essentially an inverse of T4 (see Table 2), and we have the sequence 
$$
\rm P4 \to T5 \to P3.
$$
Finally, the region VI has no transition and its phase space always corresponds to the pattern P4.

When we drop the stellar potential term ${\cal H}_{\rm SP}$ with $\rho_1=0$, the combination $\eta\rq{}\equiv \eta p\le 0$
now becomes the appropriate parameter to characterize the contraction of the outer orbit (see \S 2.2).  
We can easily confirm that, in this case, only the two patterns P3 and P4 are realized, as for the standard KL-mechanism.
As we see later, the patterns P1 and P2 cause interesting effects for the inner orbit, but these appear only with the stellar potential term ${\cal H}_{\rm SP}$.

\section{Bifurcation at separatrix crossing}
In the previous section, we discussed how the phase-space structure evolves along with the contraction of the outer orbit. 
We paid special attention to the profiles of the separatrixes. In this section, we study the evolution of individual trajectories
in the time-varying phase space, such as Figs. 8 and 9.  In \S 5.1, we introduce the idea of the adiabatic invariant and then, in \S 5.2, apply it 
to the numerical demonstrations in Figs. 3 and 4.
In \S 5.3, we make somewhat formal arguments on the probabilistic bifurcations at the separatrix crossings for the pattern P2. In \S 5.4, we discuss the  probabilistic bifurcations 
for our hierarchical triple systems. In \S 5.5, for individual orbits, we examine the maximum eccentricities observed in a certain time interval.

\subsection{ adiabatic invariant}
Firstly, we explain the adiabatic invariant for a one-dimensional Hamiltonian $\mathcal{H}(q,p;\lambda)$ that contains a time-varying parameter $\lambda$.  In the phase space $(q,p)$,  we consider the time evolution of a periodic trajectory described by this Hamiltonian.  If the timescale of the variation of $\lambda$ is much larger than the rotation period of the trajectory, the following integral is conserved
\beqa
S(\lambda)\equiv\oint p(q,\lambda) dq, \label{ada}
\eeqa
and known as an adiabatic invariant \citep{landau1969,peale1987,ssd}.  Here the integral is taken for the trajectory that can be 
effectively regarded as periodic.  In the phase space, this integral corresponds to the area inside the periodic trajectory. 
This geometrical interpretation allows us to intuitively follow the time evolution of a trajectory in the phase space. We just need to track contours whose relevant areas are the same.

For our Hamiltonian (7), the characteristic time-scale of the orbit is $O(1)$, and thus we should have $|\eta/(d\eta/dt)|\gg1$ for applying the adiabatic invariant.

Here, we should comment on a technical detail about the definition of the adiabatic invariants for circulating trajectories. 
In Fig. 8, unlike a librating trajectory, a circulating trajectory is not literary periodic
in the $(g_1,G_1)$ coordinate. But, it becomes periodic with the regular coordinate $(x\rq{},y\rq{})$ defined in Eq. (\ref{ct1}), and the adiabatic invariant can be straightforwardly defined. Inversely, in the original coordinate $(g_1,G_1)$, this adiabatic invariant for a circulating  trajectory corresponds to the area between the line $G_1=J_1$ and the trajectory, and  we employ this geometrical interpretation below.  Note also that the area for a librating trajectory 
is identical in both coordinates, as they are related by a canonical transformation.  Here, we implicitly assume to appropriately handle the contracted range for the angular variable $g_1$ ($[0,2\pi)$ to $[0,\pi)$ as explained in \S 4).

For using the adiabatic invariant, we need a careful analysis when a trajectory crosses a separatrix.
 A separatrix crossing  is a quite interesting phenomenon and is the underlying mechanism behind the differences between Figs. 3 and 4.  Since the orbital period of a separatrix is infinite, the conservation of the integral  (\ref{ada}) is no longer guaranteed, and,  indeed, the adiabatic invariant could have a jump at a separatrix crossing \citep{ssd}. Still, we can estimate the post-crossing adiabatic invariant, using the continuity of the trajectory.

Now we concretely discuss the separatrix crossings for patterns P1, P2 and P3 shown in Fig. 8.
For the pattern P1, let us consider a trajectory in a retrograde circulation  (above the red separatrix) in Fig. 8b, with its adiabatic invariant $S_0$. The separatrix crossing occurs when the area of the librating region (inside the red separatrix) increases to $S_0$.  After the crossing, the trajectory smoothly become a librating trajectory around the fixed point at $g_1=\pi/2$. Due to the continuity, its adiabatic invariant is same as the original value $S_0$.

Next, for the pattern P3 in Fig. 8d,  we examine a librating trajectory above the blue separatrix, with its adiabatic invariant $S_0\rq{}$
 (the area inside the trajectory around the stable fixed point at $g_1=\pi/2$).
When the area of the whole librating region bounded by the blue separatrix decreases down to $S_2\rq{}$, the trajectory crosses the separatrix and starts prograde circulation.  At the crossing,  the adiabatic invariant has a gap and becomes 
\beq
(1-J_1)\pi-S_2\rq{}.  \label{sa}
\eeq
 Here, $(1-J_1)\pi$ is the total area of the phase space.

As discussed above, the separatrix crossing for the two patterns P1 and P3 (and also P4) can be easily understood. Therefore, hereafter, we concentrate on the crossing for the pattern P2 that has two separatrixes and three distinct regions, as in Figs. 9a-9c.

To begin with, we define the following two integrals
\beqa
S_{+}(\eta)&=&\int_{0}^{\pi} (\tilde{G}_{1,+}(\eta) - J_1) dg_{1},\\
S_{-}(\eta)&=&\int_{0}^{\pi} (\tilde{G}_{1, -}(\eta) - J_1) dg_{1}.
\eeqa
respectively corresponding to the areas below the upper (red) and lower (blue) separatrixes of the pattern P2.

As an example of a separatrix crossing for the patten P2,  in Fig. 9a, 
we consider a retrogradely circulating trajectory above the red separatrix with its adiabatic invariant $S_0\rq{}\rq{}$.  As $\eta$ increases, the area $S_+(\eta)$ grows and the separatrix crossing occurs at $\eta=\eta_c$ where we have
\beq
S_0\rq{}\rq{}=S_+(\eta_c). \label{scross}
\eeq
After the crossing, the trajectory  shifts to either of the following two trajectories. One is a librating motion inside the two separatrixes and the adiabatic invariant becomes 
\beq
S_+(\eta_c)-S_-(\eta_c). \label{lib}
\eeq
The other is the circulating one below the blue separatrix, with the post-crossing value
\beq
S_-(\eta_c). \label{und}
\eeq
The branching ratio of these two will be discussed in \S 5.3. 

\subsection{Tracing the evolution of trajectories}

In this subsection, we discuss the time evolution for the two runs, R3 and R4 already introduced in \S 3 (see also Figs. 3 and 4). These are given for the parameters $(J_1,p)=(0.2,0.2)$, and the phase space has the single transition T3 at $\eta_3=-1.25$ (see Table 2 and Fig. 7).

At the initial epoch $\eta=-20$, the two trajectories commonly have $G_1=0.95$ and $g_1\sim 0$, and thus their adiabatic invariants are effectively the same $2.37$.  From this value and  the condition (\ref{scross}),   we can predict the epoch $\eta_c=-1.66$ for the separatrix crossing and  can also evaluate the  areas at that time
\beq
S_+(\eta_c)=2.37,~~S_-(\eta_c)=0.68.
\eeq

In Figs. 10 and 11, at $\eta=-20$, -1.68, -1.64 and -1.20, we present  the snapshots of the two runs R3 and R4 obtained by numerically integrating the canonical equations (as described in \S 3), along with the separatrixes. The boundaries of the green regions are the contours of our Hamiltonian,  determined analytically from the relevant adiabatic invariants.  Here, we appropriately included the predicted changes at the separatrix crossings, as explained in \S 5.1. 
More specifically, in Fig. 10, the  areas for the green regions are respectively, (a) $S_+(\eta_c)=2.37$, (b) 2.37, (c) $S_-(\eta_c)=0.68$ and (d) 0.68.  Meanwhile, in Fig. 11, the areas are (a) $2.37$, (b) 2.37, (c) $S_+(\eta_c)-S_-(\eta_c)=1.69$ and (d) $(1-J_1)\pi-1.69=0.82$.

 From the good agreements between the numerical results (cyan points) and the predictions (the boundaries of the green regions), we can  confirm the usefulness of the adiabatic invariant and its transitions at separatrix crossings.

In Figs. 10 and 11, the two trajectories have almost the same evolution before the separatrix crossing around the predicted value $\eta_c=-1.66$. After the crossing, the two trajectories show distinct bifurcation.  As shown in Fig. 10, the trajectory of the run R3 starts a prograde circulation below the blue separatrix,  and its eccentricity $e_1=\sqrt{1-G_1^2}$  suddenly increases, consistent with Fig. 3.
 Its later evolution is well predicted by the new adiabatic invariant $0.68$  with no additional separatrix crossing.

On the other hand, in Fig. 11c,  after $\eta=\eta_c$, the trajectory of the run R4  has a librating motion between the two separatrixes, and the range of its eccentricity oscillation becomes larger, including the original range (as observed in Fig. 4).  
This trajectory has the secondary separatrix crossing at $\eta=-1.22$, and temporarily takes  $G_1 \simeq 1$, corresponding to $e_1\simeq 0$ (also seen in  Fig. 4). 
 In this manner, we can understand the notable differences between Figs. 3 and 4, though the structure of the separatrixes.



\begin{figure*}
\begin{center}
\subfigure[$\eta=-20$]{
\includegraphics[width=4cm]{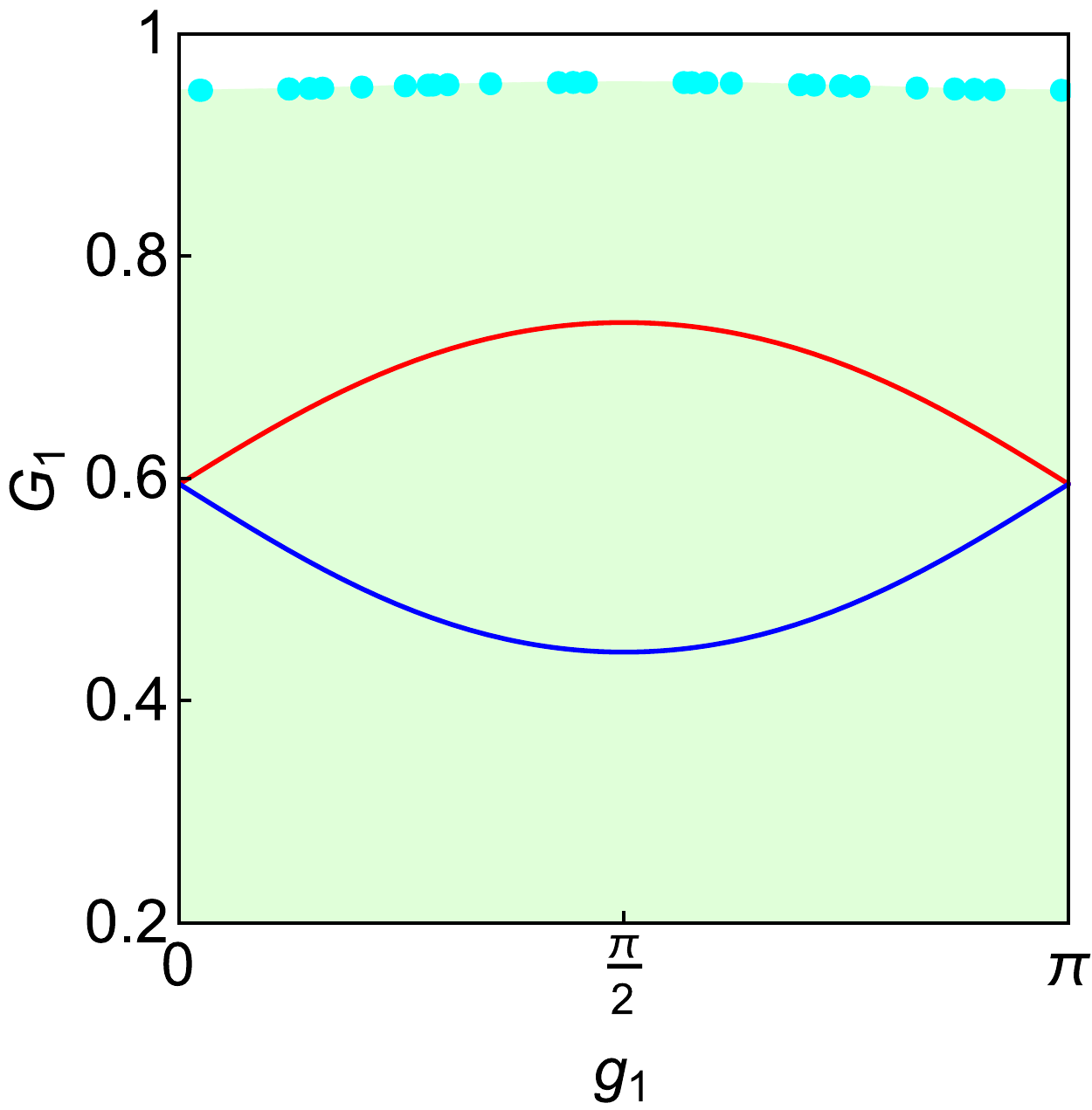}}
\subfigure[$\eta=-1.68$]{
\includegraphics[width=4cm]{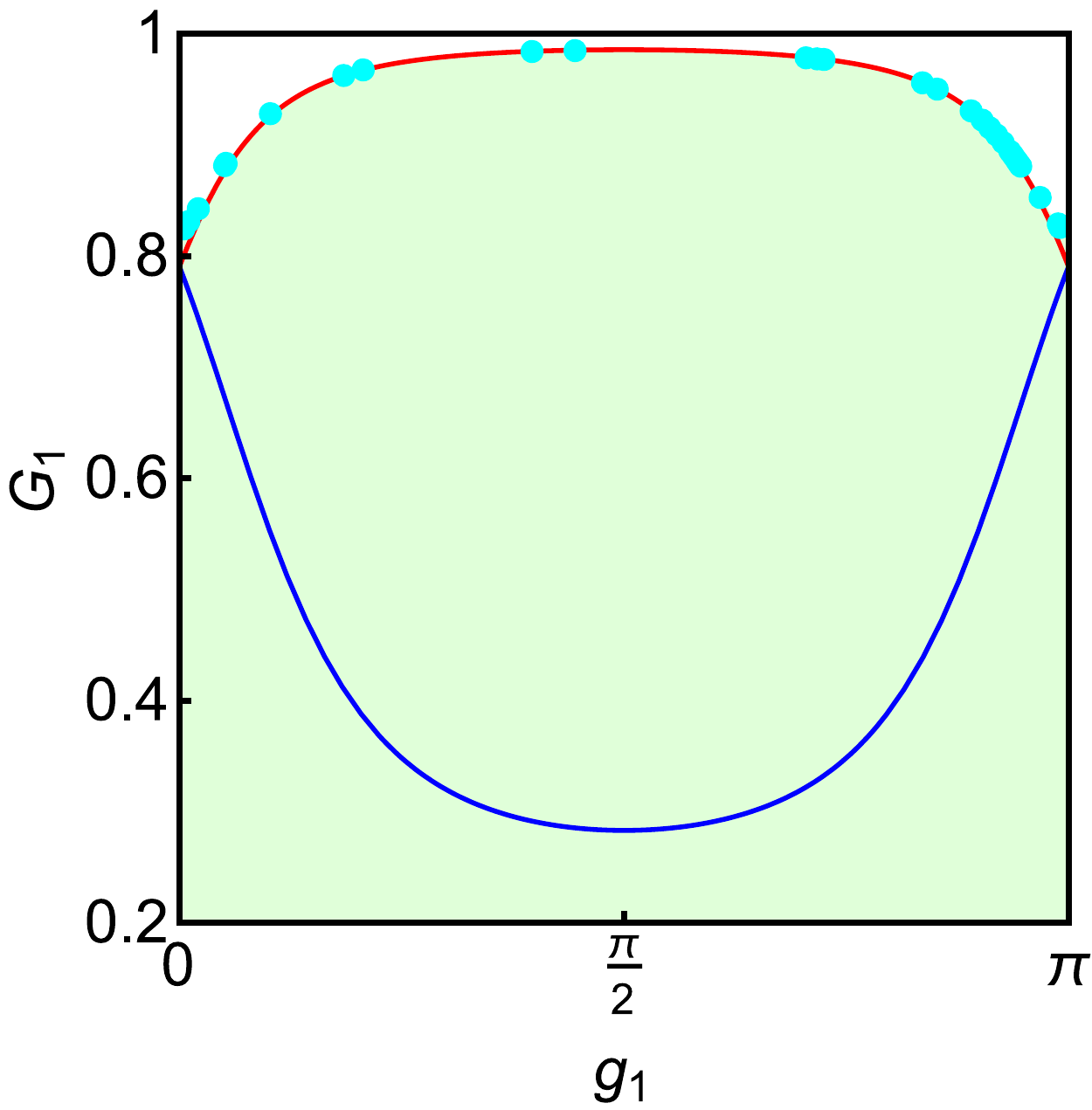}}
\subfigure[$\eta=-1.64$]{
\includegraphics[width=4cm]{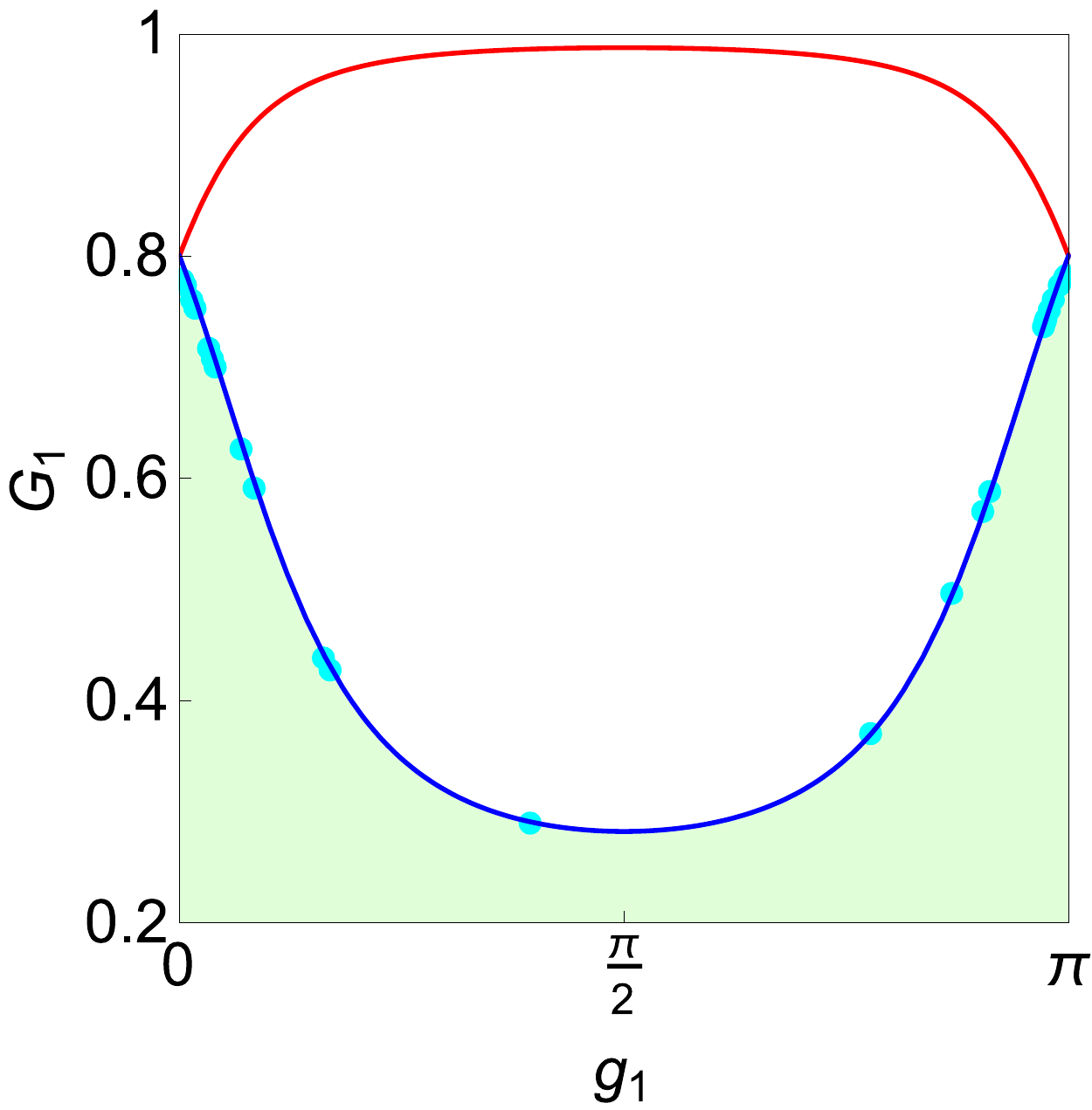}}
\subfigure[$\eta=-1$]{
\includegraphics[width=4cm]{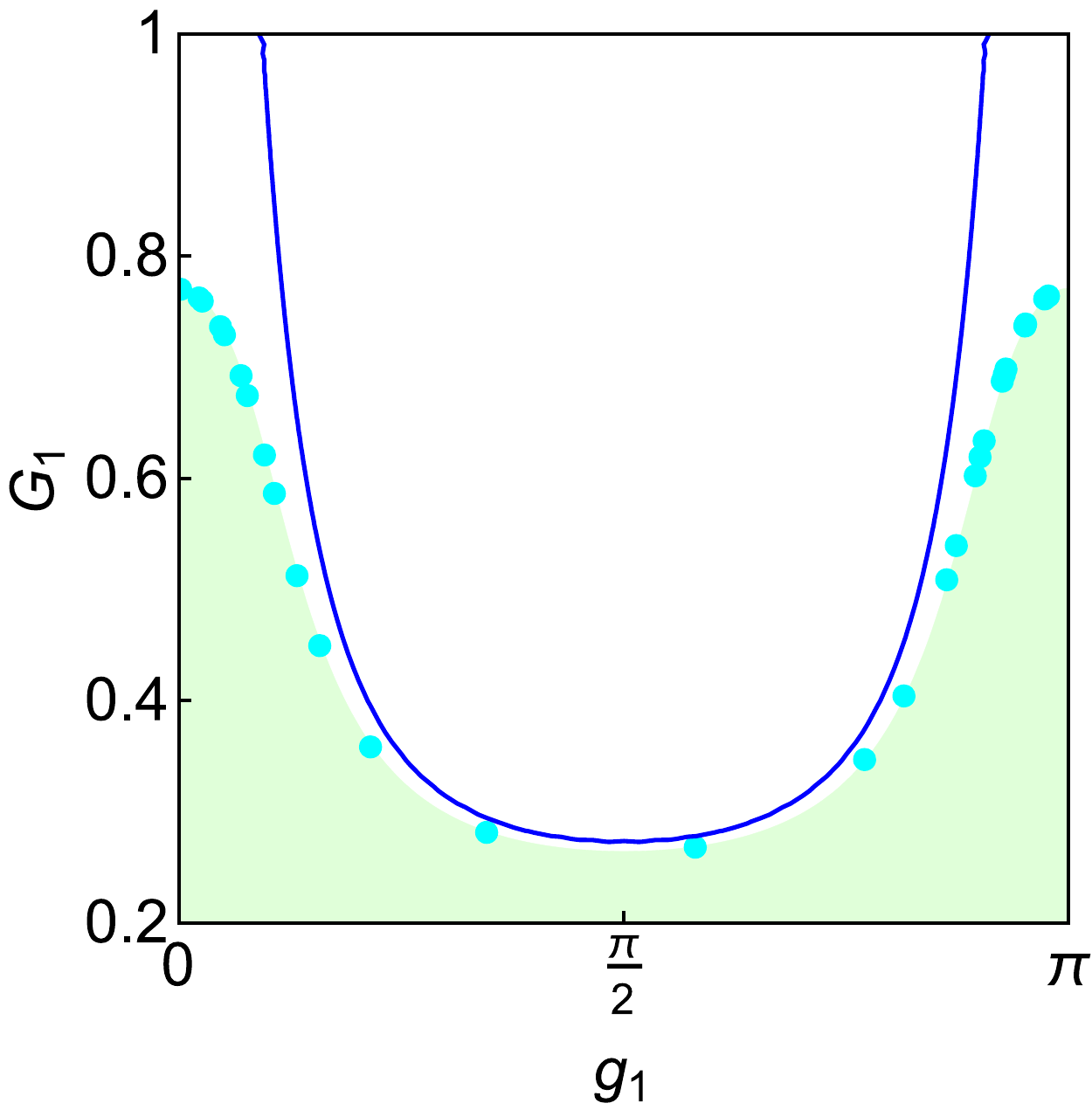}}
\caption{Evolution of a trajectory from the initial condition $(G_1,g_1)=(0.950,0.04)$ at $\eta=-20$ with $(J_1,p)=(0.2,0.2)$.
 The cyan points represent results obtained from the numerical runs R3 (also shown in  Fig.3), and the areas of the green regions show the corresponding adiabatic invariants.
 The red and blue lines are the upper and lower
separatrixes $\tilde{G}_{1.+}$ and $\tilde{G}_{1,-}$. After the encounter with the red separatrix $\tilde{G}_{1.+}$ at $\eta\simeq-1.66$, 
the trajectory moved to the lower circulating region under the blue separatrix $\tilde{G}_{1.-}$.  The green regions of (a) and (b)  have the identical area. The same is true for (c) and (d). 
}
\label{fig:ad1}
\end{center}
\end{figure*}

\begin{figure*}
\begin{center}
\subfigure[$\eta=-20$]{
\includegraphics[width=4cm]{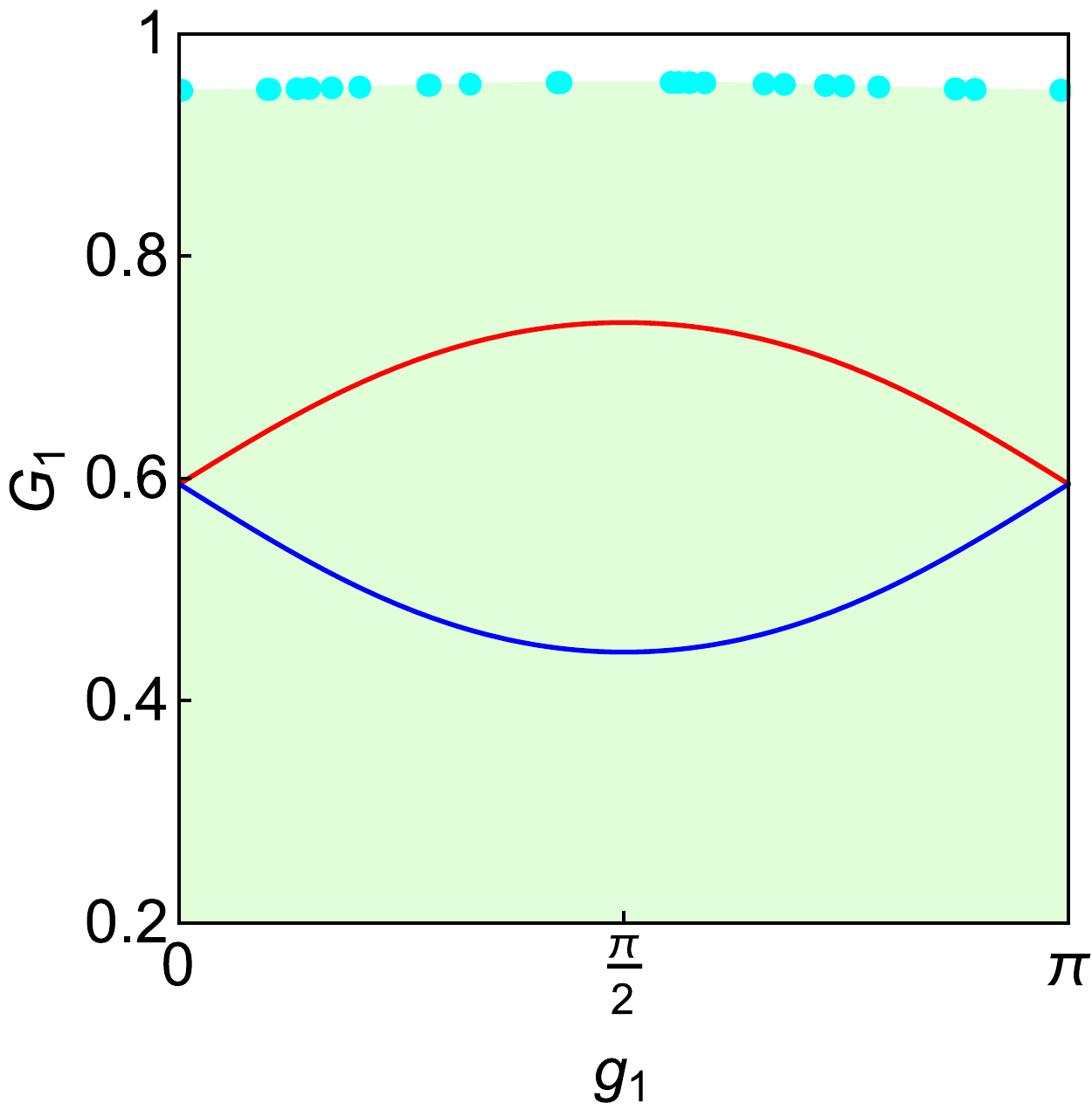}}
\subfigure[$\eta=-1.68$]{
\includegraphics[width=4cm]{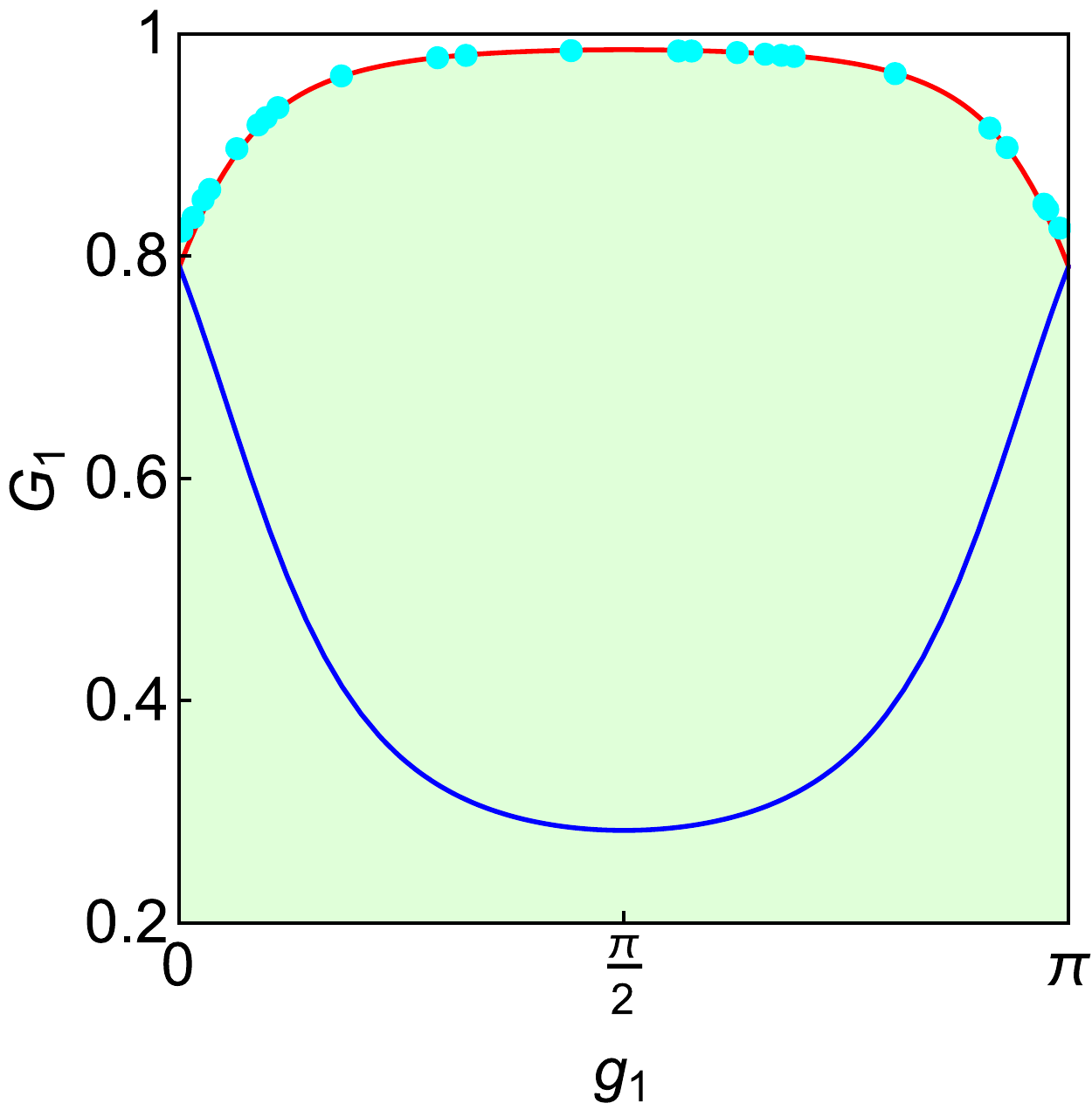}}
\subfigure[$\eta=-1.64$]{
\includegraphics[width=4cm]{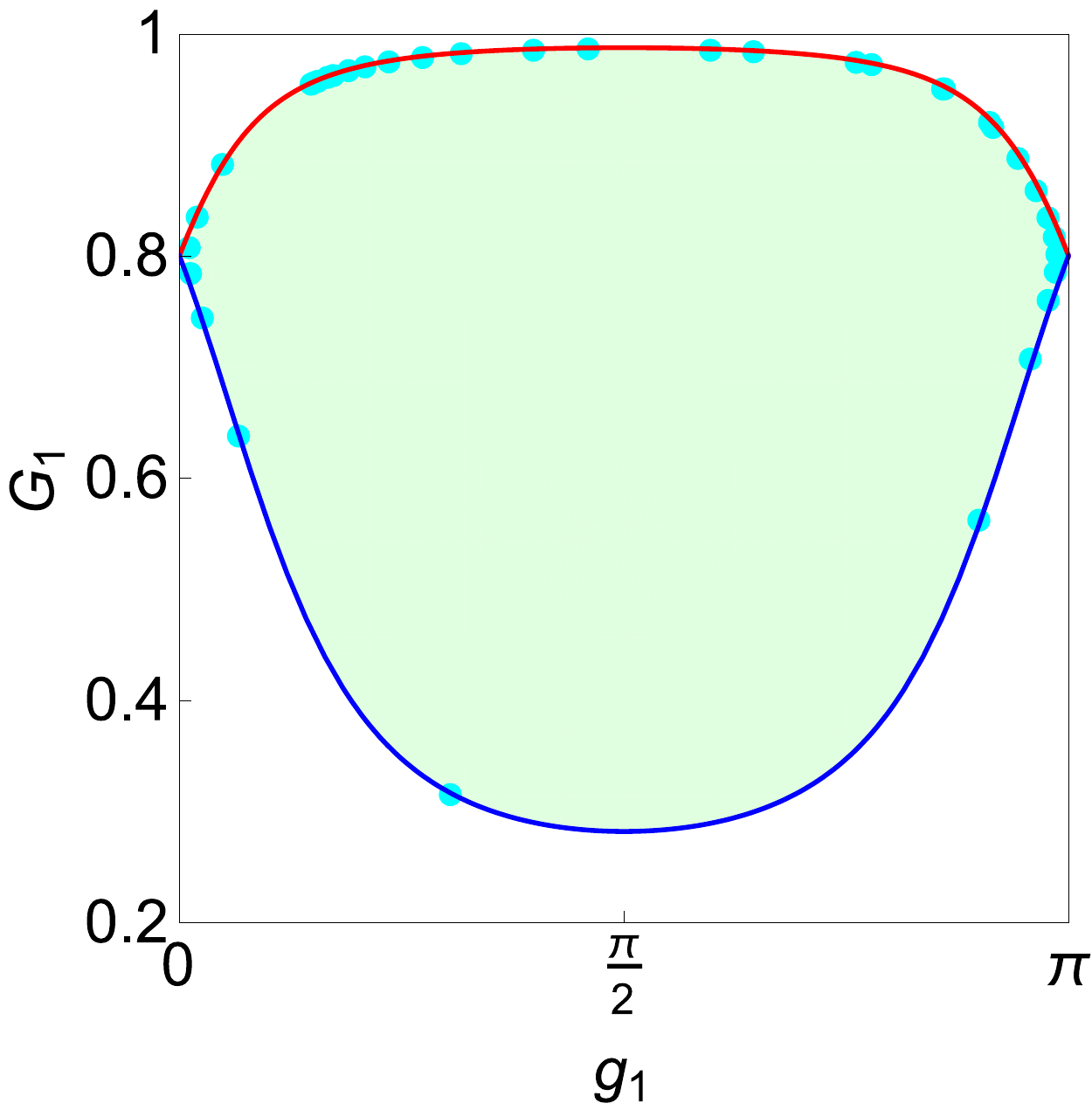}}
\subfigure[$\eta=-1$]{
\includegraphics[width=4cm]{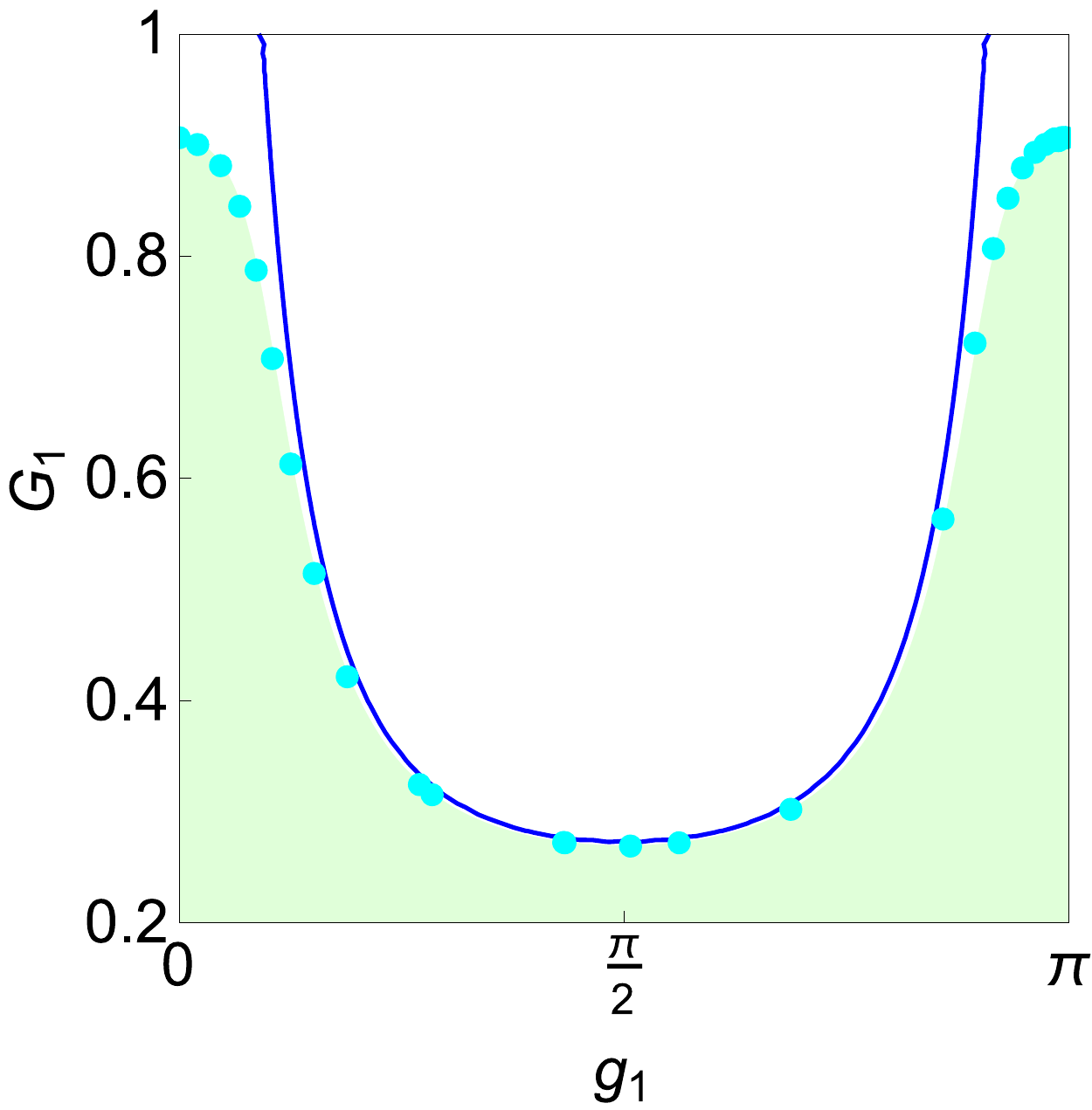}}
\caption{Same as Fig. \ref{fig:ad1}, but from the initial
conditions are $(G_{1}, \, g_1)=(0.950, \,0.01)$.  The cyan points are results from the run R4 (also shown in  Fig.4).
 The trajectory is temporarily 
captured into libration region after the first separatrix
crossing at $\eta\simeq-1.66$. Subsequently, it moves to the lower circulating region, crossing the blue separatrix at $\eta\simeq-1.22$.  The green regions of (a) and (b)  have the identical area. But (c) and (d) have different areas. }
\label{fig:ad2}
\end{center}
\end{figure*}

\subsection{Probability of bifurcation}

As discussed so far, the phase-space pattern P2 has the two circulating regions  and the intermediate librating region. In Figs. 10 and 11, when an upper circulating trajectory crosses the red
separatrix, it could either move to the lower circulating region or the librating region, depending sensitively on the initial conditions.  Even though the canonical equations (\ref{can}) are purely deterministic, we can effectively regard this bifurcation process as probabilistic, given the strong dependence of the initial conditions. 

In this subsection, concentrating on the pattern P2, we discuss the branching ratio specifically for the upper circulating trajectory at the separatrix crossings (see \citealt{henrard1982,borderies1984,ssd,gd} for related analysis on mean motion resonances). We can easily extend our arguments for the separatrix crossings from other two regions (as briefly mentioned at the end of this subsection). 

We  consider the time evolution of the  phase space associated with given parameters $(J_1,p)$. To begin with, we define the following two quantities
\beq
v_{+}(\eta)\equiv\frac{dS_{+}(\eta)}{d\eta},~~v_{-}(\eta)\equiv\frac{dS_{-}(\eta)}{d\eta},
\eeq
representing the variation rates of the areas below the two separatrixes.  Additionally,  we define $P_L(\eta)$ as the transition probability of an upper circulating trajectory into the librating region, just crossing the upper separatrix  at the epoch $\eta$.
This definition should be correctly kept in mind, for the arguments below.  Our goal in this subsection is provide the simple expression for $P_L(\eta)$.

Actually, for our system, we generally have $v_{+}(\eta)>0$ for the pattern P2 with $v_{+}(\eta_3)=0$ at the transition T3. 
During the time interval between $\eta$ and $\eta+\Delta \eta$, the area $v_+(\eta)\Delta \eta>0$ newly crosses the upper separatrix downwardly.  The key issue here is how this eroded upper phase-space element is redistributed to the lower circulating or the intermediate  librating regions. Considering the Liouville's theorem, the probability $P_L(\eta)$ is given by the fraction of the original area  $v_+(\eta)\Delta \eta$  redistributed to the intermediate librating region. 
  For our system with $v_+>0$, depending on  $v_-$, we have the following three cases C1-C3 (see Problem 3.43 in \citealt{gd}).

(C1) $v_+(\eta)\ge 0\ge v_-(\eta)$. The area of the intermediate librating region increases by $(v_+-v_-)\Delta \eta>0$, but the lower circulating region decreases by $v_- \Delta \eta<0$.  Therefore, the upper circulating trajectory will be always absorbed into the librating region, and we identically have  $P_L=1$.

(C2)  $v_+(\eta)\ge v_-(\eta)\ge 0$. The area of the lower circulating region increases by $v_-\Delta \eta>0$ and, at the same time, that of the lmiddle  librating region increases by $(v_+-v_-)\Delta \eta>0$. Both increments are compensated by the decrement of the  upper circulating region.  Therefore,  the  transition probability is given as
\beqa
P_L(\eta)=\frac{v_+(\eta)-v_-(\eta)}{v_+(\eta)}=1-\frac{v_{-}(\eta)}{v_{+}(\eta)}
\label{eq:prob}.
\eeqa

(C3)   $v_-(\eta)\ge v_+(\eta)\ge 0$.  Only the lower circulating region increases,  and thus it always absorbs the upper circulating trajectory, resulting in $P_L=0$.

In order to clarify the boundaries between these three cases, we define the two epochs $\eta_{\rm eq}$ and $\eta_{\rm z}$ with the following conditions
\beq
v_+(\eta_{\rm eq})=v_-(\eta_{\rm eq}),~~~
v_-(\eta_{\rm z})=0.
\eeq
The boundary between C2 and C3 is at $\eta=\eta_{\rm eq}$, and that between C1 and C3 is at $\eta=\eta_{\rm z}$.

In the next subsection, we concretely evaluate the probability $P_L$ as a function of the initial eccentricity $e_{\rm 1,i}$ given at a large negative $\eta$($\ll -1$). For $\eta \ll -1$, the $g_1$ dependence can be ignored for our Hamiltonian (\ref{ham}), and thus its contour line is nearly parallel  to the $g_1$-axis (see Fig. 8a).  This allows us to simply evaluate the initial adiabatic invariant $S_i$ as follows
\beq
S_i(e_{\rm 1,i})\equiv \pi \lmk\sqrt{1-e_{\rm 1,i}^2}-J_1\rmk. \label{eli}
\eeq
Then, we can  relate the epoch of the separatrix crossing $\eta_c$ with the initial eccentricity $e_{\rm1,i}$, using the following equation
\beq
S_i(e_{\rm 1,i})=S_+(\eta_c). \label{trans}
\eeq
We formally express their relation by $e_{\rm 1,i}(\eta_c)$ and $\eta_c(e_{\rm 1,i})$.
For example, as a function of the initial eccentricity 
$e_{\rm 1,i}$, the transition probability is simply given by
\beq
P_L[\eta_c(e_{\rm 1,i})].
\eeq

Here, it should be noted that, with the identity $v_+>0$ in the phase P2, the probabilistic bifurcation corresponding to C2 can be realized only for the separatrix crossing from the upper circulating region. This is because the trajectories in the lower circulating region and the middle librating region cannot move into the upper circulating region and are not probabilistic.

\subsection{Application of the probability formula to our systems}

Now, we provide some examples for the bifurcation probability $P_L(e_{\rm 1,i})$ as a function of the initial eccentricity $e_{\rm 1,i}$  defined at $\eta \ll -1$.  We should recall that, in \S 5.3 and 5.4, we deal with the separatrix crossing and associated bifurcation only for the pattern P2. 

The pattern P2 is realized for the parameters $(J_1,p)$ in the regions I-IV in Fig. 7. As mentioned earlier, in these regions, we always have $v_+>0$ for P2 with $v_+(\eta_3)=0$ at the final epoch $\eta=\eta_3$  (the transition T3).  Since the upper separatrix converges to $G_1=1$ at $\eta=\eta_3$, we also have the corresponding eccentricity $\lim_{\eta \to \eta_3}e_{\rm 1,i} (\eta)=0$.

\begin{table}
\caption{The transitions realized  for our three examples of  $(J_1,p)$.  The important transition epochs $\eta$ and the associated eccentricities $e_{\rm 1,i}(\eta)$ (through Eq. (\ref{trans})) are presented. The eccentricities with asterisk are given by Eq. (\ref{eas}) for $\eta=-\infty$.}
\label{tab:4}
\begin{tabular}{cc|c|c|c|c|l}
\hline
\multicolumn{1}{ |c  }{\multirow{2}{*}{$(J_1,p)$} } &
\multicolumn{1}{ |c| }{$\eta_2$} & $\eta_{\rm z}$ & $\eta_{\rm eq}$ & $\eta_{3}$     \\ \cline{2-5}
\multicolumn{1}{ |c  }{}                        &
\multicolumn{1}{ |c| }{$e_{\rm 1,i}(\eta_2)$} & $e_{\rm 1,i}(\eta_{\rm z})$ & $e_{\rm 1,i}(\eta_{\rm eq})$ & $e_{\rm 1,i}(\eta_3)$     \\ \cline{1-5}
\multicolumn{1}{ |c  }{\multirow{2}{*}{$(0.8,0.2)$} } &
\multicolumn{1}{ |c| }{$-$1.64} & -- & $-$1.35 & $-$1.25     \\ \cline{2-5}
\multicolumn{1}{ |c  }{}                        &
\multicolumn{1}{ |c| }{0.39} & -- & 0.17 & 0     \\ \cline{1-5}
\multicolumn{1}{ |c  }{\multirow{2}{*}{$(0.2,0.2)$} } &
\multicolumn{1}{ |c| }{--} & $-$2.07 & $-$1.37 & $-$1.25 \\ \cline{2-5}
\multicolumn{1}{ |c  }{}                        &
\multicolumn{1}{ |c| }{$0.81^*$} & 0.40 & 0.15 & 0 \\ \cline{1-5}
\multicolumn{1}{ |c  }{\multirow{2}{*}{$(0.1,0.9)$} } &
\multicolumn{1}{ |c| }{--} & -- & -- & $-$10 \\ \cline{2-5}
\multicolumn{1}{ |c  }{}                        &
\multicolumn{1}{ |c| }{$0.26^*$} & -- & -- & 0  \\ \cline{1-5}
\end{tabular}

\end{table}

Below, we analyze the three representative models with $(J_1,p)=(0.2,0.8),(0.2,0.2)$ and (0.1,0.9).  Their basic parameters are summarized in Fig. 4.  

\subsubsection{$(J_1,p)=(0.2,0.8)$ in the region II}

In Fig.  12, we provide the rates $v_+$ and $v_-$ for $(J_1,p)=(0.2,0.8)$ that has the pattern P2 during the finite interval $-1.64<\eta<-1.25$ (see Table 4). During this interval, the trajectories with initial eccentricities $0<e_{\rm 1,i}<0.39$ cross the upper separatrix downwardly.  
In Fig. 12, we have the transition epoch $\eta_{\rm eq}=-1.35$ with $e_{\rm 1,i}(\eta_{\rm eq})=0.17$.  Then, as discussed in the previous subsection and shown in Fig. 13,  we have 
\begin{eqnarray}
P_L(e_{\rm 1,i})=\left\{ \begin{array}{ll}
{\rm Eq}.(\ref{eq:prob}) & (0.17<e_{\rm 1,i}<0.39) \\
0 & (0<e_{\rm 1,i}\le0.17). \\
\end{array} \right.
\end{eqnarray}
 Every point in the regions II and IV in Fig. 7 has a bifurcation probability $P_L(e_{\rm 1,i})$ whose profile is similar to Fig. 13.

\begin{figure}
\begin{center}
\includegraphics[width=8cm]{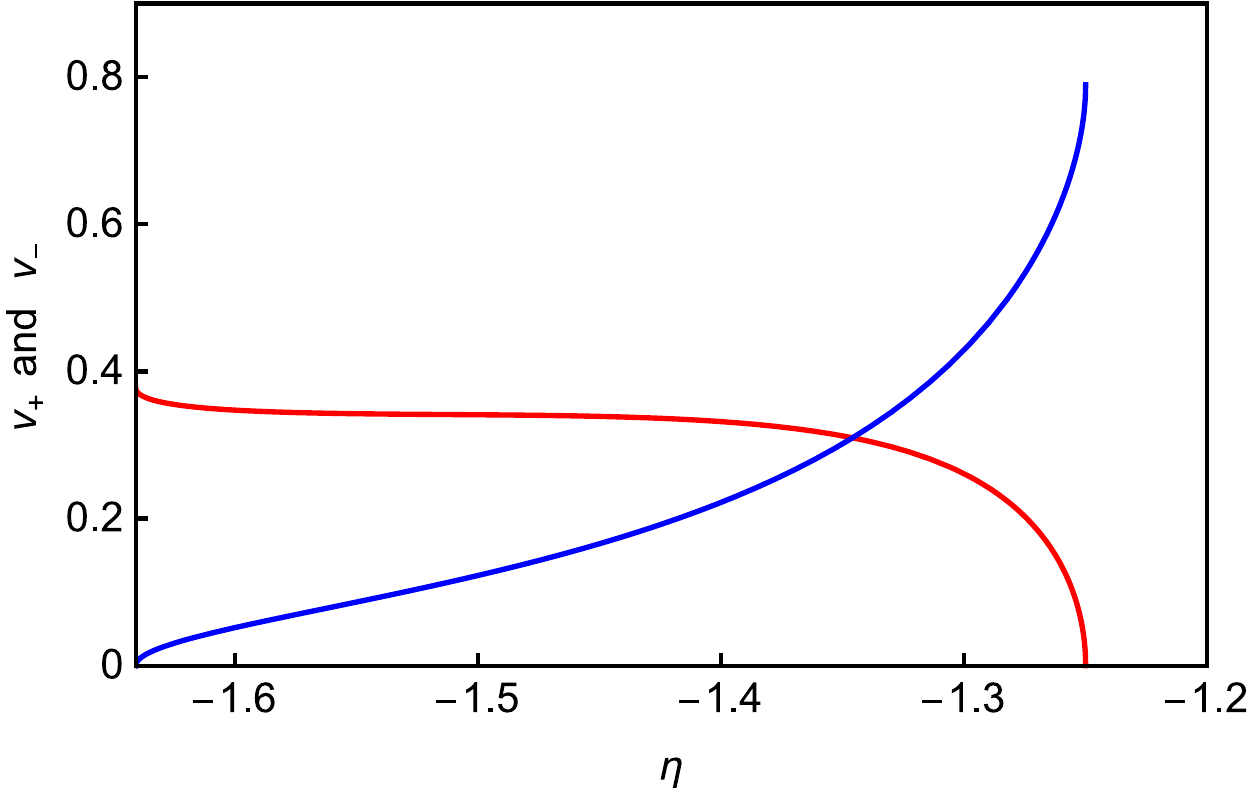}
\caption{The rates $v_+(\eta)$ (red line) and $v_-(\eta)$ (blue line) for  ($J_1$,
$p$)=(0.8, 0.2) in the region II (see Fig.7). }
\label{fig:vel1}
\end{center}
\end{figure}

\begin{figure}
\begin{center}
\includegraphics[width=8cm]{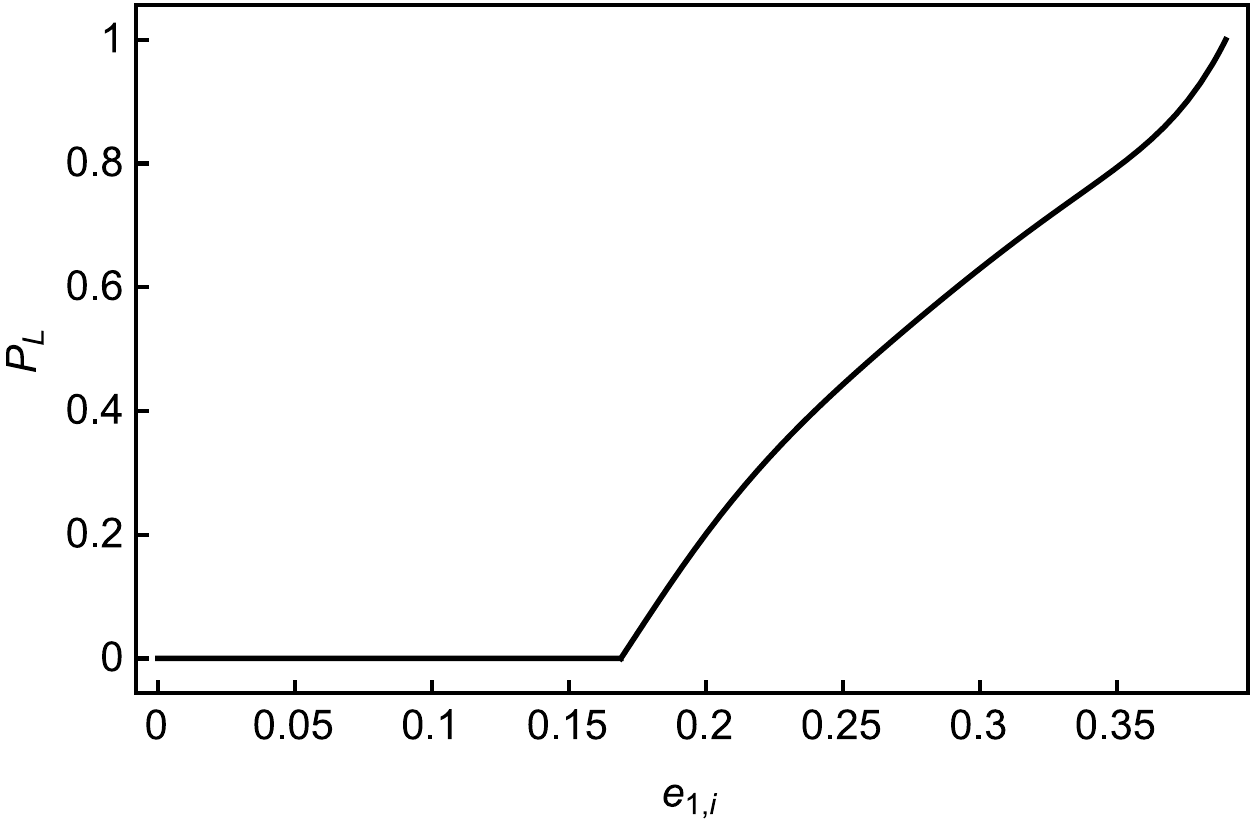}
\caption{The transition probability $P_L$ into libration for ($J_1$, $p$)=(0.8, 0.2).  The horizontal axis represents 
the inner eccentricity $e_{\rm l,i}$ at $\eta \ll -1$.}
\label{fig:prob1}
\end{center}
\end{figure}

\subsubsection{$(J_1,p)=(0.2,0.2)$ in the region III}

Meanwhile, in the regions I and III in Fig. 7, the pattern P2 is realized for $\eta<\eta_3$ without a lower bound, and we have $\lim_{\eta\to -\infty} \tilde{G}_{1, +}(g_1,\eta)=p^{1/3}$ for the upper separatrix (see \S 4.2).  This asymptotic value corresponds to the initial eccentricity 
\beq
e_{\rm 1,i}=\sqrt{1-p^{2/3}}, \label{eas}
\eeq
and is provided in Table 4 with the asterisk $*$. 

In Fig. 14, for $(J_1,p)=(0.2,0.2)$,   we show the rates $v_+$ and $v_-$ at $\eta<\eta_3$.  
Now, Eq. (\ref{eas}) is given as $e_{\rm 1,i}=0.81$, and the trajectories with initial eccentricity $0<e_{\rm 1,i}<0.81$ cross the upper separatrix, during $\eta<\eta_3=-1.25$ (see Table 4).

In Fig. 15,  we provide the probability $P_L(e_{\rm 1,i})$ for these eccentricities. With respect to the cases C1, C2 and C3 explained  in the previous subsection,  we have the two critical epochs $\eta_{\rm z}=-2.07$ and $\eta_{\rm eq}=-1.37$, corresponding to $e_{\rm 1,i}=0.40$ and 0.15 respectively.   Therefore, as shown in Fig. 15, we have
\begin{eqnarray}
P_L(e_{\rm 1,i})=\left\{ \begin{array}{ll}
1 & (0.40\le e_{\rm 1,i}<0.81) \\
{\rm Eq}.(\ref{eq:prob}) & (0.15<e_{\rm 1,i}<0.40) \\
0 & (0<e_{\rm 1,i}\le0.15). \\
\end{array} \right.
\end{eqnarray}

We also mention that the  area of the lower circulating region in the pattern P2 (see Fig. 9) becomes minimum at $\eta_{\rm z}=-2.07$.  This area corresponds to the initial eccentricity $e_{\rm 1,i}=0.91$.

For the parameters $(J_1,p)=(0.2,0.2)$, the characteristic initial eccentricities $e_{\rm 1,i}=0.15,0.40,0.81$ and 0.91 play important roles later in \S 5.5.

\subsubsection{$(J_1,p)=(0.1,0.9)$ in the region III}

For $(J_1,p)=(0.1,0.9)$, in contrast to $(0.2,0.2)$, we identically have $v_-<0$ for $\eta<\eta_3$, as shown in Fig. 16. Therefore, the bifurcation probability becomes $P_L(e_{\rm 1,i})=1$ throughout  $\eta<\eta_3$.

Actually, for the parameters $(J_1,p)$ in the regions I and III,  the characteristic profiles of the rates $(v_+,v_-)$ are either like Fig. 14 or Fig. 16.
In  fact,  it was numerically confirmed that we have at most one solution $\eta_{\rm eq}$ for the equation $v_-(\eta_{\rm eq})=0$.

Then, additionally considering the general profile of the function $v_+(\eta)$ (namely $v_+(\eta)>0$ for $\eta<\eta_3$ and $v_+(\eta_3)=0$), the existence of the probabilistic bifurcation C2 is determined only by the sign of $v_-(\eta_3)$.  For $v_-(\eta_3)<0$, we always have $P_L(e_{\rm 1,i})=1$ at $\eta<\eta_3$, as demonstrated for the example $(J_1,p)=(0.1,0.9)$.  But for $v_-(\eta_3)>0$, we have the probabilistic bifurcation C2, as shown in Figs. 14 and 15 for  $(J_1,p)=(0.2,0.2)$. In the end, based on this criteria $v_-(\eta_3)<0$, we numerically found that, in the regions I-IV shown in  Fig. 7, only the shaded region does not contain the probabilistic bifurcation C2.

\begin{figure}
\begin{center}
\includegraphics[width=8cm]{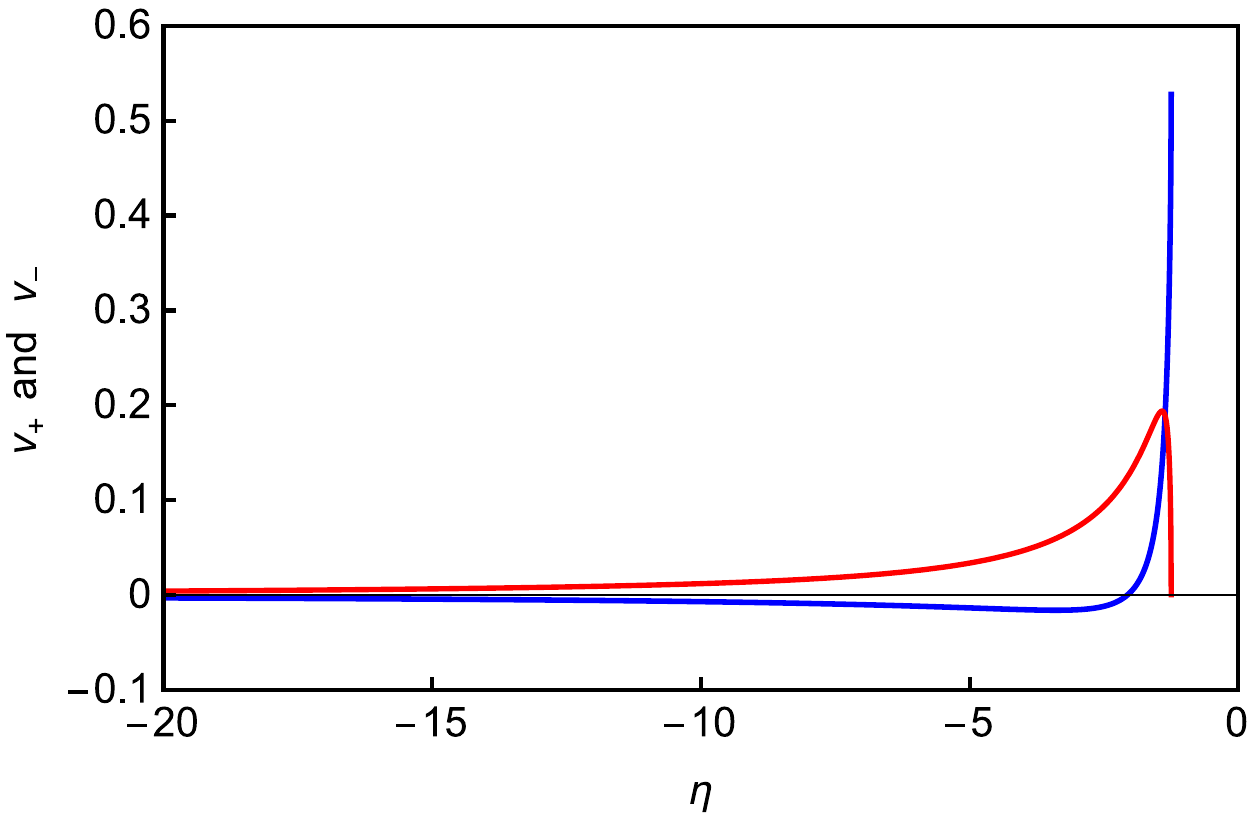}
\caption{Similar to Fig. 12 but for ($J_1$,
$p$)=(0.2, 0.2). The red line is for $v_+(\eta)$  and the blue line is for  $v_-(\eta)$.}
\label{fig:vel2}
\end{center}
\end{figure}

\begin{figure}
\begin{center}
\includegraphics[width=8cm]{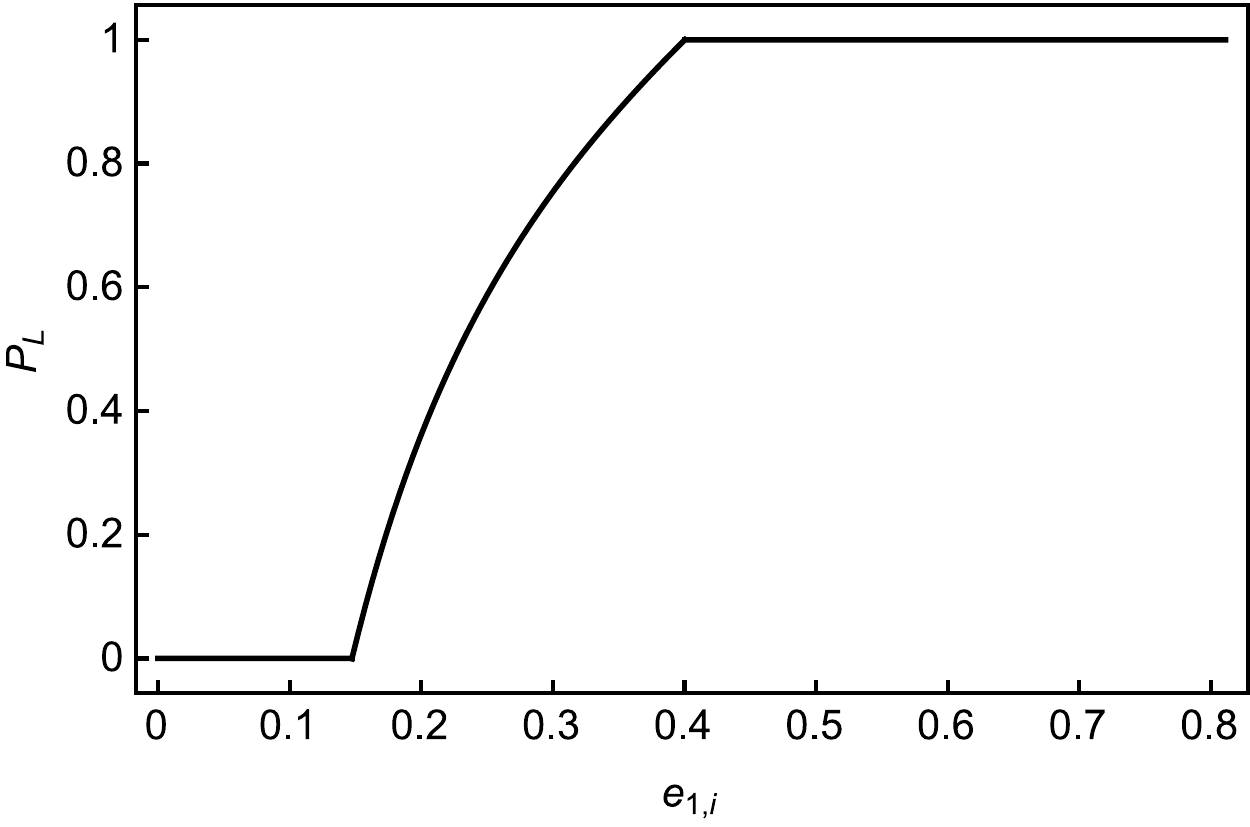}
\caption{The transition probability into libration similar to Fig. 13 but for  ($J_1$, $p$)=(0.2, 0.2).}
\label{fig:prob2}
\end{center}
\end{figure}

\begin{figure}
\begin{center}
\includegraphics[width=8cm]{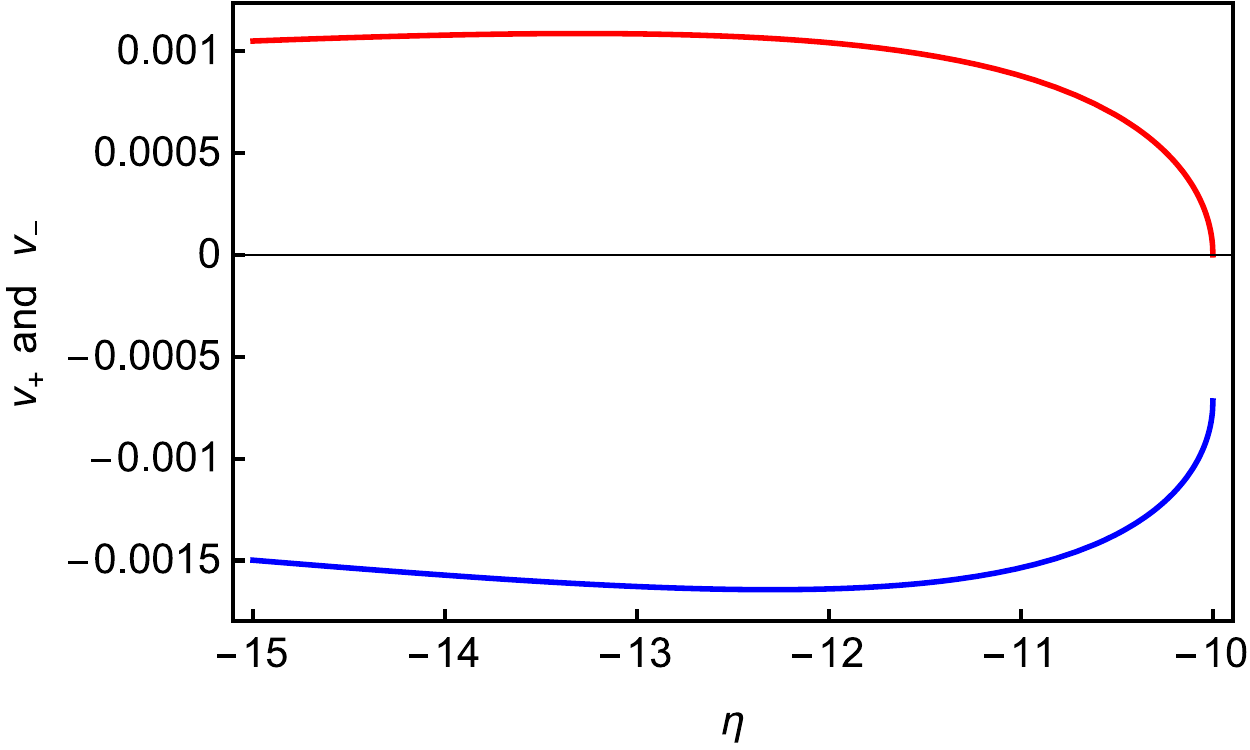}
\caption{Similar to Fig. 12 but for  ($J_1$,
$p$)=(0.1, 0.9). The red line is for $v_+(\eta)$  and the blue line is for  $v_-(\eta)$. We identically have $v_{-}(\eta)<0$ for the pattern P2, resulting in $P_L(\eta)=1$.}
\label{fig:vel3}
\end{center}
\end{figure}


\subsection{maximum eccentricity}

As mentioned earlier, realization of a large eccentricity could result in  astrophysically intriguing phenomenon such as  tidal disruption events and gravitational wave bursts. In this subsection, for each trajectory,  we examine its maximum eccentricity observed in a certain time interval. 

For the two sets of model parameters $(J_1,p)=(0.2,0.2)$ and $(0.15,0.001)$, we basically follow the evolution of the whole trajectories in the phase space from  $\eta=-20$ down to  $-0.7$,  fixing the infall rate at $\dot{\eta}=10^{-2}$. 

At the initial epoch $\eta=-20$, we start the orbital evolution from $(g_1,G_1)=(\pi/2, \sqrt{1-e_{\rm 1,ini}})$ with the input parameter $e_{\rm 1,ini}$.

For each trajectory,  we read the maximum eccentricity  $e_{\rm 1,max}$ during the final rotation period at the termination epoch $\eta=-0.7$.  Both of  the two models have the phase-space pattern P3 (see {\it e.g.} Fig.8d) at $\eta=-0.7$, and the local maximum $e_{\rm 1,max}$ is  easily obtained from  the $G_1$-coordinate of the trajectory,   when taking the phase $g_1=\pi/2$  below the stable fixed point.

Here, we should notice that, using  $e_{\rm 1,ini}$ and  $e_{\rm 1,max}$ respectively, we can uniquely specify the orbital contour in the phase spaces at the two epochs $\eta=-20$ and $-0.7$ (see {\it e.g.} Figs.8b and 8d).  Meanwhile, for each trajectory,  we also define  $e_{\rm 1,max,his}$ as the global maximum  of the eccentricity $e_1$  recorded between $\eta=-20$ and $-0.7$.

\subsubsection{results for $(J_1,p)=(0.2,0.2)$}

In Figs. 17 and 18,  we provide the correspondence  between the initial eccentricity $e_{\rm 1,ini}$ and the final one  $e_{\rm 1,max}$ for $(J_1,p)=(0.2,0.2)$.
From $\eta=-20$ to $-0.7$, the eccentricity of the stable fixed point (at $g_1=\pi/2$) monotonically increases from $e_1=0.81$ to 0.85  (see Fig.9).  Furthermore, for this model parameters, we simply have 
$e_{\rm 1,max,his}=e_{\rm 1,max}$, because of the evolutionary profile of the phase space.  In Fig. 17, we have $e_{\rm 1,max}\ge0.85$, as easily understood from the definition of $e_{\rm 1,max}$.

As discussed in \S 5.3, at $\eta=-\infty$, a trajectory in the phase space moves on a horizontal line characterized by the eccentricity  $e_{\rm 1,i}$.
At $\eta=-20$, a trajectory is no longer a straight line, but we still have 
$e_{\rm 1,ini}\simeq e_{\rm 1,i}$ for each  trajectory.  
In fact, the characteristic eccentricities mentioned in \S 5.4.2 appears clearly in Fig. 18 where we define the end points A, B, D, E, F and the junction point C. 
More specifically, the points A and F have $e_{\rm 1,ini}\simeq e_{\rm 1,i}=0.91$, while B and F have $e_{\rm 1,ini}\simeq e_{\rm 1,i}=0.40$. At the critical epoch  $\eta=\eta_z=-2.07$, these two eccentricities correspond to the circulating trajectories just below the lower separatrix and just above the upper separatrix, respectively (see Fig. 9). 

In Fig. 18,   the point C has  $e_{\rm 1,ini}\simeq e_{\rm 1,i}=0.15$, related to the upper separatrix at $\eta=\eta_z=-1.37$.  The two branches BC and EC are the components of the probabilistic bifurcation discussed in \S 5.4.2.  After $\eta>\eta_z$, the former moved below the lower separatrix.  Meanwhile, the segment EC was captured into the middle libration regime, encircling the branch EF that had  already entered the libration regime at $\eta=\eta_z$.  Therefore, the vertical gaps AF and BE are the same.

Next, by studying the inverse mapping $e_{\rm 1,max}\to e_{\rm 1,ini}$, we can see how the final phase-space is constituted by the initial phase-space elements.  For example, the EF branch is a double value function of  the final quantity $e_{\rm 1,max}$.    This part is originally caused by the blending of two distinct regions in the phase-space, along with the expansion of the middle circulating regime (see Figs. 9a and 9b).\footnote{Strictly speaking, the range $0.85<e_{\rm 1,max}<  0.95$ is already inside the libration region at $\eta=-20$.  The blending between $-20<\eta<-0.7$ is only for the upper part $0.95<e_{\rm 1,max}<  0.964$. }


\begin{figure}
\begin{center}
\includegraphics[width=8cm]{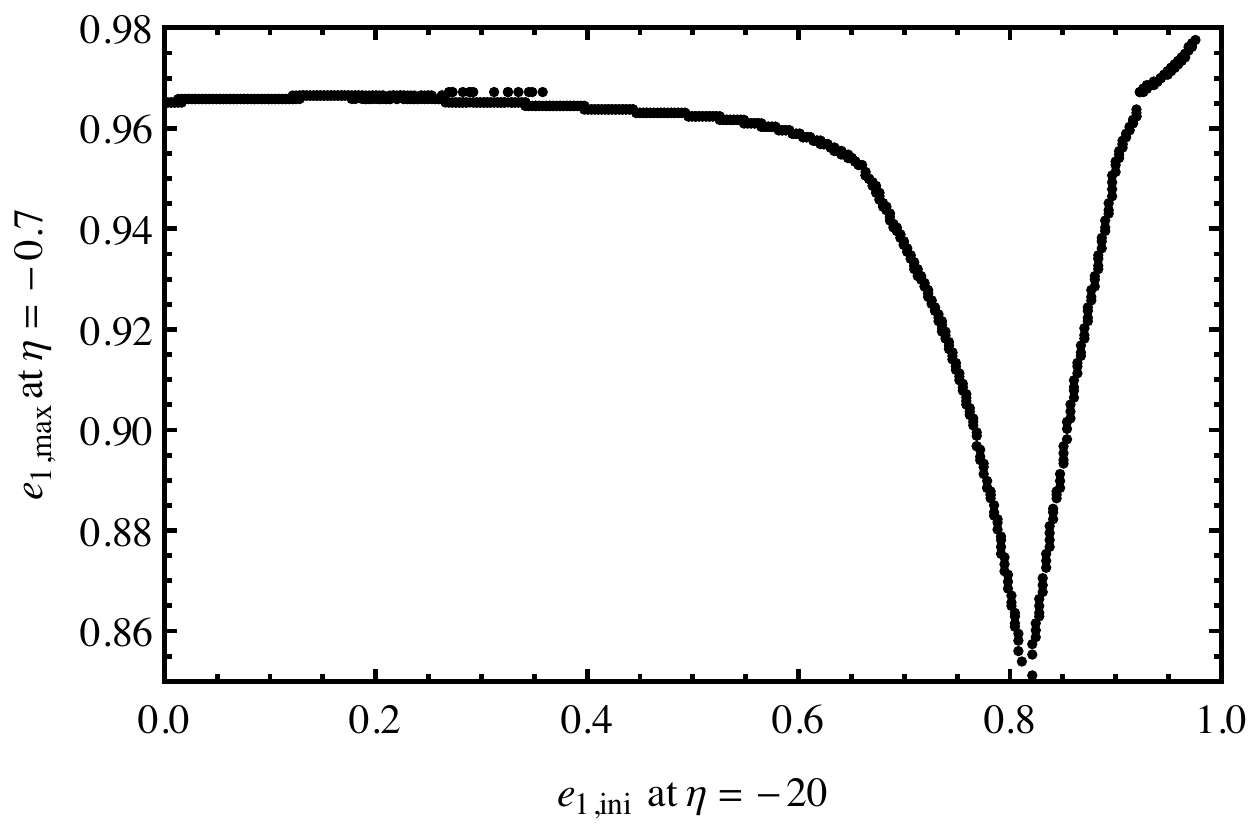}
\caption{The correspondence between the initial eccentricity $e_{\rm 1,ini}$ at $\eta=-20$ and the final eccentricity $e_{\rm 1,max}$  at $\eta=-0.7$.  The initial phase of  the trajectory is $g_1=\pi/2$. The final eccentricity $e_{\rm 1,max}$ is the local maximum that is realized during the final rotation cycle in our phase space.  The model parameters are $(J_1,p)=(0.2,0.2)$.  For this model, the global maximum $e_{\rm 1,max,his}$ is identical to the local one $e_{\rm 1,max}$.}
\label{fig:17}
\end{center}
\end{figure}

\begin{figure}
\begin{center}
\includegraphics[width=8cm]{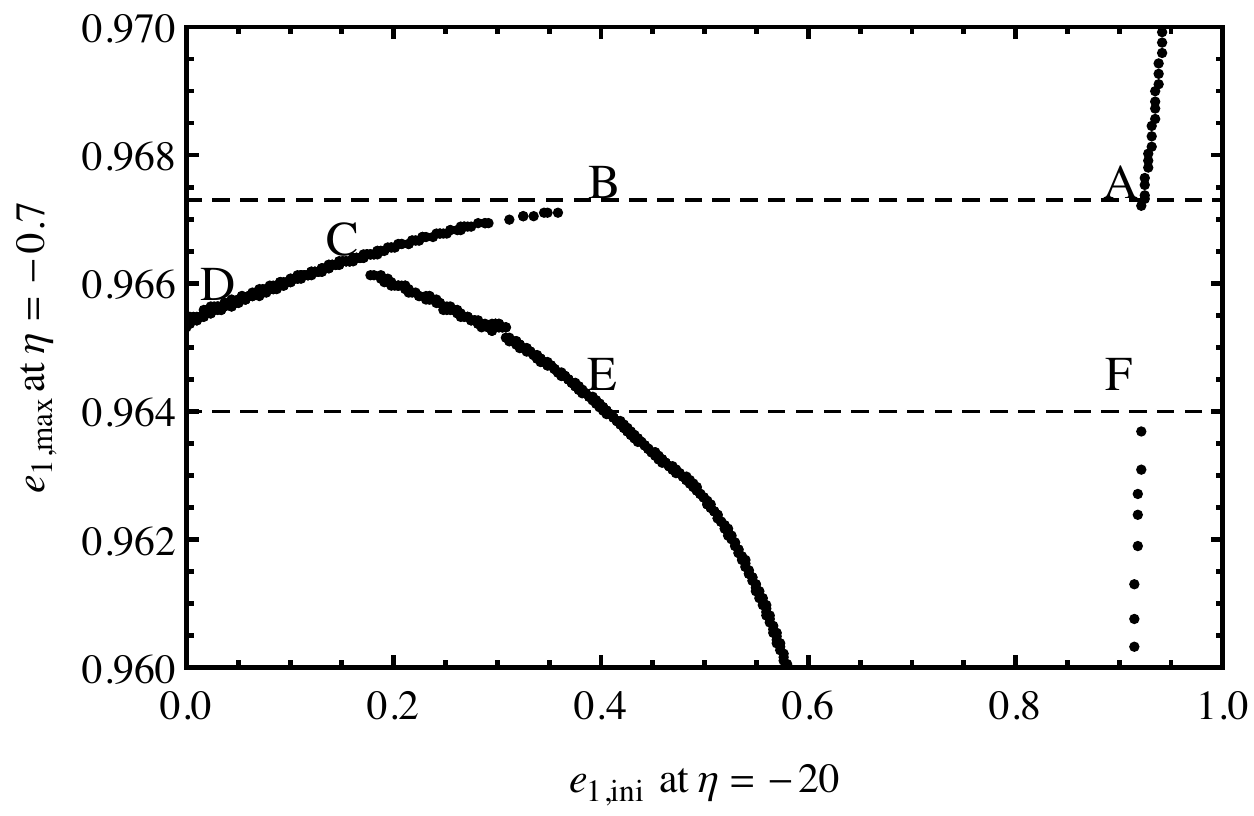}
\caption{The detailed  figure for Fig. 17 around $e_{\rm 1,max}=0.966$. We introduce the labels A,B,D,E and F for the end points, and C for the junction point. }
\label{fig:18}
\end{center}
\end{figure}

\subsubsection{results for $(J_1,p)=(0.15,0.001)$}

In Figs. 19 and 20, we present the locally maximum value $e_{\rm 1,max}$ (black points) at $\eta=-0.7$ and   the globally maximum value $e_{\rm 1,max,his}$ (green points) recorded between $\eta=-20$ and $-0.7$. These are given for $(J_1,p)=(0.15,0.001)$.

For the simpler case $p=0$,  the phase space has the pattern P1 (see {\it e.g.} Fig. 8) in the period $\eta_1<\eta<-1(=\eta_2=\eta_3)$.  The red (upper) separatrix sweeps the whole phase space  in this period. At the separatrix crossing, because of  the profile of the  upper (red) separatrix during P1,  we have $G=J_1$, and all the trajectories temporarily take  $e_1=\sqrt{1-J_1^2}$ which is the allowed maximum value.  The main reason for adopting the present parameter $p=0.001$ here is to examine how this simple result for $p=0$ is modified for a small but finite $p$.

As demonstrated in Fig.19, we actually have
\beq
e_{\rm 1,max,his}\simeq \sqrt{1-J_1^2}=0.9887 
\eeq
for initial eccentricity $ e_{\rm 1,ini}<0.93$.  This result clearly shows that the characteristic motion associated with a separatrix could be an efficient mechanism to realize a large eccentricity and promote the strong interaction between stars and the central black hole. 
In Fig. 19, we should notice that, at $\eta=-20$,  trajectories with  $ e_{\rm 1,ini}>0.93$ are already inside the libration zone around the stable fixed point, and could not preferably cross the separatrix during our calculation.  We also have
\beq
e_{\rm 1,max,his}> e_{\rm 1,max},
\eeq
since the global maximum $e_{\rm 1,max,his}$ is recorded at the separatrix crossing, in contrast to the previous example with $(J_1,p)=(0.2,0.2)$.

In Fig. 20, we take a closer look at $e_{\rm 1,max,his}$ around $\sqrt{1-J_1^2}=0.9887$. We can see a break at 
\beq
e_{\rm 1,ini}\sim 0.57.
\eeq

Actually, for $e_{\rm 1,ini}< 0.57$, the trajectory cross the upper separatrix during the pattern P2 (after $\eta=\eta_2=-1.42$), not P1.  But the lower (blue) separatrix during P2 does not pass the lowest end $G_1=J_1$ of our phase-space \footnote{Fig. 8c is not clear-cut about this. See {\it e.g.} Fig 9c for a more illustrative example.}, resulting in 
\beq
e_{\rm 1,max,his}<\sqrt{1-J_1^2}.
\eeq

 The scatter in Fig. 20 is mainly caused by the finiteness of $\dot \eta$, not by numerical errors.  In fact,  we confirmed that  the scatter is decreased for a slower rate $\dot \eta$.

\begin{figure}
\begin{center}
\includegraphics[width=8cm]{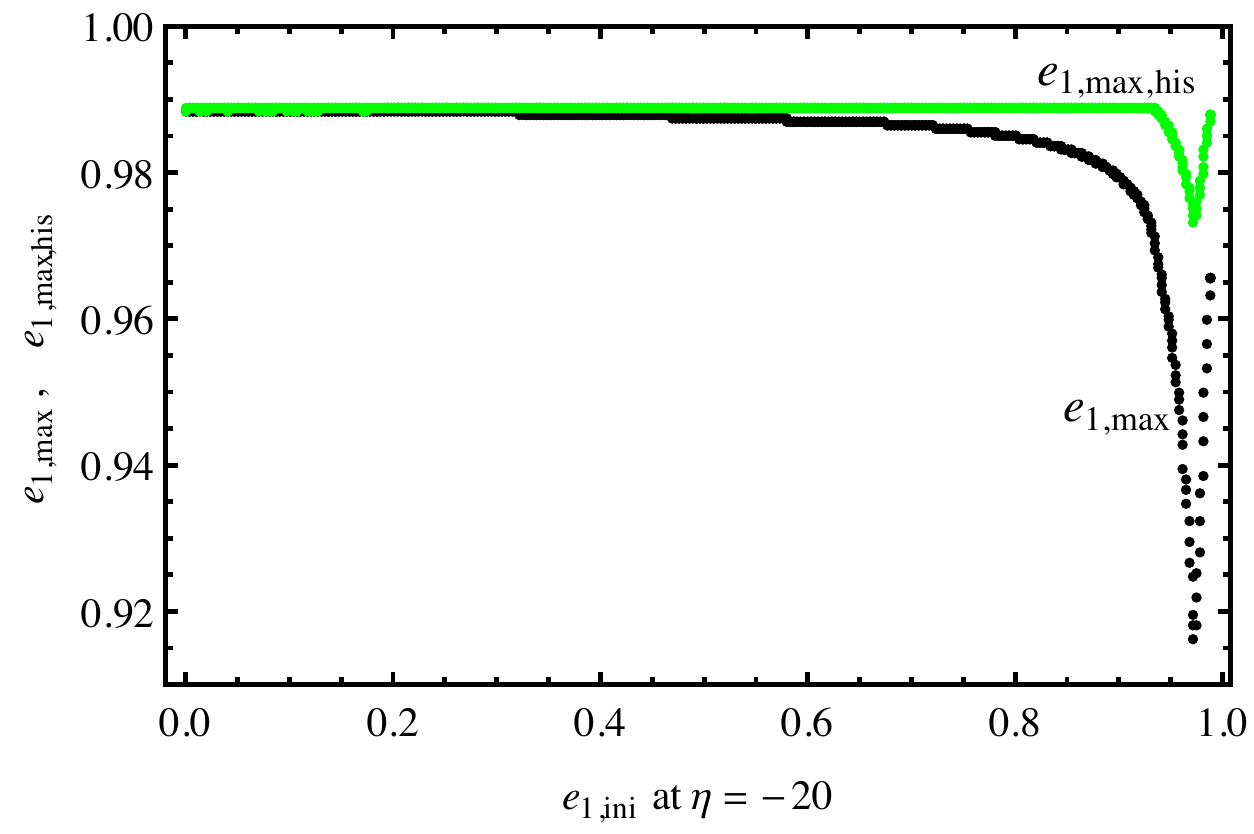}
\caption{Similar to Fig. 19, but for the model parameters  $(J_1,p)=(0.15,0.001)$.  The black points are for the local maximum $e_{\rm 1,max}$ at $\eta=-0.7$, while the green points represent the global maximum $e_{\rm 1,max,his}$ recorded between $\eta=-20$ and $-0.7$.  In contrast to Fig. 17, we now have $e_{\rm 1,max,his}>e_{\rm 1,max}$.}
\label{fig:19}
\end{center}
\end{figure}

\begin{figure}
\begin{center}
\includegraphics[width=8cm]{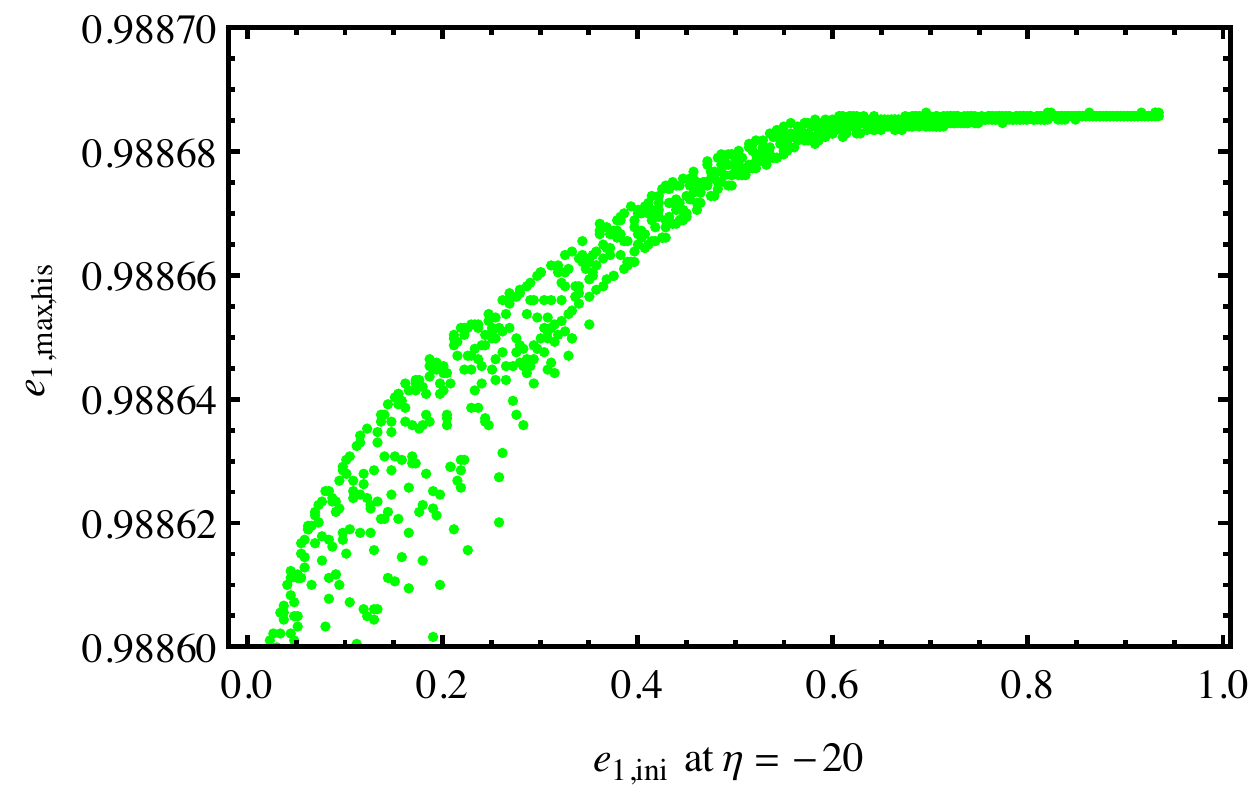}
\caption{The detailed  figure for Fig. 19 around $e_{\rm 1,max}=\sqrt{1-J_1^2}=0.9887$. Only the global maximum $e_{\rm 1,max,his}$ are shown. We have a break around $e_{\rm 1,ini}=0.58$.}
\label{fig:20}
\end{center}
\end{figure}

\section{summary}

Using the framework of the secular theory for a hierarchical triple system, we have studied the long-term orbital evolution of individual stars in a galactic nuclear star cluster to which an secondary MBH is gradually infalling with vanishing eccentricity. Our secular Hamiltonian $\mathcal{H}_{\rm T}(g_{1}, G_{1};\eta)$ is composed by the three terms; the quadrupole gravitational field 
$\mathcal{H}_{\rm qp}$ induced by the outer MBH, the gravitational potential $\mathcal{H}_{\rm SP} $ of the cluster itself,  and the post-Newtonian correction $\mathcal{H}_{\rm 1PN}$ due to the central MBH. This Hamiltonian has two  constant  parameters $(J_1,p)$ and is described by the effective time variable $\eta$.  

As demonstrated in Figs. 3 and 4, 
the eccentricities of  stars in the cluster could show  sharp transitions that depend strongly on their initial conditions. Our primary goal in this paper was to understand the mechanism behind these interesting behaviors, through the phase-space evolution of our Hamiltonian induced by the infalling outer MBH.

To closely examine the phase-space evolution, we first analyzed distribution of fixed points and identified the five critical transitions at  $\eta_i$ ($i=1,\cdots,5$) when their basic properties ({\it e.g.} number, stability) change (see Table 2).   As shown in Fig. 7, the  parameters $(J_1,p)$ determine the combinations of the transitions that are realized during the infall of the secondary MBH. 
Then, we showed that, in the phase-space,  the profile of the separatrixes can be divided into the five types P0 to P4 (see Fig. 8).  The particularly important one P2 is generated by a competition between the prograde apsidal precession enforced by the two terms $\mathcal{H}_{\rm qp}$ and $\mathcal{H}_{\rm 1PN}$ and the retrograde one by the remaining term $\mathcal{H}_{\rm SP}$.

Next, we traced the evolution of individual orbits in the time varying phase-space. Here, we applied a geometrical approach using the adiabatic invariant, and confirmed its validity. Taking a step further, we calculated the branching ratio of a bifurcation at a separatrix crossing that plays a crucial role for the notable behaviors in Figs. 3 and 4.  

 Our analytical studies have been somewhat abstract. But the characteristic behaviors  ({\it e.g.} the sharp probabilistic bifurcations  and the transient realizations of large eccentricities) would be identified in N-body simulations. These are the clear  signatures of the separatrix crossings that are originally induced by the decay of the outer orbit.  As mentioned earlier, in the numerical simulations in \cite{bode2014}, the outer orbit decays faster than the KL oscillations. Meanwhile, in  \cite{li2014}, the outer orbit is fixed. Therefore, it is not surprising that the characteristic behaviors  were not reported in these papers.

In the field of celestial mechanics, geometrical studies similar to this paper have long been made for mean-motion resonances,  including probabilistic bifurcations at the resonant capture \citep{ssd}.  We expect that our analysis for the  secular theory would help us to develop a deep understanding of orbital dynamics related to separatrix crossing, form a wider perspective. 

Our Hamiltonian is a one-dimensional system with the dynamical variables $(g_1,G_1)$. When the outer orbit is eccentric,  the octupole term can enrich the system,  involving the additional  set of conjugate variables $(\Omega_1,J_1)$ for the inner orbit. Here $\Omega_1$ is the longitude of ascending node.  For example, it is well known that the octupole term can generate chaotic behaviors  \citep{naoz2016}. Since a separatrix is also closely related to chaos, it would be interesting to study the effects of the outer eccentricity (for related processes, see \citealt{iwasawa2011,sesana2011,madigan2012, merritt2013,vasiliev2015}).

Our study is based on the secular theory that introduces the averaging operations for the inner and outer orbits. But this prescription is known to break down for highly eccentric inner orbits that would be especially important for astrophysical phenomenon, such as the tidal disruption events or gravitational wave emissions \citep{katz2012,bode2014}.  Direct N-body simulations would be useful to quantitatively examine the related issues and also evaluate the relaxation effects.

\section*{Acknowledgements}

This work is supported by JSPS Kakenhi Grant-in-Aid for  for Scientific Research (No.~15K05075)
and for Scientific Research on Innovative Areas (Nos.~24103006 and 17H06358).

\bibliographystyle{mn2e}
\bibliography{KL}

\appendix
\section{Cluster Potential}

The stellar potential term ${\cal H}_{\rm sp}$ in Eq. (\ref{hs}) is derived under the assumptions that the nuclear stellar cluster is stationary and has isotropic density and velocity distributions up to infinite distance $r=\infty$. However, along with the contraction of the tertiary MBH $m_2$ from the outer part of the cluster, these assumptions are violated, {\it e.g.} due to orbital instabilities.  Consequently,  the stellar density (and thus potential) profile could become non-stationary and  anisotropic.  In this appendix, we briefly discuss how this outer boundary condition affects our Hamiltonian analysis for the inner orbital evolution.

For the stellar cluster, we define $a_c$ as the semimajor axis above which the isotropies are violated. This length scale would be roughly proportional to $a_2$ the distance to the tertiary MBH,  and decrease with time.  Below, without loss of generality, we consider the situation (more specifically the potential term ${\cal H}_{\rm sp}$) at a specific outer distance $a_2$, and thereby fix the characteristic  semimajor axis $a_c$.

First, we should notice that, even if the outer density profile had changed with time, the stars initially at a semimajor axis $x<a_c$ could keep isotropic density and velocity profile, because of the Liouville\rq{}s theorem.  In other words, the number density of stars in our phase-space $(g_1,G_1)$ were initially homogeneous for an isotropic velocity profile, and the density of these stars has not changed with time both in the phase space and the positional space.  

Meanwhile, the stars initially at a semimajor axis $x>a_c$ would now have anisotropic density profile that is axisymmetrical and also plane symmetric with respect to the outer orbital plane.  For simplicity, we ignored the scatterings of semimajor axis from $x>a_c$ to $x<a_c$.

Given the inclination dependence of the orbital stability ({\it e.g.} \citealt{eggleton1995}), the stellar density toward the equatorial direction ($I=\pi/2$) would be different from that of  the polar directions ($I=0$ and $\pi$).
For an inner orbit at $x\ll a_c$, the gravitational effect of the outer anisotropic density profile at $x$ would be approximately given by a positive or negative mass ring on the equator at $x$. Similar to the quadrupole effect of the outer MBH, contributions of these rings at $x>a_c$ would be effectively absorbed into the parameter $\eta$  defined in Eq. (\ref{eta}) for our normalized Hamiltonian ${\cal H}_{\rm T}$. Therefore, our Hamiltonian (\ref{ham}) would work better than natively expected. But, in any cases,  detailed numerical simulations would be helpful  to check validity of discussions in this appendix.

\end{document}